\documentclass[11pt,a4paper]{article}
\pdfoutput=1

\usepackage{dcolumn}
\usepackage{bm}
\usepackage{amsthm}
\usepackage{setspace}
\usepackage[normalem]{ulem}
\usepackage{relsize}
\usepackage{cancel}
\usepackage{verbatim}
\usepackage{enumitem}
\usepackage{mathtools}
\usepackage{empheq}
\usepackage{array}
\usepackage{booktabs}
\usepackage{jheppub}

\allowdisplaybreaks[1]

\usepackage{amsthm}

\newcommand{\del}{\partial}
\newcommand{\Lie}[1]{\mathcal{L}_{#1}\,}

\newcommand{\wn}{\widetilde\nabla}

\newcommand{\delw}{\delta_{{}_W}^B}
\newcommand{\dels}{\delta_\sigma^W}
\newcommand{\delwt}{\tilde \delta_{{}_W}^B}
\newcommand{\Lien}{\Lie{n}}

\newcommand{\tb}{\widetilde \Box}
\newcommand{\te}{\tilde \epsilon}
\newcommand{\teab}{\tilde \epsilon ^{\alpha\beta}}
\newcommand{\temn}{\tilde \epsilon^{\mu\nu}}
\newcommand\numberthis{\addtocounter{equation}{1}\tag{\theequation}}

\newcommand{\tr}{\operatorname{Tr}}

\newcolumntype{I}[1]{>{\centering\arraybackslash$}m{#1}<{$}}
\newlength{\mycolwd}

\title{Non-Relativistic Scale Anomalies}

\author{Igal Arav$^a$,}
\author{Shira Chapman$^b$,}
\author{Yaron Oz$^a$}
\affiliation{
$^a$ Raymond and Beverly Sackler School of Physics and Astronomy, Tel-Aviv University, 55 Haim Levanon street, Tel-Aviv, 69978, Israel, \\
$^b$ Perimeter Institute for Theoretical Physics
31 Caroline Street North, ON N2L 2Y5, Canada
}
\emailAdd{aravigal@post.tau.ac.il}
\emailAdd{schapman@perimeterinstitute.ca}
\emailAdd{yaronoz@post.tau.ac.il}

\abstract{
We extend the cohomological analysis in arXiv:1410.5831 of anisotropic Lifshitz scale anomalies. We consider non-relativistic theories with a dynamical critical exponent $z=2$ with or without non-relativistic boosts and a particle number symmetry.
We distinguish between cases depending on whether the time direction does or does not induce a foliation structure.
We analyse both $1+1$ and $2+1$ spacetime dimensions. In $1+1$ dimensions we find no scale anomalies with Galilean boost symmetries.
The anomalies in $2+1$ dimensions with  Galilean boosts and a foliation structure are all B-type and are identical to the Lifshitz case in the purely spatial sector.
With Galilean boosts and without a foliation structure we find also an A-type scale anomaly.
There is an infinite ladder of B-type anomalies in the absence of a foliation structure with or without Galilean boosts.
We discuss the relation between the existence of a foliation structure and  the causality of the  field theory.
}


\begin{document}

\maketitle
\flushbottom

\section{Introduction}

Scale anomalies in non-relativistic theories have been the subject of recent studies.
Specifically, in the context of Lifshitz field theories satisfying
 an anisotropic scale symmetry of the form:
\begin{equation}\label{eq:AnisotropicScaleSymmetry}
t \rightarrow \lambda^z t, \qquad x^i \rightarrow \lambda x^i, \qquad i=1,\ldots,d,
\end{equation}
it has been found that the allowed structures for scale anomalies satisfying the Wess-Zumino consistency conditions vary depending on the values of the dynamical exponent $z$ and the dimension $d$ \cite{Arav:2014goa}. In all cases studied only B-type anomalies were found \cite{Arav:2014goa}. The anomaly coefficients for $z=2$ in $2+1$ dimensions were computed in particular examples using heat kernel and holography in \cite{Baggio:2011ha,Griffin:2011xs}. The authors of \cite{Jensen:2014hqa,Auzzi} extended the study to non-relativistic theories with Galilean boost symmetry through a null reduction of a relativistic theory in one higher dimension.

A complete classification of scale anomalies consistent with the Wess-Zumino consistency conditions in non-relativistic theories is valuable,
and is likely to have theoretical as well as experimental manifestations (e.g. at quantum critical points). It could, for instance, lead to non-relativistic RG flow theorems for anomaly coefficients \cite{Zamolodchikov:1986gt,Komargodski:2011vj}, or a non-relativistic generalization of the relation between scale and conformal invariance \cite{Dymarsky:2013pqa}.

In this paper we present a complete classification of non-relativistic scale anomalies in various setups, by analysing the appropriate cohomology in a curved  background.
This includes a separate analysis for cases, in which the time 1-form does not induce a foliation structure, i.e., when it does not satisfy the Frobenius condition. In some of the setups studied, we take into account an additional Galilean boost symmetry accompanied by a background $U(1)$ gauge field associated to particle number. We focus our analysis on the case of $z=2$ in $1+1$ and $2+1$ dimensions, however the prescription we present is valid also for other dimensions and the analysis can be extended in a straightforward manner. The generalizations to other values of $z$ are non-trivial since in the case of $z\neq 2$ the gauge and scale transformations do not commute. We leave this for future study.

In order to couple our theory to a curved background and perform the cohomological analysis we use both a spacetime metric $g_{\mu\nu}$ and a 1-form
$t_\mu$ representing the time direction to build our cohomological invariants. Our description is equivalent to the Newton-Cartan geometry \cite{Jensen:2014hqa,Jensen:2014aia, Jensen:2014wha, Son:2005rv, Son:2008ye,
Son:2013rqa, Geracie:2014nka,Bergshoeff:2014uea,Hartong:2015wxa} which we review in subsection \ref{subsec:compNewtonCartan}.

We find that in $1+1$ spacetime dimensions, when we impose Galilean boost symmetry the theory has no anomalies for $z=2$. In $2+1$ dimensions for a $z=2$ Galilean theory we find the following results. When the time direction induces a foliation there are no boost invariants outside the purely spatial
sector.\footnote{Recently, there has been a calculation in \cite{Auzzi} that seems to be in contradiction with our findings. As we will
discuss in section 5.1, the number of independent expressions in our analysis is less than in \cite{Auzzi}, and we therefore expect that not all expressions used for the calculation in \cite{Auzzi} are independent.} In the purely spatial sector the anomalies are identical to the Lifshitz case \cite{Arav:2014goa}. In the absence of a foliation structure there are no boost invariants with less than four derivatives.
In \cite{Jensen:2014hqa} it has been claimed that for the case of $z=2$ Galilean theories, the results can be derived from a null reduction of a relativistic theory in one higher dimension. This implies the existence of an A-type anomaly (in the terminology of \cite{Deser:1993yx}) in $2+1$ dimensions. We indeed establish this using our cohomological analysis directly in 2+1 terms rather than using the null reduction (see equation \eqref{Anoamlies_BNF}). This allows us to directly compare our results to the Lifshitz case.
In all the other cases we consider all anomalies are B-type.
In the cases without a foliation structure (with or without Galilean boosts) there is an infinite ladder of anomalies generated by multiplying B-type anomalies by the spatial anti-symmetric part of the derivative of the normalized time direction.
We discuss possible consequences of an absence of a foliation structure.

This paper is organised as follows. In section \S \ref{sec:CurvedSetup} we describe the various choices that have to be made when coupling our theory to curved spacetime. We include a comparison with the Newton-Cartan geometry of \cite{Jensen:2014aia}. In section \S \ref{sec:CohomSetup} we present the Ward identities and describe the cohomological setup which we use to study the anisotropic scale anomalies. We include in this section a classification by sectors. In sections \S \ref{sec:Results_1+1} and \S \ref{sec:Results_2+1} we detail our results for the various setups in $1+1$ and $2+1$ dimensions with $z=2$. We conclude in section \S \ref{sec:summary} with a list of possible future directions. Most of the technical details are left for appendixes.

\section{Coupling to Curved Spacetime}\label{sec:CurvedSetup}
We consider a non-relativistic theory in $d+1$ dimensions with the anisotropic scale symmetry \eqref{eq:AnisotropicScaleSymmetry}. On a curved spacetime, the relevant geometric data is the metric $g_{\mu\nu}$
or alternatively the vielbeins $e^a{}_\mu$, a 1-form $t_\alpha$ that represents the time direction\footnote{As in the Frobenius case, the 1-form $t_\alpha$ is not unique since a locally rescaled 1-form $f t_\alpha$ represents the same time direction.}
(or the normalized $n_\alpha$: $n_\alpha n^\alpha = -1$) and possibly a $U(1)$ gauge field $A_\mu$ (that corresponds to a conserved particle number).
We extend our previous results from \cite{Arav:2014goa} regarding Lifshitz anomalies both to cases that have in addition non-relativistic Galilean
boost invariance, and to cases in which the time direction is possibly not hypersurface orthogonal, i.e. does not satisfy the Frobenius condition:
\begin{equation}
n_{[\alpha} \nabla_\beta n_{\gamma]} = 0,
\label{F}
\end{equation}
which in the differential form language reads $n\wedge d n = 0$. Note, that even when (\ref{F}) is satisfied we cannot specialize in our analysis to the case
$d n=0$ since this condition is not invariant under anisotropic Weyl transformations.

In our analysis we distinguish between four cases:

\begin{enumerate}
\item {}With Frobenius and Galilean boost invariance -- This case was considered in \cite{Auzzi}. As mentioned in the introduction we disagree with their results.

\item {}With Frobenius and no Galilean boost invariance -- this is the case of Lifshitz field theories,  studied in our previous paper \cite{Arav:2014goa} as well as
\cite{Baggio:2011ha,Griffin:2011xs} and most of past literature on scaling anomalies in Lifshitz field theories. In most of these cases the ADM decomposition is used, which implies that the Frobenius condition is satisfied.

\item {}Without Frobenius and with Galilean boost invariance -- This case was considered in
\cite{Jensen:2014hqa} by relating it to a relativistic case in $d+2$ dimensions via a null reduction.

\item {}Without Frobenius and with no Galilean boost invariance -- This case has not been considered in the literature so far.
\end{enumerate}
Our aim is to compare the cohomology of scaling anomalies in these four cases and consider how the aforementioned choices influence the results.

\subsection{Implications of the Absence of a Foliation Structure}\label{subsec:ImplicationNoFrobeius}

In analysing the cases 3 and 4 outlined above, we will be considering a curved background where the 1-form
$n_{\mu}$ does not satisfy the Frobenius condition (\ref{F}).
Thus, the curved background lacks a foliation structure.
It has been noted in the literature (see e.g. \cite{Geracie:2014nka}), that such cases should be avoided since they imply a breakdown of causality in the non-relativistic
field theory.
The argument is based on Caratheodory's theorem (see e.g. \cite{Frankel} theorem 6.13).
The theorem asserts  that if the Frobenius condition is not satisfied at a point $x$, then there is a neighborhood of $x$ where any point in the
neighborhood can be reached from $x$ by a future directed curve. This implies a lack of causal structure.

On the other hand, the non-relativistic field theories whose scale anomalies we wish to analyse, are defined in flat space which does have
a foliation and therefore has a natural causal structure. The curved background structure to which we couple the theories is only providing
sources to the  various field theory currents. In particular, the 1-form $n_{\mu}$ couples to the field theory energy current.
Imposing the Frobenius condition on the source $n_{\mu}$ means that we do not allow a calculation of correlations functions of all the components
of the energy current. There is, however, no a priori reason for such a requirement and it is not obvious how it follows from any causality requirement
imposed on the non-relativistic field theory in flat space.

What Caratheodory's theorem certainly implies is that one should take care when attempting
to mathematically formulate and calculate the correlation functions of all the components
 of the energy current using the background 1-form source $n_{\mu}$, and it is possible that there is no such consistent framework.
However, we see no such mathematical difficulty when using this source that does not satisfy the Frobenius condition
in the cohomological calculation. As pointed out above, there is a major difference between the scale anomalies depending
on whether the 1-form source satisfies the Frobenius condition or not. While with a Frobenius condition one has only B-type anomalies, we find that
without a Frobenius condition one can have an A-type non-relativistic scale anomaly as well as an infinite number of B-type anomalies.

In studying the non-Frobenius case, we closely follow \cite{Arav:2014goa} in terms of notations.
Although the time 1-form no longer represents a foliation as it does in the Frobenius case, many of the definitions used in the Frobenius case may be easily extended to this case. In particular, one can still decompose any tensor to components which are tangent and normal to the space directions (using the time direction $n_\mu$ and the projector on the space directions $P_{\mu\nu} = g_{\mu\nu} + n_\mu n_\nu$).

Generally, the covariant derivative of the 1-form $n_\alpha$ can be decomposed as follows:
\begin{equation}\label{eq:basic_decomp_normal}
\nabla_\alpha n_\beta = (K_S)_{\alpha\beta}+ (K_A)_{\alpha\beta}-a_\beta n_\alpha,
\end{equation}
where $(K_S)_{\mu\nu}$, $(K_A)_{\mu\nu}$ and $a_\alpha$ are space tangent tensors (normal to the time direction) that satisfy:
\begin{enumerate}
\item $(K_S)_{\mu\nu} = \frac{1}{2}\Lie{n} P_{\mu\nu}$ is symmetric and reduces to the extrinsic  curvature of the foliation in case the Frobenius condition is satisfied.
\item $(K_A)_{\mu\nu} = P_\mu^{\mu'} P_\nu^{\nu'} \nabla_{[\mu'} n_{\nu']}$ is anti-symmetric and vanishes in the Frobenius case.
\item $a_\alpha = \Lie{n} n_\alpha$ is the acceleration vector.
\end{enumerate}
As in the Frobenius case we can still define a space tangent derivative as follows:
\begin{equation}\label{SpaceTangentDerDef}
\wn_\mu \widetilde T_{\alpha\beta\ldots} \equiv P^{\mu'}_\mu P^{\alpha'}_\alpha P^{\beta'}_\beta \ldots \nabla_{\mu'} \widetilde T_{\alpha'\beta'\ldots},
\end{equation}
for $\widetilde T_{\alpha\beta\ldots}$ a space tangent tensor. This definition still satisfies $ \wn_\rho P_{\mu\nu} = 0 $.
Using the commutation of two such space derivatives one can show that for any space tangent vector $\widetilde V_\alpha$:
\begin{equation}
 \left[\wn_\mu, \wn_\nu\right] \widetilde V_{\alpha}  =
 \widetilde R_{\alpha\rho\mu\nu}  {{\widetilde V}^\rho} +
 2K_{\mu\nu}^A\Lie{n} \widetilde V_{\alpha},
\end{equation}
where we have defined:
\begin{equation}
\widetilde R_{\alpha\rho\mu\nu} \equiv P_\alpha^{\alpha'} P_\rho^{\rho'} P_\mu^{\mu'} P_\nu^{\nu'} R_{\alpha'\rho'\mu'\nu'} - 2K^A_{\mu\nu}K_{\alpha\rho}-K_{\mu\alpha}K_{\nu\rho}+K_{\nu\alpha}K_{\mu\rho},
\end{equation}
and:
\begin{equation}
K_{\alpha\beta} \equiv (K_S)_{\alpha\beta}+(K_A)_{\alpha\beta}.
\end{equation}
In the Frobenius case, the tensor $\widetilde R_{\alpha \rho\mu\nu}$ reduces to the intrinsic Riemann tensor of the foliation, however generally it does not have all of the regular symmetries of the Riemann tensor.
It is therefore useful to define a  modified tensor:
\begin{equation}\label{eq:RhatDefinition}
\widehat R_{\alpha\rho\mu\nu} \equiv \widetilde R_{\alpha\rho\mu\nu}
+2K^A_{\mu\nu}K^S_{\alpha\rho}
+K^A_{\mu\alpha}K^S_{\nu\rho}
+K^S_{\mu\alpha}K^A_{\nu\rho}
-K^A_{\nu\alpha}K^S_{\mu\rho}
-K^S_{\nu\alpha}K^A_{\mu\rho},
\end{equation}
which satisfies the usual Riemann tensor symmetries except for the second Bianchi identity, and coincides with $\widetilde R_{\alpha\rho\mu\nu}$ in the Frobenius case.

Many of the identities derived in our previous paper \cite{Arav:2014goa} can be generalized to the non-Frobenius case (see appendix \ref{app:general_ids}). In general these identities will be modified with terms involving $K^A_{\mu\nu}$.
One example is the following set of relations between $a_\alpha$ and $K^A_{\mu\nu}$:
\begin{equation}
 \wn_{[\alpha} a_{\beta]} = \Lie{n} K^A_{\alpha\beta},
 \qquad
\wn_{[\alpha} K^A_{\beta\gamma]} = K^A_{[\alpha\beta}\, a_{\gamma]}.
\end{equation}
Note that $\wn_\mu a_\nu$ is no longer symmetric in this case.

We thus conclude that the main implications of the lack of a foliation structure on our cohomological analysis is in the addition of the extra 2-form $K^A_{\alpha\beta}$ to the list of basic tangent tensors as it appears in \cite{Arav:2014goa}, as well as the appropriate modifications to the various identities satisfied by the basic tangent tensors.

\subsection{Implications of Galilean Boost Symmetry}

The symmetry group of Lifshitz theories is composed of time and space translations, rotations and Lifshitz scaling. The generalization of these symmetries to a curved background is given by symmetries under time-direction-preserving diffeomorphisms (TPD) and anisotropic Weyl transformations (see subsection \ref{subsec:relsymms}).
However, in many cases the non-relativistic theory satisfies the full Schr\"odinger algebra. In curved spacetime we have to consider two additional symmetries. In the terminology of \cite{Jensen:2014aia} these are the Milne boosts and a $U(1)$ gauge symmetry.\footnote{The $U(1)$ symmetry appears as a central extension of the Galilean algebra in the commutation relations of the boosts and translation generators. In the absence of such a gauge field the theory would be a massless non-relativistic theory, which we do not consider physical.}

In the literature, the coupling of Galilean-invariant theories to a curved spacetime is usually implemented using the Newton-Cartan geometry (see \cite{Jensen:2014hqa,Jensen:2014aia, Jensen:2014wha, Son:2005rv, Son:2008ye,
Son:2013rqa, Geracie:2014nka,Bergshoeff:2014uea,Hartong:2015wxa}). However, since our goal in this  work is to compare the anisotropic scaling cohomologies of Galilean and non-Galilean-invariant theories, we find it useful to have a joint framework for the description of both types of theories on a curved background. We do this by including the gauge field associated with the conserved particle number, along with the Milne boost and $U(1)$ gauge symmetries, in our previously developed framework. For our purposes, this description is equivalent to the Newton-Cartan one. We compare our terminology with the Newton-Cartan one in subsection \ref{subsec:compNewtonCartan}.

The first implication of the Galilean symmetry is the presence of an additional gauge field $A_\mu$ as in \cite{Jensen:2014hqa}, associated to the particle number current.
We can decompose the gauge field into space tangent and normal components as follows:
\begin{equation}
A_0 \equiv n^\mu A_{\mu}, \qquad
\tilde A_\mu \equiv P_\mu^{\mu'} A_{\mu'}.
\end{equation}
The gauge invariant data is encoded in the field-strength tensor $F_{\mu\nu}$, or alternatively in the electric and magnetic fields, defined by:
\begin{equation}\label{eq:ElectroMagnetic}
E_\mu \equiv F_{\mu\nu} n^\nu, \qquad B_{\mu\nu} \equiv P_\mu^{\mu'} P_\nu^{\nu'} F_{\mu'\nu'},
\end{equation}
both are space tangent.

The second implication of the Galilean symmetry is the presence of the gauge symmetry and the Milne boost symmetry. In cohomological terms, in this case we are looking at the relative cohomology of the anisotropic Weyl operator with respect to Milne boosts and gauge transformations, in addition to TPD. We are therefore required to restrict the possible terms in the cohomology to ones which are both gauge invariant and Milne boost invariant. The restriction to gauge invariant terms is easily achieved by using the electric and magnetic fields rather than the gauge field itself, but the restriction to boost invariant terms is less obvious. We accomplish it here by starting with all TPD and gauge invariant terms, performing a Milne boost transformation on each of them and finding the combinations which are boost invariant.\footnote{One could also start with basic boost invariant tensors such as $ P_{\mu\nu} + n_\mu A_\nu + n_\nu A_\mu $ and use them to build boost invariant scalars, but these would not be automatically gauge invariant, and one would have to find the gauge invariant combinations. We have taken the other route here.}

In conclusion, the implications of the Galilean symmetry in our analysis is the addition of the electric and magnetic fields to the list of basic tangent tensors as it appears in \cite{Arav:2014goa}, and the restriction of the various terms considered in the analysis to ones which are gauge and Milne boost invariant.

\subsection{The Relevant Symmetries}\label{subsec:relsymms}

In this subsection, we detail the relevant symmetries for our problem and the way in which they act on the various background fields: $g_{\mu\nu}, t_\mu, A_\mu$ or alternatively for theories that require a vielbein formalism $e^a{}_\mu, t^a, A_\mu$. In the following sections,  whenever it is possible, we refer to the most general case - the one that includes the gauge field and does not assume the Frobenius condition -
 and the other cases are inferred by setting the gauge field or
 $ K^A_{\alpha\beta} $
to zero. The cohomological analysis will then be performed for each case separately.

\begin{enumerate}
\item Galilean Boost invariance. That is, invariance under infinitesimal Milne boosts (in the terminology of \cite{Jensen:2014hqa}) which are given by:
\begin{equation}\label{BoostTrans}
\begin{split}
\delw n^\mu &= W^\mu,
\qquad
\delw n_\mu = 0, \\
\delw A_\mu &= -W_\mu,
\qquad
\delw g_{\mu\nu} = W_\mu n_\nu+W_\nu n_\mu,
\end{split}
\end{equation}
where $W^\mu$ is a space tangent ($W^\mu n_\mu = 0$) parameter of the transformation.
Equivalently in terms of the vielbeins:
\begin{equation}
\delw e^a  {}_\mu = W^a n_\mu, \qquad
\delw e_a{} ^\mu  = - W^\mu n_a , \qquad
\delw t^a = \delw n^a = 0.
\end{equation}

This is the curved space version of Galilean boosts:
\begin{equation}
\begin{split}
& x^i \rightarrow x^i + v^i t ,
\qquad
t \rightarrow t ,
\\
& \del_i \rightarrow \del_i ,
\qquad
\del_t \rightarrow \del_t - v^i \del_i .
\end{split}
\end{equation}

\item Gauge invariance. That is, invariance under a standard $U(1)$ gauge symmetry associated with particle number, given by:
\begin{equation}
\delta^G_\Lambda A_\mu = \del_\mu \Lambda, \qquad \delta^G_\Lambda g_{\mu\nu} = \delta^G_\Lambda t_\mu = \delta^G_\Lambda e^a  {}_\mu = 0.
\end{equation}

\item Anisotropic Weyl invariance. That is, invariance under anisotropic Weyl transformations:
\begin{align}\label{eq:RelSymmWeylTranform}
\begin{split}
& \dels t_\alpha    = 0 ,
\qquad
 \dels (g^{\alpha \beta} t_\alpha t_\beta)  = -2\sigma z (g^{\alpha \beta} t_\alpha t_\beta),
\\
& \dels P_{\alpha\beta}  = 2 \sigma P_{\alpha\beta},
\qquad
 \dels n_\alpha  = z  \sigma n_\alpha, \qquad \delta^W_\sigma n^\alpha  = - z  \sigma n^\alpha,
\\
&\dels A_\mu = (2-z) \sigma A_\mu, \\
\end{split}
\end{align}
where $P_{\mu\nu}= g_{\mu\nu} +n_\mu n_\nu$ is the spatial projector and the weight of the gauge field can be determined from the Galilean algebra (more specifically from the commutator of a translation and a Galilean boost).
Alternatively, using the vielbeins:
\begin{align}
\begin{split}
& \dels (n_a e^a{}_\mu) = z \sigma n_a e^a{}_\mu,
\qquad \dels (P_b^a e^b{}_\mu) = \sigma P_b^a e^b{}_\mu,
\\
& \dels t^b = -z\sigma t^b,
\qquad  \dels n^b=0 .
\end{split}
\end{align}

\item Invariance under time direction preserving diffeomorphisms (TPD). These are diffeomorphisms that preserve the time direction, that is, diffeomorphisms with a parameter $\xi$ that obeys $ \Lie{\xi} t_\alpha \propto t_\alpha $. This can be extended to any diffeomorphism by having the time direction 1-form transform appropriately:
\begin{equation}
\begin{split}
\delta^D_\xi g_{\mu\nu} & = \nabla_\mu \xi_\nu + \nabla_\nu \xi_\mu,
\quad
\delta^D_\xi t_\alpha = \Lie{\xi} t_\alpha = \xi^\beta \nabla_\beta t_\alpha + \nabla_\alpha \xi^\beta t_\beta ,
\\
\delta_\xi^D A_\alpha & =  \Lie{\xi} A_\alpha = \xi^\beta \nabla_\beta A_\alpha + \nabla_\alpha \xi^\beta A_\beta,
\end{split}
\end{equation}
or in vielbein formalism (we also include here local Lorentz transformations):
\begin{align}
\begin{split}
\delta^D_\xi e^a{}_\mu & = \xi^\nu \nabla_\nu e^a{}_\mu + \nabla_\mu \xi^\nu e^a{}_\nu,
\qquad
\delta^D_\xi t^a = \xi^\nu \nabla_\nu t^a,
\\
\delta_\xi^D A_\alpha & =  \Lie{\xi} A_\alpha = \xi^\beta \nabla_\beta A_\alpha + \nabla_\alpha \xi^\beta A_\beta,
\\
\delta^L_\alpha e^a{}_\mu & =  - \alpha^a{}_b e^b{}_\mu,
\qquad
\delta^L_\alpha t^a = -\alpha^a{}_b t^b .
\end{split}
\end{align}
\end{enumerate}

Note, that when using the BRST description, one also has to define the action of $\delw$, $\delta^G_\Lambda$, $\delta^W_\sigma$, $\delta^D_\xi$ and $\delta^L_\alpha$ on the Grassmannian parameters $W^\mu$, $\Lambda$, $\sigma$, $\xi^\mu$ and $\alpha^a{}_b$ such that $\delta = \delw+\delta^G_\Lambda+\delta^W_\sigma+\delta^D_\xi+\delta^L_\alpha$ is nilpotent. We detail them here only for the $z=2$ case which is the case we consider in this work:
\begin{align}
\begin{split}
&
\begin{alignedat}{5}
&\delw W^\mu = 0, \quad &
&\delta^G_\Lambda W^\mu = 0, \quad&
&\dels W^\mu = -2 \sigma W^\mu, \quad&
&\delta^D_\xi W^\mu = \Lie{\xi} W^\mu,\quad &
&\delta^L_\alpha W^\mu = 0,
\end{alignedat}
\\
&
\begin{alignedat}{5}
&\delw \Lambda =0, \quad&
&\delta^G_\Lambda  \Lambda = 0, \quad&
&\dels  \Lambda = 0, \quad&
~~~&\delta^D_\xi  \Lambda = \xi^\nu \nabla_\nu \Lambda, \quad&
&\delta^L_\alpha  \Lambda = 0,
\\
&\delw \sigma = 0, \quad&
&\delta^G_\Lambda  \sigma = 0, \quad&
&\dels  \sigma = 0, \quad&
&\delta^D_\xi  \sigma = \xi^\nu \nabla_\nu \sigma, \quad&
&\delta^L_\alpha  \sigma = 0,
\\
&\delw \xi^\mu = 0, \quad&
&\delta^G_\Lambda  \xi^\mu = 0, \quad&
&\dels  \xi^\mu = 0, \quad&
&\delta^D_\xi  \xi^\mu = \xi^\nu \nabla_\nu \xi^\mu , \quad&
&\delta^L_\alpha  \xi^\mu = 0,
\\
&\delw \alpha^a{}_b = 0,\quad &
&\delta^G_\Lambda  \alpha^a{}_b = 0,\quad &
&\dels  \alpha^a{}_b = 0,\quad &
&\delta^D_\xi  \alpha^a{}_b = \xi^\nu \nabla_\nu \alpha^a{}_b,\quad &
&\delta^L_\alpha  \alpha^a{}_b = -\alpha^a{}_c \alpha^c{}_b,
\end{alignedat}
\end{split}
\end{align}
where $\Lie{\xi} W^\mu = \xi^\nu \nabla_\nu W^\mu - \nabla_\nu \xi^\mu W^\nu$.

\subsection{Comparison with the Newton-Cartan Geometry}\label{subsec:compNewtonCartan}
In this subsection we compare our notations and conventions to those of \cite{Jensen:2014aia} in which the non-relativistic setup used a Newton-Cartan (NC) geometry.

The NC geometry is defined in terms of the  spatial metric without a priori referring to an external full spacetime metric (which is ambiguous). The relevant curved data is the spatial metric $h^{\mu\nu}_{NC}$, the local time direction $n_\mu^{NC}$ and the velocity vector $v^\mu_{NC}$ satisfying:
\begin{equation}\label{eq:NC_constrs}
n^{NC}_\mu h_{NC}^{\mu\nu} = 0, \qquad v_{NC}^\mu n^{NC}_\mu =1 ,
\end{equation}
along with the gauge field $A^{NC}_\mu$.
This uniquely defines $h^{NC}_{\mu\nu}$ such that:
\begin{equation}
h_{NC}^{\mu\alpha} h^{NC}_{\alpha\nu} = P^\mu_\nu = \delta^\mu_\nu-v_{NC}^\mu n^{NC}_\nu, \qquad h^{NC}_{\mu\alpha} v_{NC}^\alpha=0 .
\end{equation}
These NC structures relate to our definitions as follows:
\begin{equation}
\begin{split}
h^{NC}_{\mu\nu} &= P_{\mu\nu}, \qquad
n^{NC}_\mu = n_\mu, \qquad
v_{NC}^\mu = -n^\mu, \\
A^{NC}_\mu &= A_\mu, \qquad
g_{\mu\nu} = h^{NC}_{\mu\nu} - n^{NC}_\mu n^{NC}_\nu.
\end{split}
\end{equation}
The infinitesimal Milne boost transformations in the NC framework are given by:
\begin{equation}
\begin{split}
v_{NC}^\mu &\rightarrow v_{NC}^\mu + h_{NC}^{\mu\nu}\psi_\nu,
\\
h^{NC}_{\mu\nu} &\rightarrow h^{NC}_{\mu\nu} - (n^{NC}_\mu P_\nu^\rho + n^{NC}_\nu P_\rho^\mu) \psi_\rho + O(\psi^2),\\
A^{NC}_\mu &\rightarrow A^{NC}_\mu + P_\mu^\nu \psi_\nu + O(\psi^2).
\end{split}
\end{equation}
This transformation corresponds to \eqref{BoostTrans} with $W^\mu \equiv - h_{NC}^{\mu\nu} \psi_\nu$.

Next, we turn to the definition of the covariant derivative. Two choices are common in the NC literature for the affine connection, both of which have non-vanishing torsion. First, the gauge invariant connection, given by:
\begin{equation}
\left(\Gamma^{NC}_{GI}\right)^{\mu}_{~\nu\rho}  = v_{NC}^\mu \del_\rho n^{NC}_\nu + \frac{1}{2} h_{NC}^{\mu\sigma} (\del_\nu h^{NC}_{\rho\sigma} + \del_\rho h^{NC}_{\nu\sigma} - \del_\sigma h^{NC}_{\nu\rho}) +h_{NC}^{\mu\sigma} n^{NC}_{(\nu} F^{NC}_{\rho)\sigma},
\end{equation}
where $F^{NC}_{\rho \sigma}$ is the field strength associated with the $U(1)$ gauge field. The second is the boost invariant connection, given by:
\begin{equation}
\left(\Gamma^{NC}_{BI}\right)^{\mu}_{~\nu\rho}  =
\left(\Gamma^{NC}_{GI}\right)^{\mu}_{~\nu\rho}
+
h_{NC}^{\mu\sigma} \left(-A^{NC}_\sigma \del_{[\rho} n^{NC}_{\nu]} + A^{NC}_\nu \del_{[\rho} n^{NC}_{\sigma]} + A^{NC}_\rho \del_{[\nu} n^{NC}_{\sigma]} \right).
\end{equation}
We, however, use the standard torsionless Levi-Civita connection associated with the metric $g_{\mu\nu}$. The relation between this connection and the gauge invariant NC connection is given by:
\begin{equation}
\left(\Gamma\right)^{\mu}_{~\nu\rho}  =
\left(\Gamma^{NC}_{GI}\right)^{\mu}_{~\nu\rho}
+n^\mu \nabla_\rho n_\nu - P^{\mu\sigma} n_{(\nu} F_{\rho) \sigma}
 - P^{\mu\sigma} n_\rho \nabla_{[\nu} n_{\sigma]} - P^{\mu\sigma} n_\nu \nabla_{[\rho} n_{\sigma]},
\end{equation}
where $\nabla_\mu$ is the covariant derivative associated with the Levi-Civita connection.
Note also that when projected on space tangent directions (as in \eqref{SpaceTangentDerDef}), the Levi-Civita and the gauge invariant NC connections coincide:
\begin{equation}
\tilde\Gamma^\mu_{~\nu\rho} \equiv P^\mu_{\mu'} P^{\nu'}_\nu P^{\rho'}_\rho \Gamma^{\mu'}_{~\nu'\rho'} = P^\mu_{\mu'} P^{\nu'}_\nu P^{\rho'}_\rho \left(\Gamma^{NC}_{GI} \right)^{\mu'}_{~\nu'\rho'},
\end{equation}
whereas the space projected boost invariant NC connection is given by:
\begin{equation}
P^\mu_{\mu'} P^{\nu'}_\nu P^{\rho'}_\rho \left(\Gamma^{NC}_{BI} \right)^{\mu'}_{~\nu'\rho'} = \tilde\Gamma^\mu_{~\nu\rho} - \tilde{A}^\mu K^A_{\rho\nu} + (K^A)_\rho^{~\mu}\tilde{A}_\nu + (K^A)_\nu^{~\mu}\tilde{A}_\rho .
\end{equation}

We would like to stress that the choice of connection is not essential to the cohomological analysis we are performing:
The algebra of the various symmetries we are considering does not depend on the choice of connection. In addition, since the difference between the Levi-Civita connection and the NC ones is a tensor that is composed of the basic background fields (the metric, the time direction and the gauge field), any scalar written using one connection can be decomposed as a combination of scalars written using the other. The number of independent invariant expressions therefore does not depend on the choice of connection either. While the Levi-Civita connection may be considered a less ``natural'' choice for the Galilean case, it is nevertheless more convenient for our calculations, and for the purpose of comparing them to the non-Galilean cases. The results can always be translated to the NC formalism using the above formulas.

Finally, we compare the various field theory currents that couple to the background fields, as defined in \cite{Arav:2014goa} and in subsection \ref{subsec:WardIdents}, to the NC ones as defined \cite{Jensen:2014aia} (see also \cite{Hartong:2015wxa}). The NC currents are defined from the variation of the action as follows:
\begin{equation}
\delta S = \int \sqrt {-g}
\left[
\delta A^{NC}_\mu J_{NC}^\mu
-\delta \bar v_{NC}^\mu \mathcal{P}^{NC}_\mu
-\delta n^{NC}_\mu \mathcal{E}_{NC}^\mu - \frac{\delta \bar h_{NC}^{\mu\nu}}{2} T^{NC}_{\mu\nu} \right],
\end{equation}
where $J_{NC}^\mu$, $\mathcal{P}^{NC}_\mu $,$\mathcal{E}_{NC}^\mu $ and $ T^{NC}_{\mu\nu} $ represent the particle number current, the momentum density, the energy current and the spatial stress tensor respectively.\footnote{In this definition, the variation of $n^{NC}_\mu$ is unconstrained, whereas $v_{NC}^\mu$ and $h_{NC}^{\mu\nu}$ are constrained to satisfy  equation \eqref{eq:NC_constrs}. $\delta \bar v_{NC}^\mu$ and $\delta \bar h_{NC}^{\mu\nu}$ represent variations of $v_{NC}^\mu$ and $h_{NC}^{\mu\nu}$ in the spatial directions, which are unconstrained.}
The corresponding currents in our conventions are defined via:
\begin{equation}
\delta S = \int \sqrt {-g}
\left[
\frac{1}{2} T^{\mu\nu}_{(g)} \delta g_{\mu\nu} + J^\alpha \delta t_\alpha +J_m^\alpha \delta A_\alpha \right].
\end{equation}
The relations between the currents in our conventions and the NC ones are then given by:
\begin{equation}
\begin{split}
T^{\mu\nu}_{(e)} =& T_{NC}^{\mu\nu}
-n^\mu P^{\nu\rho} \mathcal{P}^{NC}_\rho - n^\nu \mathcal{E}_{NC}^\mu,
\\
\hat{J}^\mu =&  P^{\mu\rho} \mathcal{P}^{NC}_\rho - P^\mu_\rho \mathcal{E}_{NC}^\rho ,
\\
J_m^\mu =& J_{NC}^\mu ,
\end{split}
\end{equation}
where $T^{\mu\nu}_{(e)} = T^{\mu\nu}_{(g)} + J^\mu t^\nu$ is the stress energy tensor associated with the vielbeins, and $ \hat{J}^\alpha \equiv \sqrt{-g^{\mu\nu}t_\mu t_\nu}\,J^\alpha \ $ as defined in subsection \ref{subsec:WardIdents}.

 \section{The Cohomological Problem}\label{sec:CohomSetup}
Our main goal is to find the possible anomalous contributions to the Ward identity that corresponds to Lifshitz scale symmetry in the various cases outlined in section \ref{sec:CurvedSetup}, by finding the non-trivial solutions of the Wess-Zumino consistency conditions. As in \cite{Arav:2014goa}, we use the cohomological description of the problem, in terms of a BRST-like ghost. In this description one studies the relative cohomology of the nilpotent anisotropic Weyl operator $\delta_
\sigma^W $ with respect to the other symmetries of the problem. The possible anomalies are terms $A_\sigma$ of ghost-number 1 and with the right scaling dimension which are cocycles (i.e. satisfy the WZ consistency conditions):
\begin{equation}
\delta_\sigma^W A_\sigma = 0,
\end{equation}
with $\sigma$ a Grassmannian transformation parameter, and are not coboundaries (i.e. cannot be canceled  by an appropriate counterterm):
\begin{equation}
A_\sigma \neq \delta_\sigma^W G(\{F\}),
\end{equation}
for $G$ a local functional of the background fields, where both $A$ and $G$ are invariant under the rest of the symmetries of the problem.

\subsection{Ward Identities}\label{subsec:WardIdents}
We start by studying the relevant Ward identities associated with the symmetries of subsection \ref{subsec:relsymms}. Here again we refer to the most general case,
with non-relativistic boost invariance, a gauge field and in which the time direction is not hypersurface orthogonal.
Assume a classical action $ S(g_{\mu\nu}, t_\alpha, A_\alpha, \{\phi\}) $  or alternatively $ S(e^a{}_\mu, t^b, A_\alpha, \{\phi\}) $, where $\{\phi\}$ are the dynamic fields.
Define the various currents as follows. The stress energy tensor:
\begin{equation}
T^{\mu\nu}_{(g)} \equiv \left. \frac{2}{\sqrt{-g}} \frac{\delta S}{\delta g_{\mu\nu}} \right|_{t_\alpha, A_\alpha} ,
\qquad
T_{(e)} {}^\mu {}_a \equiv \left. \frac{1}{e} \frac{\delta S}{\delta e^a{}_\mu} \right|_{t^a, A_\alpha}
.
\end{equation}
The variation of the action with respect to the time direction 1-form:
\begin{equation}
J^\alpha \equiv \left. \frac{1}{\sqrt{-g}}\frac{\delta S}{\delta t_\alpha} \right|_{g_{\mu\nu}, A_\alpha} =
\frac{1}{e} e^{b\alpha} \left. \frac{\delta S}{\delta t^b} \right|_{e^a{}_\mu,A_\alpha} ,
\end{equation}
as well as its normalized version:
\begin{equation}
\hat{J}^\alpha \equiv \sqrt{|g^{\mu\nu}t_\mu t_\nu|}J^\alpha \, ,
\end{equation}
and the mass current, given by:
\begin{equation}
J_m^\alpha \equiv \left. \frac{1}{\sqrt{-g}}\frac{\delta S}{\delta A_\alpha} \right|_{g_{\mu\nu}, t_\alpha} =
\frac{1}{e} \left. \frac{\delta S}{\delta A_\alpha} \right|_{e^a{}_\mu,t_\alpha} .
\end{equation}
Note that $ J^\alpha $ is space tangent, i.e.
$ J^\alpha t_\alpha = 0 $, since the action is invariant under local rescaling of the time direction 1-form.
In cases where one can use either the metric or the vielbein descriptions, the following relation exists between $ T_{(g)}^{\mu\nu} $, $ T^{\mu\nu}_{(e)} \equiv T_{(e)}{}^\mu{}_a e^{a\nu} $ and $ J^\alpha $:
\begin{equation}
T_{(e)}^{\mu\nu} = T_{(g)}^{\mu\nu} + J^\mu t^\nu .
\end{equation}

For time direction preserving diffeomorphisms (TPD), the corresponding Ward identities are given by:
\begin{equation}
\begin{split}
\nabla_\mu T_{(g)}^\mu{}_\nu &=
J^\mu \nabla_\nu t_\mu - \nabla_\mu(J^\mu t_\nu)
+J^\mu_m \nabla_\nu A_\mu - \nabla_\mu(J_m^\mu A_\nu) \\
&= \hat{J}^\mu \nabla_\nu n_\mu - \nabla_\mu(\hat{J}^\mu n_\nu)
+J^\mu_m \nabla_\nu A_\mu - \nabla_\mu(J_m^\mu A_\nu) ,
\end{split}
\end{equation}
or equivalently in terms of $T_{(e)}^{\mu\nu}$:
\begin{align}
\begin{split}
T_{(e)[\mu \nu]} &= J_{[\mu}t_{\nu]} = \hat J_{[\mu}n_{\nu]} ,\\
\nabla_\mu T_{(e)}{}^\mu{}_\nu &= J^\mu \nabla_\nu t_\mu + J^\mu_m \nabla_\nu A_\mu - \nabla_\mu(J_m^\mu A_\nu)\\
&= \hat{J}^\mu \nabla_\nu n_\mu + J^\mu_m \nabla_\nu A_\mu - \nabla_\mu(J_m^\mu A_\nu).
\end{split}
\end{align}
The Ward identity corresponding to the $U(1)$ gauge invariance is simply the conservation of the current $J^\mu_m$:
\begin{equation}
\nabla_\mu J^\mu_m = 0.
\end{equation}
Using it we get a simplified version of the previous ward identities:
\begin{align}
\begin{split}
\nabla_\mu T_{(g)}^\mu{}_\nu &=
\hat{J}^\mu \nabla_\nu n_\mu - \nabla_\mu(\hat{J}^\mu n_\nu) +J^\mu_m F_{\nu\mu},
\\
\nabla_\mu T_{(e)}{}^\mu{}_\nu &= \hat{J}^\mu \nabla_\nu n_\mu +J^\mu_m F_{\nu\mu}.
\end{split}
\end{align}
The Ward identity corresponding to anisotropic Weyl symmetry is given by:
\begin{equation}\label{ward:WeylWardIdent}
\begin{split}
D & \equiv T_{(g)}^{\mu\nu} P_{\mu\nu} - z T_{(g)}^{\mu\nu}n_\mu n_\nu +  \frac{2-z}{2} J^\mu_m A_\mu
\\
& = T_{(e)}^{\mu\nu} P_{\mu\nu} - z T_{(e)}^{\mu\nu}n_\mu n_\nu  + \frac{2-z}{2} J^\mu_m A_\mu= 0 ,
\end{split}
\end{equation}
And finally, for the boost invariant cases, the Ward identity corresponding to Milne boosts is given by:
\begin{equation}
P_{\mu\alpha} T^{\mu\nu}_{(g)} n_\nu = P_{\alpha \beta} J^\beta_m,
\end{equation}
or alternatively:
\begin{equation}
P_{\nu\alpha} T^{\mu\nu}_{(e)} n_\mu = P_{\alpha\beta} J^\beta_m,
\end{equation}
which is the famous statement of equality between particle number current and momentum density.

In this work we study the possible form of the anomalous corrections to the anisotropic Weyl Ward identity \eqref{ward:WeylWardIdent}, assuming the other symmetries are not anomalous.

\subsection{Constructing Time Direction Preserving Diffeomorphism Invariants}
\label{subsec:TPD}
As explained in our previous paper \cite{Arav:2014goa}, the cohomological analysis starts by constructing all possible TPD invariant expressions of a certain Lifshitz scaling dimension. This can be accomplished by taking all possible contractions of a set of basic space tangent tensors.
A tensor $ \widetilde{T}_{\alpha\beta\gamma\ldots}$ is called space tangent if
\begin{equation}
 n^\alpha \widetilde{T}_{\alpha\beta\gamma\ldots} = n^\beta \widetilde{T}_{\alpha\beta\gamma\ldots} = \ldots = 0 .
\end{equation}
In the case without a foliation structure we have to add $K^A_{\mu\nu}$ to the list of basic tangent tensors.
In the cases with Galilean boost-invariance, the electric and magnetic fields $E_\mu$, $B_{\mu\nu}$ associated with the $U(1)$ gauge symmetry are also included. The list of basic tangent tensors then becomes:
\begin{enumerate}
\item The spatial metric $ P_{\mu\nu} = g_{\mu\nu} + n_\mu n_\nu $.
\item The acceleration vector $ a_\mu \equiv \Lie{n} n_\mu = n^\nu \nabla_\nu n_\mu$.
\item The tensors $K^A_{\mu\nu}$, $K^S_{\mu\nu}$ as defined in subsection \ref{subsec:ImplicationNoFrobeius}.
\item The modified ``intrinsic'' Riemann tensor $\widehat R_{\mu\nu\rho\sigma}$ as defined in  equation \eqref{eq:RhatDefinition}.
\item The space tangent Levi-Civita tensor:
$\tilde \epsilon^{\mu\nu\rho \ldots} =
n_\alpha \epsilon^{\alpha\mu\nu\rho\ldots}$.
\item Lie derivatives (temporal derivatives) in the direction of $ n^\alpha $: $ \Lie{n}$. Note that if some tensor $\widetilde T_{\alpha\beta\ldots} $ is space tangent, then $ \Lie{n} \widetilde T_{\alpha\beta\ldots} $ is also space tangent.
\item Space tangent covariant derivatives: $\wn_\mu \widetilde T_{\alpha\beta\ldots} \equiv P^{\mu'}_\mu P^{\alpha'}_\alpha P^{\beta'}_\beta \ldots \nabla_{\mu'} \widetilde T_{\alpha'\beta'\ldots}$ for $\widetilde T_{\alpha\beta\ldots}$ a space  tangent tensor, as defined in equation \eqref{SpaceTangentDerDef}.
\item The electric field $E_\mu$ and the magnetic field $B_{\mu\nu}$ as defined in equation \eqref{eq:ElectroMagnetic}.
\end{enumerate}

The various tensors were chosen such that they scale uniformly under anisotropic Weyl scaling transformation  with a  scaling dimension $d_\sigma$:
\begin{equation}\label{FPD_Objs:unif}
\delta^W_\sigma \mathcal{O} = (d_\sigma) \sigma \mathcal{O} + (\del \sigma)  ,
\end{equation}
where $\sigma(x)$ is the Grassmannian local parameter of the anisotropic Weyl transformation and $\del \sigma$ stands for any term proportional to derivatives of the ghost $\sigma$.
The basic tangent tensors have the following scaling dimensions:
\begin{align}\label{FPD_Objs:scalingdims}
&d_\sigma [P_{\mu\nu}]  =2,
& &d_\sigma [P^{\mu\nu}]  =-2,
& &d_\sigma [\widetilde{R}_{\mu\nu\rho\sigma}] = d_\sigma [\widehat{R}_{\mu\nu\rho\sigma}]  =2, \notag
\\
&d_\sigma [\tilde\epsilon_{\alpha\beta\ldots}] = d,
& & d_\sigma [\tilde\epsilon^{\alpha\beta\ldots}] = -d,
& &d_\sigma [K^S_{\mu\nu}] = 2-z,
\\
&
d_\sigma [K_{\mu\nu}^A] = z,
&& d_\sigma [\Lie{n} \widetilde T_{\alpha\beta\ldots}] = d_\sigma[\widetilde T_{\alpha\beta\ldots}] - z,
& & d_\sigma [\wn_\alpha \widetilde T_{\alpha\beta\ldots}] = d_\sigma [\widetilde T_{\alpha\beta\ldots}], \notag
\\
&
d_\sigma [a_\alpha] = 0,
 &&
d_\sigma [E_\mu] = 2-2z, &&
\notag
d_\sigma [B_{\mu\nu}] = 2-z,
\notag
\end{align}
where $ \widetilde T_{\alpha\beta\ldots} $ is any space tangent tensor with uniform scaling dimension.

Various relevant identities for the basic tangent tensors are listed in appendix \ref{app:general_ids}.
The complete boost and Weyl transformation rules for these tensors are listed in appendixes \ref{App:BoostTrans}-\ref{app:WeylTrans}.

\subsection{Classification by Sectors}\label{subsec:ClassSectors}

The various terms in the cohomology all have the form $ \int \sqrt{-g}\phi $, where $ \phi $ is a scalar of uniform scaling dimension $ -(d+z) $, built from contractions of the basic tangent tensors of subsection \ref{subsec:TPD}. Suppose that $ n_{K_S} $, $ n_{K_A} $, $n_a$, $n_R$, $n_\epsilon$, $ n_\nabla$, $ n_\mathcal{L} $, $n_B$ and $n_E$ are the number of instances of the various basic tangent tensors (as indicated by the subscript)
that appear in $\phi$, and $ n_P $ the number of spatial metric instances required to contract them.
For the scaling dimension to be correct we require:
\begin{equation}\label{rest:scalingreq}
(2-z)n_{K_S}+zn_{K_A}-z n_\mathcal{L} +2n_R + d n_\epsilon+(2-z) n_B +(2-2z) n_E -2  n_P = -d-z .
\end{equation}
For all indices in $\phi$ to be contracted in pairs we require:
\begin{equation}\label{rest:indexreq}
2 n_{K_A} + 2 n_{K_S} + n_a + n_\nabla + 4n_R + d n_\epsilon +n_E +2n_B = 2 n_P  .
\end{equation}
From requirements \eqref{rest:scalingreq} and \eqref{rest:indexreq} we obtain the conditions:
\begin{gather}\label{rest:constraint1}
z [ n_{K_S} - n_{K_A} +n_\mathcal{L}  +n_B +2n_E] +2 n_{K_A} + n_a + n_\nabla + 2 n_R  - n_E = d+z,
\\
\label{rest:constraint2}
n_a+n_\nabla + d n_\epsilon +n_E \quad \text{is even} .
\end{gather}
If we define $ n_T \equiv n_{K_S} - n_{K_A} +n_\mathcal{L}  +n_B +2n_E $ as the total number of time derivatives and $ n_S \equiv 2 n_{K_A} + n_a + n_\nabla + 2 n_R  - n_E $ as the total number of spatial derivatives in the expression,\footnote{In the non-Frobenius case, the numbers $n_T$ and $n_S$ do not have an obvious interpretation as the total number of derivatives in the time and space directions (they may even become negative). However, it is still useful to define them for the classification to sectors. With a slight abuse of language we will keep referring to them as the total number of time and space derivatives respectively.} we  get the following form for these conditions:
\begin{gather}\label{rest:constraint1b}
z n_T + n_S = d+z, \\
\label{rest:constraint2b}
n_S + d n_\epsilon \quad \text{is even} .
\end{gather}
The numbers $ n_T $, $ n_S $ and $ n_\epsilon $ remain unchanged when applying the Weyl operator $ \delta^W_\sigma $ to any tangent tensor or when using identities relating different tangent tensors.
We therefore classify expressions  according to sectors, each corresponding to specific values of
$ (n_T,n_S,n_\epsilon) $. When studying the cohomological problem we may focus on each sector separately. We also define for convenience the total number of derivatives:
\begin{equation}
n_D = n_T + n_S,
\end{equation}
which is always positive, unlike $n_T$ and $n_S$.

Note that the time reversal and parity properties of the expressions in a certain sector are given by:
\begin{equation}\label{rest:pt_props}
T = (-1)^{n_T}, \qquad P = (-1)^{n_\epsilon}.
\end{equation}

An important difference in the classification to sectors from the case studied in \cite{Arav:2014goa} is that the negative contributions to equation \eqref{rest:constraint1b} allow in some cases for an infinite number of sectors. For $z$ integer, the total contribution of the electric field is always positive. This is not the case for the contribution of $K_A$.
In the case of $z=2$ for instance, the total contribution of $K_A$ is vanishing hence allowing for an infinite number of sectors, each corresponding to a different number of derivatives $n_D$. Each of these sectors may (and in fact does, as we show in the following sections) contain a different set of possible independent anomalies. Thus we conclude that a direct consequence of discarding the Frobenius condition is the possibility of having an infinite set of independent anomalous contributions to the Ward identities of the theory.

A final remark is in order, regarding the cases with Galilean boost invariance. As mentioned in section \ref{sec:CurvedSetup}, these cases require finding the combinations of expressions which are invariant under Milne boosts. In order to make use of the classification to sectors for this type of analysis, we define the scaling dimension of the boost transformation parameter $W^\mu$ to be: $d_\sigma [W^\mu] = -z $, so that the scaling dimension of a scalar expression remains invariant under $\delw$. As a consequence, for boost-ghost number 1 expressions, the boost parameter $W^\mu$ contributes $(n_T,n_S,n_\epsilon) = (1,-1,0)$ to equation \eqref{rest:constraint1b}.

\subsection{A Prescription for Finding the Anomalous Terms} \label{ssb:prescription}

In this subsection we give a detailed prescription for finding the anomalous terms in the relative cohomology of the anisotropic Weyl operator for $z=2$ and any $d$.
The prescription is as follows:

\begin{enumerate}
\item Identify the sectors. Those are the sets $n_T$, $n_S$ and $n_\epsilon$ satisfying \eqref{rest:constraint1b}, \eqref{rest:constraint2b}. The cohomological analysis can be performed for each sector separately.
\item Build all TPD and gauge invariant expressions  in each sector by contracting the basic tangent tensors of subsection  \ref{subsec:TPD}. Use the electric and magnetic field rather than the gauge field itself to obtain gauge invariant expressions. We denote the independent basis of expressions $\phi_i$, taking into account the relevant identities from  appendix \ref{app:amazing} and additional dimensionally dependent identities.

\item For the cases with Galilean boost symmetry identify Milne boost invariant expressions: Denote by $\chi_j$ all the independent TPD and gauge invariant expressions with one $W_\mu$ in that sector. These span the possible results of the Milne boost transformations. Find the boost transformation $\delw\phi_i$ and express it as a linear combination of $\chi_j$. Suppose this linear combination is given by: $\delw\phi_i = B_{ij} \chi_j$. Find the boost invariant combinations $\phi^{BI} = D_{i} \phi_i $ by studying the null space of $ B_{ij} D_{i} = 0$. From this point on we keep $\phi^{BI}_k$, $k=1,\ldots, n_{BI}$ as our new independent basis of expressions, where $n_{BI}$ is the number of independent boost invariant expressions.

\item To find the cocycles of the relative cohomology of the anisotropic Weyl  operator:
	\begin{itemize}[label= --]
	\item Build the integrated expressions of ghost number one: $ I_i = \int \sqrt{-g}\,\sigma \phi_i $.
	\item Apply the Weyl operator $\delta^W_\sigma$ to each of these terms to obtain ghost number two expressions.
Reduce each of them to a linear combination of independent expressions of the form $L_j = \int \sqrt{-g}\, \sigma \psi_j$ where $\psi_j$ are ghost number one expressions.
Suppose these linear combinations are given by: $ \delta^W_\sigma I_i = - M_{ij} L_j $.
	\item Find all linear combinations of the basic ghost number one expressions $ E = C_i I_i $ (where $C_i$ are constants) that satisfy $ \delta^W_\sigma E = 0 $, by solving the linear system of equations:
$M_{ij} C_i = 0$.
The space of solutions is the cocycle space. Let $ E_i,\ i=1,\ldots,n_{cc}$ be some basis for this space, where $ n_{cc} $ is its dimension.
	\end{itemize}
	
\item To find the coboundaries of the relative cohomology:
\begin{itemize}[label= --]
\item Build the integrated expressions of ghost number zero: $ G_i = \int \sqrt{-g}\phi_i $.
\item Apply the Weyl operator $ \delta^W_\sigma $ to each of them to obtain ghost number one expressions. Reduce each of them to a linear combination of the expressions $ I_i $. Suppose these combinations are given by: $ \delta^W_\sigma G_i = S_{ij} I_j $. The span of these combinations is the coboundary space. Let $ F_i,\ i=1,\ldots,n_{cb} $ be some basis for this space, where $n_{cb}$ is its dimension.
\end{itemize}
\item Finally, to find the anomalous terms in the cohomology, check which of the cocycles $ I_i $ are not in the span of the coboundaries $ F_i$. We denote these by $ A_i,\ i=1,\ldots,n_{an} $, where $ n_{an} = n_{cc} - n_{cb} $ is the number of independent anomalies.
\end{enumerate}

For distinguishing between A-type and B-type anomalies we use the same definitions as in our previous paper. B-type anomalies are defined as Weyl invariant scalar densities (trivial descent cocycles up to coboundary terms) whereas A-type anomalies have non-trivial descent. This does not necessarily align with the definition of A-type anomalies as topological invariants. For further discussion on this topic see \cite{Arav:2014goa}.

\section{Scale Anomalies for 1+1 Dimensions with z=2}\label{sec:Results_1+1}
In $1+1$ dimensions the Frobenius condition is always satisfied, and the magnetic field $ B_{\mu\nu} $ and modified ``intrinsic'' Riemann tensor $ \widehat R_{\alpha\rho\mu\nu} $ vanish identically.
We study the Galilean case here (as opposed to the Lifshitz one that was studied in the previous paper).

The classification to  sectors is done according to equation \eqref{rest:constraint1} which takes here the form:
\begin{equation}
z[n_{K_S} + n_\mathcal{L} + 2n_E+n_W] + [n_a+n_\nabla-n_E-n_W]=1+z,
\end{equation}
where we have added the boost parameter for convenience when classifying the results of the boost transformations.

For $z=2$ we obtain:
\begin{equation}
2 n_{K_S} + 2 n_\mathcal{L} + 3n_E+n_a+n_\nabla = 3.
\end{equation}

For the condition \eqref{rest:constraint2} to be satisfied we must have $n_\epsilon=1$. We have the following sectors: $(n_T,n_S,n_\epsilon)=(0,3,1)$, $(1,1,1)$ or $(2,-1,1)$.
We find that there are no possible anomalies in any of them.
Note that in $1+1$ dimensions there is no need to keep track of indexes, and so we suppress them throughout this section.

\subsection{The (0,3,1) Sector}
The independent ghost number zero expressions in this sector are given by:
\begin{equation}
\phi_1 = \epsilon a a^2, \qquad
\phi_2 = \epsilon a \wn a, \qquad
\phi_3 = \epsilon \wn^2 a.
\end{equation}
All of them are invariant under boosts. The cohomological analysis is therefore identical to the corresponding one in the previous paper \cite{Arav:2014goa}, where we found no possible anomalies.

\subsection{The (1,1,1) Sector}
Here, the independent ghost number zero expressions are given by:
\begin{equation}
\phi_1 = K_S \epsilon a, \qquad
\phi_2 = \epsilon  \wn K_S, \qquad
\phi_3 = \epsilon \Lie{n} a.
\end{equation}
We first look for boost invariant combinations.
The independent boost-ghost number one expressions are given by:
\begin{equation}
\chi_1=\epsilon a^2 W, \qquad
\chi_2 = \epsilon \wn a W, \qquad
\chi_3 = \epsilon a \wn W, \qquad
\chi_4 = \epsilon \wn^2 W.
\end{equation}
Performing boost transformations on the ghost number zero expressions gives:
\begin{equation}
\begin{split}
&\delw \phi_1 = \chi_1 +\chi_3,
\\
&\delw \phi_2 = \chi_2 +\chi_3 +\chi_4,
\\
&\delw \phi_3 = \chi_1 +\chi_2 +\chi_3,
\end{split}
\end{equation}
and it is easy to check that there are no boost invariants in this sector.

\subsection{The (2,-1,1) Sector}
This sector contains just one ghost number zero expression $\phi_1 = \epsilon E$. This expression is not boost invariant.

\section{Scale Anomalies for 2+1 Dimensions with z=2}
\label{sec:Results_2+1}

The main case we study in this work is the one of $2+1$ dimensions with a dynamical exponent of $z=2$. As detailed in previous sections, we compare 4 different cases in our analysis:
\begin{enumerate}
\item The case with the Frobenius condition satisfied and Galilean boost invariance,
\item The case with the Frobenius condition satisfied and no Galilean boost invariance,
\item The case without the Frobenius condition and with Galilean boost invariance,
\item The case without the Frobenius condition and with no Galilean boost invariance.
\end{enumerate}
While the second case was already studied in our previous paper \cite{Arav:2014goa}, the others are new. As noted in subsection \ref{subsec:ClassSectors}, cases 3 and 4 (the ones in which the Frobenius condition is not satisfied) contain an infinite number of sectors with increasing total number of derivatives $n_D$, while cases 1 and 2 contain only a finite number of sectors with $n_D \leq 4 $. In this work we restrict our attention to the sectors with $n_D < 4$ and the parity even sector with $n_D = 4$ in all 4 cases, leaving other sectors to future work. In some of the cases we offer some conclusions regarding other sectors as well.

For calculations in $2+1$ dimensions it is convenient to define the scalars $B$, $K_A$ and the tensors $\tilde K_{\alpha\beta}$, $\tilde K^S_{\alpha\beta} $ as follows:
\begin{equation}
 B_{\mu\nu} \equiv B \tilde \epsilon_{\mu\nu}
,\qquad
 K^A_{\mu\nu} \equiv K_A \tilde \epsilon_{\mu\nu},
 \qquad
 \tilde K_{\alpha\beta} \equiv \tilde \epsilon_\alpha{}^\gamma K^S_{\gamma\beta},
 \qquad
 \tilde K^S_{\alpha\beta} \equiv \tilde K_{(\alpha\beta)}.
\end{equation}

In general, we find that cases 1,2 and 4 contain only B-type anomalies (in the sectors we study), whereas case 3 contains both an A-type anomaly and B-type anomalies.  We also find that cases 3 and 4 allow for an infinite number of B-type anomalies.

\subsection{With Frobenius and Galilean Boost Invariance}\label{subsec:WithFrobeniusWithBoost}

This case contains only a finite number of sectors:
\begin{itemize}[label=--]
\item (2,0,0) - Details of calculations can be found in Appendix \ref{FrobBoost2_0_0},
\item (2,0,1) - Details of calculations can be found in Appendix  \ref{FrobBoost2_0_1},
\item (1,2,0) - Details of calculations can be found in Appendix
 \ref{FrobBoost1_2_0},
\item (1,2,1) - Details of calculations can be found in Appendix  \ref{FrobBoost1_2_1},
\item (0,4,0) and (0,4,1), which are left unchanged compared to the Lifshitz case as proven below in subsection \ref{subsubsec:ProofBoostInvSpatial}.
\end{itemize}

We find in general that boost invariant expressions only exist in the purely spatial sectors (that is the ones with $n_T=0$, $n_S=4$), which are left unchanged compared to the Lifshitz case studied in \cite{Arav:2014goa}. We therefore find only 1 possible anomaly in this case, which is B-type and given in \eqref{anomWFWB}.

This result is in contradiction with \cite{Auzzi}, where it is claimed that there is an A-type anomaly in this case.
The discrepancy can probably be traced to the fact that while there are 12 independent ghost number zero expressions in the (0,4,0) sector (see equation (4.34) in \cite{Arav:2014goa}), there are 16 expressions in
\cite{Auzzi} (equation (3.18)) that are being treated as independent.
We suspect the that these 16 terms are not
independent and that this leads to an incorrect result.

\subsubsection{Proof of Boost Invariance of the Purely Spatial Sectors}\label{subsubsec:ProofBoostInvSpatial}
In this subsection we prove that when the time direction is hypersurface orthogonal (that is, the Frobenius condition is satisfied), all of the possible independent TPD invariant expressions in the purely spatial sectors (the ones with $n_T=0$) are invariant under Galilean boosts.

We start by noting that in this case $K^A_{\mu\nu}=0$. We therefore have for the purely spatial sectors:
\begin{equation}
n_T = n_{K_S} + n_\mathcal{L} + n_B + 2n_E  = 0,
\end{equation}
and therefore $n_{K_S} = n_\mathcal{L} = n_B = n_E = 0$. We are left with expressions containing only $a_\mu$, $\wn_\mu$, $\widehat R_{\alpha\beta\gamma\delta}$ (which is in this case the same as $\widetilde R_{\alpha\beta\gamma\delta}$) and the spatial projector $P_{\mu\nu}$.
The boost transformation properties of these tensors can be found in appendix \ref{App:BoostTrans} and reduce to the following in the Frobenius case:
\begin{equation}
\delwt P_{\mu\nu} = 0, \qquad
\delwt a_\mu = 0, \qquad
[\delwt,\wn_\mu]=0, \qquad
\delwt \widetilde R_{\alpha\beta\gamma\delta} = 0,
\end{equation}
where we have defined $\delwt$ to be the space projected boost transformation of a space tangent tensor:
\begin{equation}
\delwt \widetilde{T}_{\alpha\beta\gamma\ldots}
= P_\alpha^{\alpha'} P_\beta^{\beta'} \ldots \delw
 \widetilde{T}_{\alpha'\beta'\gamma'\ldots}\ .
\end{equation}
Note that since $\delw P^{\mu\nu}=0$, this projected boost transformation operator satisfies the Leibniz product rule:
\begin{equation}
\delwt \left[\widetilde T_{\alpha\beta\ldots\mu\nu\ldots} \widetilde S_{\gamma\delta\ldots}{}^{\mu\nu\ldots}\right] = \delwt \widetilde T_{\alpha\beta\ldots\mu\nu\ldots} \widetilde S_{\gamma\delta\ldots}{}^{\mu\nu\ldots} + \widetilde T_{\alpha\beta\ldots\mu\nu\ldots} \delwt \widetilde S_{\gamma\delta\ldots}{}^{\mu\nu\ldots},
\end{equation}
where $\widetilde T_{\alpha\beta\ldots\mu\nu\ldots}$ and $\widetilde S_{\gamma\delta\ldots\mu\nu\ldots} $ are space tangent tensors. Note also that for a scalar expression $\phi$:
\begin{equation}
\delw \phi = \delwt \phi.
\end{equation}
From these properties we conclude that for any scalar $\phi$ built only from $a_\mu$, $\wn_\mu$ and $\widehat R_{\alpha\beta\gamma\delta}$ we have $\delw \phi = 0$. Therefore any scalar in the purely spatial sectors is boost invariant.

This implies that the purely spatial sectors of our previous paper \cite{Arav:2014goa} are left unchanged since all the terms in these sectors are automatically boost invariant, and the rest of the cohomological analysis for these sectors is the same.
We are therefore left with the only anomaly found there in the $(0,4,0)$ sector:
\begin{equation}\label{anomWFWB}
A^{(0,4,0)} = \int \sqrt{-g}\, \sigma \left( \widehat{R} + \wn_\alpha a^\alpha \right)^2,
\end{equation}
which is B-type. The cohomology of the $(0,4,1)$ sector contained no cocycles.

\subsection{With Frobenius and No Galilean Boost Invariance}
This case was studied in full in our previous paper on Lifshitz cohomology.
We repeat our results here for completeness. The possible anomalies we found for this case were:
\begin{align}
A_1^{(2,0,0)} = & \int \sqrt{-g}\,\sigma \left[ \tr(K_S^2) - \frac{1}{2} K_S^2 \right],\label{Lifshitz_Anomalies_2_0_0}
\\
A_2^{(0,4,0)} = & \int \sqrt{-g}\, \sigma \left( \widehat{R} + \wn_\alpha a^\alpha \right)^2,
\label{Lifshitz_Anomalies_0_4_0}
\end{align}
where the superscript indicates the sector each of them belongs to, and we define:
\begin{equation}
K_S \equiv (K_S)^\mu_\mu,
\qquad
Tr(K_S^2) \equiv (K_S)^{\mu\nu}(K_S)_{\mu\nu}.
\end{equation}
Both of these anomalies are B-type.	
	
\subsection{Without Frobenius and with Galilean Boost Invariance}\label{subsec:WithoutFrobeniusWithBoost}

As explained in subsection  \ref{subsec:ClassSectors}, this case contains an infinite number of sectors. As previously stated, we restrict our analysis to the sectors with $n_D<4$ and the parity even $n_D=4$ sector. These are the following sectors:
\begin{itemize}[label=--]
\item (2,0,0) - Details of the calculations can be found in Appendix \ref{app:NFWB2_0_0},
\item (2,0,1) - Details of the calculations can be found in Appendix \ref{app:NFWB2_0_1},
\item (1,2,0) - Details of the calculations can be found in Appendix \ref{app:NFWB1_2_0},
\item (1,2,1)  - Details of the calculations can be found in Appendix \ref{app:NFWB1_2_1},
\item (0,4,0) - Details of the calculations can be found in Appendix \ref{app:NFWB_4_0_0}. A summary of the results is given in subsection \ref{subsubsec:NFWB0_4_0}.
\end{itemize}
In general we find no boost invariant expressions in the sectors with $n_D<4$. The sector (0,4,0),  however, does contain various boost invariant expressions. We find two possible anomalies in this sector: one A-type anomaly and one B-type anomaly. The structure of the cohomology in this sector mirrors the cohomology of relativistic conformal anomalies in 3+1 dimensions. This is a consequence of the null-reduction as discussed in subsection \ref{subsubsec:nullRed}. Note that out of all four different cases this is the only case that contains a possible A-type anomaly. Therefore in order to find an A-type anomaly one has to both give up the Frobenius condition and impose Galilean boost invariance.

We did not study in detail the sectors with $n_D>4$. However we can demonstrate that these sectors contain an infinite number of B-type anomalies. The argument is detailed in subsection \ref{subsubsec:MoreThanFour}.

\subsubsection{Comparison with the Null Reduction}\label{subsubsec:nullRed}

One can relate the Newton-Cartan structure defined on a ($d+1$)-dimensional manifold $M_{d+1}$ to the geometric structure of a ($d+2$)-dimensional Lorentzian manifold $M_{d+2}$ with a null isometry via the null-reduction procedure (see \cite{Jensen:2014hqa} and  references therein). One considers a ($d+2$)-dimensional manifold with a Lorentzian metric $G_{AB}$ and a null Killing vector $n^M$, and decomposes $G$ along coordinates $(x^-, x^\mu)$, where $x^-$ is a coordinate along the integral curves of the null vector $n^M$. The various NC structures then arise as components of the $d+2$ metric decomposed along these coordinates:

\begin{equation}
G = 2 n_\mu^{NC} dx^\mu (dx^- + A^{NC}_\nu dx^\nu) + h^{NC}_{\mu\nu}dx^\mu dx^\nu.
\end{equation}
It can then be shown that diffeomorphism invariance on $M_{d+2}$ is equivalent to the combination of diffeomorphism, Milne boost and $U(1)$ gauge invariance on the NC manifold $M_{d+1}$.\footnote{The $U(1)$ gauge symmetry in the NC geometry appears as a symmetry under the reparameterization of the $x^-$ coordinate on $M_{d+2}$, while Milne boost symmetry appears as an ambiguity in the decomposition of the metric $G$ along these coordinates.} Thus the problem of finding Milne boost and gauge invariant scalars on $M_{d+1}$ is mapped to the problem of finding scalars on $M_{d+2}$ built from the metric $G_{AB}$ and the null vector $n^M$.\footnote{There is a subtlety involved in this statement regarding the Frobenius condition: When the condition is satisfied by $n_M$, scalars that are built from contractions of the metric $G_{AB}$, the curvature, $n_M$ and their covariant derivatives do not exhaust all possible diffeomorphism invariant expressions, since there are other scalars, such as $a^2$ where $a$ is defined via $dn=a \wedge n$, which cannot be written this way. Therefore including only these scalars is equivalent to assuming the Frobenius condition is not satisfied.}
 The anisotropic Weyl transformation can also be introduced on the $M_{d+2}$ manifold by defining (for $z=2$):
\begin{equation}
\dels G_{AB} = 2 \sigma G_{AB},\qquad \dels n_M = 2 \sigma n_M.
\end{equation}

Using the null reduction we can derive several expectations for our cohomological analysis for the ($2+1$)-dimensional, $z=2$ case without Frobenius and with Galilean boost invariance. First, note that both the total number of derivatives $n_D$ in an expression and its Weyl scaling dimension $d_\sigma$ are preserved when reducing expressions from the Lorentzian ($3+1$)-dimensional manifold to the ($2+1$)-dimensional manifold. We can therefore expect the following:
\begin{enumerate}
\item Since there are no scalar expressions in the ($3+1$)-dimensional manifold with less than 4 derivatives ($n_D<4$) and scaling of $d_\sigma = 4$, we expect to find no boost and gauge invariant expressions in the sectors with $n_D<4$.

\item Scalar expressions in the ($3+1$)-dimensional manifold that are built only from the metric $G_{AB}$ and the Riemann tensor $R_{ABCD}$ (but not from $n^M$) have $n_D = d_\sigma $. Therefore in the ($2+1$)-dimensional manifold the corresponding expressions satisfy this relation too, which implies that such scalar expressions with the correct dimension of $d_\sigma=4$ belong to the sectors with $n_D = d_\sigma = 4$ in our analysis (that is the sectors with 4 space derivatives). On the other hand, these expressions transform under the Weyl transformation exactly like the corresponding expressions from the well-known $3+1$ relativistic conformal case (since the metric transforms the same). Therefore we expect the anisotropic Weyl cohomology in the $n_D=4$ sectors to mirror the one from the $3+1$ conformal case: The $(0,4,0)$ is expected to contain one A-type anomaly corresponding to the ($3+1$)-dimensional Euler density $E_4$, one B-type anomaly corresponding to the Weyl tensor squared $W^2$, and one coboundary corresponding to the trivial $ \Box R $. The $(0,4,1)$ sector (which we do not study in this work) is expected to contain one B-type anomaly corresponding to the $(3+1)$-dimensional Pontryagin density.
\end{enumerate}
These expectations are indeed met in our results. Note that, while we use the null reduction here to derive these expectations, our cohomological analysis is performed directly in the $2+1$ setting and does not make use of the null reduction.

\subsubsection{The (0,4,0) Sector}\label{subsubsec:NFWB0_4_0}
In this sector we find four boost invariant quantities which can be identified with the null reduction of $R^2$, $W^2$, $E_4$ and $\Box R$ of the $3+1$ dimensional theory up to multiplication by some constant coefficient (see equation \eqref{eq:BoostInvariants} and the definition of the $\phi_i$-s in \eqref{eq:ThePhis}). As expected, out of these, there are three independent Weyl cocycles, one of which is a coboundary. Hence we find two possible anomalies in this sector, given by the densities:
\begin{equation}
\begin{split}\label{Anoamlies_BNF}
\mathcal{A}_{E_4}^{(0,4,0)} = &
(\wn_\mu +a_\mu)
\left( 4 K_A B a^\mu + 8E^\mu K_A^2 +8 K_A \tilde K_S^{\mu\nu} a_\nu + 4K_A K_S \temn a_\nu \right.
\\&
\left. +2 (a_\nu \wn^\nu a^\mu -a^\mu \wn_\nu a^\nu) - 8\temn K_A\Lie{n} a_\nu \right)
\\
& + (\Lien +K_S) \left(16 K_A \Lien K_A + 8K_S K_A^2\right),
\\
\mathcal{A}_{W^2}^{(0,4,0)} = &  (\widehat R +\wn_\mu a^\mu -8 K_A B)^2  + 12 \wn^\mu K_A  (\wn_\mu + a_\mu) B  \\ &
+ 12 \temn \wn_\nu K_A (\wn_\mu+a_\mu) K_S + 12 \temn \wn_\mu K_A \Lie{n} a_\nu
 \\
&  + 24 \wn_\alpha K_A \wn_\beta \tilde K_S^{\alpha\beta}  -
 72 K_A E^\mu \wn_\mu K_A
 - 36 (\Lie{n} K_A)^2.
\end{split}
\end{equation}
Note that, as expected, $\mathcal{A}_{E_4}^{(0,4,0)}$ is a total derivative and an A-type anomaly while $\mathcal{A}_{W^2}^{(0,4,0)}$ is a B-type anomaly (that is, a Weyl-invariant density). The details of the calculations, as well as the method used for identifying the various expressions with the respective ($3+1$)-dimensional expressions, can be found in Appendix \ref{app:B_noF}.

It is important to note how these results reduce to the ones of subsection \ref{subsec:WithFrobeniusWithBoost} when we require the Frobenius condition to be satisfied, that is when setting $K_A=0$. It is easy to see that, when $K_A=0$, the B-type anomaly $\mathcal{A}_{W^2}^{(0,4,0)}$ reduces to the single anomaly \eqref{anomWFWB} of subsection \ref{subsec:WithFrobeniusWithBoost}. The anomaly, $\mathcal{A}_{E_4}^{(0,4,0)}$ reduces to the expression $	2 (\wn_\mu +a_\mu) (a_\nu \wn^\nu a^\mu -a^\mu \wn_\nu a^\nu)$. While this expression is still a cocycle of the relative cohomology, it becomes a coboundary in the $K_A=0$ case,
 and can be removed by adding the following counter term to the action:
\begin{equation}
 W_{c.t.} = \int \sqrt{-g} \left(\frac{1}{2} a^\alpha \wn_\alpha (a^2) +\frac{3}{8} a^4 \right),
\end{equation}
which is both gauge and boost invariant when $K_A=0$. This explicitly shows that in order to obtain an A-type anomaly one has to forgo the Frobenius requirement.

\subsubsection{Sectors with More than Four Derivatives}\label{subsubsec:MoreThanFour}

While we have not fully studied the relative Weyl cohomology in the sectors with $n_D>4$ (there is an infinite number of such sectors), we show here that one can find an infinite number of independent B-type anomalies in these sectors.

First, we note that the total contribution of $K_A$ to equation \eqref{rest:constraint1} is zero (for $z=2$), hence we are allowed to have as many $K_A$ instances as we want in our expressions.
More specifically, any scalar expression $\phi$ with scaling dimension 4 can be multiplied by $K_A^n$ (where $n$ is any integer number) to get another expression ($K_A)^n \phi$ with the same scaling dimension.

We also note that for $z=2$, $K_A$ is both boost and Weyl invariant (see Appendix \ref{App:BoostTrans} and \ref{app:WeylTrans}) and hence for any B-type anomaly density  in the relative cohomology $\mathcal A$, $(K_A)^n \mathcal A$ also represents a B-type anomaly (it cannot be a coboundary term as it is clearly not a total derivative, and Weyl coboundaries are always total derivatives as explained in \cite{Arav:2014goa}). For example, since $\mathcal{A}_{W^2}^{(0,4,0)}$ as defined in \eqref{Anoamlies_BNF} represents a B-type anomaly with $n_D=4$ derivatives, the expression $ \mathcal A^{(n)} = (K_A)^n \mathcal{A}_{W^2}^{(0,4,0)} $ represents a B-type anomaly with $n_D = 4+n$ derivatives in the relative cohomology for any $n$. We see that giving up the Frobenius condition implies the possibility of having an infinite number of independent anomalies. We emphasize that these are not necessarily the only anomalies in the sectors with $n_D>4$. A full cohomological analysis is required in order to obtain all possible anomalies in these sectors.

\subsection{Without Frobenius and with No Galilean Boost Invariance}\label{subsec:NFNGB}

As explained in subsection \ref{subsec:ClassSectors}, like the previous case, this case contains an infinite number of sectors. We again restrict our attention to the sectors with $n_D<4$ and the parity even $n_D=4$ sector, which are the following sectors:
\begin{itemize}[label=--]
\item (2,0,0) - Details can be found in Appendix \ref{app:NFNB_2_0_0_2_0_1_1_2_0},
\item (2,0,1) - Details can be found in Appendix \ref{app:NFNB_2_0_0_2_0_1_1_2_0},
\item (1,2,0) - Details can be found in Appendix \ref{app:NFNB_2_0_0_2_0_1_1_2_0},
\item (1,2,1) - Details can be found in Appendix \ref{app:NFNB_1_2_1},
\item (0,4,0) - Details can be found in Appendix \ref{app:NFNB_0_4_0}.
\end{itemize}

Altogether we find 6 different possible anomalies in these sectors, all of which are B-type:
The $(2,0,0)$, $(2,0,1)$ and $(1,2,1)$ sectors remain unchanged from the Lifshitz case with the Frobenius condition satisfied, which was studied in \cite{Arav:2014goa}. Therefore as in that case, the sector $(2,0,0)$ contains only the B-type anomaly given by the density:
\begin{equation}\label{NBNFanom_2_0_0}
\mathcal A_1^{(2,0,0)} = \tr(K_S^2) - \frac{1}{2} K_S^2 ,
\end{equation}
the $(2,0,1)$ sector is empty and the $(1,2,0)$ has no possible anomalies.
The $(1,2,1)$ sector is changed from the Frobenius case, and unlike that case, contains a B-type anomaly given by the density:
\begin{equation}\label{NBNFanom_1_2_1}
\mathcal A_1^{(1,2,1)} = K_A \left[\tr(K_S^2) - \frac{1}{2} K_S^2 \right] .
\end{equation}
Finally, the $(0,4,0)$ sector is changed as well, and contains 4 different possible B-type anomalies, given by the densities:
\begin{equation}
\begin{split}\label{NBNFanom_0_4_0}
\mathcal A_1^{(0,4,0)} = & \,
 K_A^2 \left[ \tr(K_S^2) - \frac{1}{2} K_S^2 \right], \\
\mathcal A_2^{(0,4,0)} = & \, K_A \Lie{n}^2 K_A + K_A K_S \Lie{n} K_A ,\\
\mathcal A_3^{(0,4,0)} = & \, \tilde K_S^{\alpha\beta} (a_\alpha \wn_\beta  K_A + \wn_\alpha \wn_\beta K_A) ,\\
\mathcal A_4^{(0,4,0)} = & \,
\left( \widehat{R} + \wn_\alpha a^\alpha \right)^2.
\end{split}
\end{equation}

Note that out of these six anomalies, the ones labeled as $\mathcal A_1^{(2,0,0)}$ and $\mathcal A_4^{(0,4,0)}$ are the same ones found in the Frobenius case, whereas the others are new and indeed vanish when $K_A=0$.

As in the subsection \ref{subsec:WithoutFrobeniusWithBoost}, the sectors with $n_D>4$ were not studied in detail, but an argument similar to the one outlined in subsection \ref{subsubsec:MoreThanFour} is valid here as well, namely if $\mathcal A$ represents a B-type anomaly in the relative Weyl cohomology then $(K_A)^n \mathcal A$ represents one as well. Therefore, for example, the expression: $ \mathcal A^{(n)} = (K_A)^n \left[ \tr(K_S^2) - \frac{1}{2} K_S^2 \right] $ represents an anomaly for any $n$ (with $n_D=2+n$ derivatives). This structure can be clearly seen in the anomalies labeled $ \mathcal A_1^{(1,2,1)}$ and $\mathcal A_1^{(0,4,0)}$  above.

\section{Conclusions and Summary}\label{sec:summary}
We have studied non-relativistic scale anomalies in various setups, using the cohomological description of the WZ consistency conditions. These include cases with or without a foliation structure (i.e. the Frobenius condition) and with or without Galilean boost invariance. The analysis was carried out explicitly for dynamical exponent $z=2$ both in $1+1$ and in $2+1$ dimensions. The results extend the analysis of Lifshitz scale anomalies in \cite{Arav:2014goa}.

In $1+1$ dimensions the Frobenius condition is automatically satisfied and we found no anomalies in the case with Galilean boost invariance. In
$2+1$ dimensions we summarize our findings in  table \ref{TheMostAmazingTableInTheWorld} below.
\begin{table}[!htbp]
\centering
\begin{tabular}{| l | c | c | c | c |}
\cline{2-5}
  \multicolumn{1}{c|}{} & \multicolumn{2}{c|}{
With Boost Invariance

 } & \multicolumn{2}{c|}{Without Boost Invariance}
\\
\cline{2-5}
  \multicolumn{1}{c|}{} & Sector & $n_{an}$ & Sector & $n_{an}$ 	
\\ \hline
\begin{tabular}
{@{}c@{}}With \\
Frobenius
\end{tabular}
		& (0,4,0) & 1 (see \eqref{anomWFWB})
		&
		\begin{tabular}
		{@{}c@{}}(2,0,0) \\
		(0,4,0)
		\end{tabular}
		&
				\begin{tabular}
		{@{}c@{}}1 (see \eqref{Lifshitz_Anomalies_2_0_0}) \\
		1 (see \eqref{Lifshitz_Anomalies_0_4_0})
		\end{tabular}
\\ \hline
\begin{tabular}
{@{}c@{}}Without \\
Frobenius \\
\end{tabular}
&
\begin{tabular}
		{@{}c@{}}(0,4,0) \\
		$n_D > 4$
		\end{tabular}
		&
		\begin{tabular}
		{@{}c@{}} 2 (see \eqref{Anoamlies_BNF}) \\
		infinitely many
		\end{tabular}
		&
		\begin{tabular}
		{@{}c@{}} (2,0,0) \\
		(1,2,1) \\
		(0,4,0) \\
		$n_D > 4$
		\end{tabular}
		 &
		\begin{tabular}
		{@{}c@{}} 1 (see \eqref{NBNFanom_2_0_0}) \\
		1 (see \eqref{NBNFanom_1_2_1}) \\
		4 (see \eqref{NBNFanom_0_4_0}) \\
	   infinitely many
		\end{tabular}
\\
\hline
\end{tabular}
\caption{Number of anomalies in the different sectors for the relative Weyl cohomology in $2+1$ dimensions and $z=2$, denoted by the number of time and space derivatives and the parity property: $(n_T,n_S,n_\epsilon)$. $n_{an}$ denotes the number of anomalies. $n_D$ denotes the total number of derivatives.}\label{TheMostAmazingTableInTheWorld}
\end{table}

The results of our cohomological analysis in $2+1$ dimensions lead to several interesting observations. First, when the Frobenius condition is imposed, there are no new possible anomalies in the boost invariant case compared to the ones found in the Lifshitz case discussed in \cite{Arav:2014goa}, and in fact one is left only with the single anomaly from the sector with 4 space derivatives. Second, when the Frobenius condition is not imposed, it is possible to have an infinite number of independent B-type anomalies. Third, the case with no Frobenius condition imposed and no boost invariance does contain new possible B-type anomalies over the Lifshitz case with Frobenius discussed in \cite{Arav:2014goa}, but we found no A-type anomalies (up to 4 derivatives). Finally, the case with no Frobenius condition imposed and with boost invariance (which is the one studied in \cite{Jensen:2014hqa}) does contain an A-type anomaly in the parity even $n_D=4$ sector, and in fact the structure of the Weyl cohomology in this sector mirrors the structure of the conformal case in $3+1$ dimensions, as expected from the null reduction and in agreement with \cite{Jensen:2014hqa}. This sector thus contains an A-type anomaly corresponding to the Euler density in $3+1$ dimensions and a B-type anomaly corresponding to the Weyl tensor squared in $3+1$ dimensions.

We therefore conclude that in order to have an A-type anomaly (at least up to four derivatives) in the anisotropic Weyl cohomology in $2+1$ dimensions and with $z=2$ one has to both impose Galilean boost invariance and give up the foliation structure of spacetime. However in doing so, one introduces the possibility of having an infinite set of independent anomalies in the cohomology. Whether this has any interesting implications or imposes any restrictions on the underlying field theories is left for future study. In particular, the issue of causality which we discussed should be addressed.

Since we have not fully studied the Weyl cohomology of the infinitely many sectors with $n_D>4$ in the cases without the Frobenius condition, it would also be interesting to study them in detail in the future, and try to prove our conjecture that there is no A-type anomaly in the non-boost-invariant case. Another possibility for future work would be comparing the various cases we have studied here for higher dimensions, as well as understanding the cohomological structures of anomalies for $z \neq 2$ for each of these cases.

Several other research directions follow from our work. In terms of field theory, a better understanding of the implications of the Frobenius condition when coupling a non-relativistic theory to a curved spacetime would be desirable, both in the boost invariant and the non-boost invariant cases. Studying the behavior of the various anomaly coefficients along RG flows, and especially the one of the A-type anomaly in the boost invariant case, could lead to RG flow theorems for non-relativistic theories.
This is particularly interesting since the value of $z$ may change along the flow.
It would also be interesting to address in this context the issue of anisotropic scale versus full Schr\"odinger invariance.

\section*{Acknowledgments}
This work is supported in part by
the Israeli Science Foundation Center of Excellence, BSF, GIF and the I-CORE program
of Planning and Budgeting Committee and the Israel Science Foundation (grant number
1937/12).
Research at Perimeter Institute is supported by the Government of Canada through
Industry Canada and by the Province of Ontario through the Ministry of Research and Innovation.

\appendix

\section{Useful Formulas}\label{app:amazing}
In this appendix we gather various formulas required for the analysis presented in this paper. The formulas are organized as follows:
\begin{itemize}[label=--]
\item General identities and definitions can be found in subsection \ref{app:general_ids},
\item Milne boost transformation rules can be found in subsection \ref{App:BoostTrans},
\item Anisotropic Weyl transformation rules can be found in subsection \ref{app:WeylTrans}.
\end{itemize}

Throughout this appendix we use $\widetilde T_{\alpha\beta\ldots}$ to denote a generic space tangent tensor and
$\stackrel{2d}{=}$ to denote equalities that only hold in $2+1$ dimensions.

\subsection{Definitions and Identities}\label{app:general_ids}

In this subsection we present some definitions related to the basic tangent tensors as discussed in subsection \ref{subsec:TPD} and identities that relate them to each other. We start by recounting the definitions of the basic space tangent tensors.

A tensor $\widetilde T_{\alpha\beta\ldots}$ is called space tangent if it satisfies:
\begin{equation}
 n^\alpha \widetilde{T}_{\alpha\beta\gamma\ldots} = n^\beta \widetilde{T}_{\alpha\beta\gamma\ldots} = \ldots = 0.
\end{equation}
Any tensor can be rendered space tangent by projecting it on the space directions using the space projector $P_{\mu\nu} = g_{\mu\nu} + n_\mu n_\nu$.
We decompose the derivative of the normalized time one-form as follows:
\begin{equation}\label{eq:AppDnDecomp}
\nabla_\alpha n_\beta = (K_S)_{\alpha\beta}+ (K_A)_{\alpha\beta}-a_\beta n_\alpha,
\end{equation}
where $K^S_{\mu\nu}$, $K^A_{\mu\nu}$ and $a_\mu$ are all space tangent tensors.
$a_\mu$ is the acceleration vector given by:
\begin{equation}
a_\mu \equiv \Lie{n} n_\mu = n^\nu \nabla_\nu n_\mu.
\end{equation}
$K^S_{\mu\nu}$ is symmetric and given by:
\begin{equation}
(K_S)_{\mu\nu} = \frac{1}{2}\Lie{n} P_{\mu\nu}.
\end{equation}
It reduces to the extrinsic curvature of the foliation if the Frobenius condition is satisfied. We denote its trace by $K_S \equiv (K^S)_\mu^\mu$.
$K^A_{\mu\nu}$ is antisymmetric and is given by:
\begin{equation}
(K_A)_{\mu\nu} = P_\mu^{\mu'} P_\nu^{\nu'} \nabla_{[\mu'} n_{\nu']}.
\end{equation}
It vanishes in the Frobenius case. We denote by $K_{\alpha\beta}$ the total space tangent part of $\nabla_\alpha n_\beta$, that is:
\begin{equation}
K_{\alpha\beta} \equiv (K_S)_{\alpha\beta}+(K_A)_{\alpha\beta}.
\end{equation}
We also define the space tangent Levi-Civita tensor as:
\begin{equation}
\tilde \epsilon^{\mu\nu\rho \ldots} =
n_\alpha \epsilon^{\alpha\mu\nu\rho\ldots}.
\end{equation}
The following identities follow immediately from the above definitions:
\begin{equation}
\begin{split}
\Lie{n} P_{\alpha\beta} &= 2K^S_{\alpha\beta} ,
\\
\Lie{n} P^{\alpha\beta} &= -2K_S^{\alpha\beta} +a^\alpha n^\beta+ a^\beta n^\alpha ,\\
\Lie{n} \tilde\epsilon_{\alpha\beta\gamma\ldots} &= K_S \tilde\epsilon_{\alpha\beta\gamma\ldots} ,
\\
\Lie{n} \tilde\epsilon^{\alpha\beta\gamma\ldots}
&= - K_S \tilde \epsilon^{\alpha\beta\gamma\ldots}  .
\end{split}
\end{equation}

In theories with Galilean boost invariance we decompose the gauge field as follows:
\begin{equation}
A_0 \equiv n^\mu A_{\mu}, \qquad
\tilde A_\mu \equiv P_\mu^{\mu'} A_{\mu'}.
\end{equation}
We also define the electric and magnetic fields as follows:
\begin{equation}\label{eq:ElectroMagnetic2}
E_\mu \equiv F_{\mu\nu} n^\nu, \qquad B_{\mu\nu} \equiv P_\mu^{\mu'} P_\nu^{\nu'} F_{\mu'\nu'},
\end{equation}
where $F_{\mu\nu}$ is the field strength of the gauge field $A_\mu$.

In $2+1$ dimensions the tensors $K^A_{\mu\nu}$ and $B_{\mu\nu}$ contain only one independent component each, and it is convenient to define:
\begin{equation}
\begin{split}
& B_{\mu\nu} \equiv B \tilde \epsilon_{\mu\nu} ,
\qquad
B_{\mu\nu} \tilde \epsilon^{\mu\nu} = 2 B,
\\
& K^A_{\mu\nu} \equiv K_A \tilde \epsilon_{\mu\nu},
\qquad
K^A_{\mu\nu} \tilde \epsilon^{\mu\nu} = 2 K_A.
\end{split}
\end{equation}
Additionally, when writing parity odd terms in $2+1$ dimensions it is sometimes useful to define the following tensors:
\begin{equation}\label{eq:tildeK}
\begin{split}
& \tilde K_{\alpha\beta} \equiv \tilde \epsilon_\alpha{}^\gamma K^S_{\gamma\beta},
\\
& \tilde K^S_{\alpha\beta} \equiv \tilde K_{(\alpha\beta)},
\end{split}
\end{equation}
so that $\tilde K^S _{\alpha\beta}$ contains the traceless part of $K^S_{\alpha\beta}$. It then follows that:
\begin{equation}
\begin{split}
&\tilde K_{[\alpha\beta]} = \frac{K_S}{2} \tilde \epsilon_{\alpha\beta} ,\\
&\tilde K_{\alpha\beta} = \tilde K_{(\alpha\beta)}
+\tilde K_{[\alpha\beta]} =
\tilde K^S _{\alpha\beta} + \frac{K_S}{2} \tilde\epsilon_{\alpha\beta}.
\end{split}
\end{equation}

Given a space tangent $p$-form $\tilde F_{\alpha\beta\ldots}$, we define the space tangent exterior derivative as follows:
\begin{equation}
(\tilde d \tilde F)_{\alpha\beta\gamma \ldots} \equiv  P_\alpha^{\alpha} P_\beta^{\beta'} P_\gamma^{\gamma'} (d \tilde F)_{\alpha'\beta'\gamma'\ldots},
\end{equation}
where $d\mathbf{\tilde{F}}$ is the standard exterior derivative of $\mathbf{\tilde F}$, and $(\tilde d \tilde F)_{\alpha\beta\gamma \ldots}$ is a space tangent $(p+1)$-form. It can then be easily shown that:\footnote{We use bold symbols to designate $p$-forms.}
\begin{equation}
\tilde d \mathbf{\tilde F} = d \mathbf{\tilde F}+ \mathbf{n} \wedge \Lie{n} \mathbf{\tilde F},
\end{equation}
where $\wedge$ is the standard wedge product between forms.
A general $p$-form $F_{\alpha\beta\ldots}$ can always be decomposed as follows:
\begin{equation}\label{eq:AppPFormDecomp}
\mathbf{F} = \mathbf{\tilde F} + \mathbf{n} \wedge \mathbf{F^n},
\end{equation}
where $\mathbf{\tilde F}$ is a space tangent $p$-form and $\mathbf{F^n}$ is a space tangent $(p-1)$-form. For example, from \eqref{eq:AppDnDecomp} we have the following decomposition for $d\mathbf{n}$:
\begin{equation}\label{eq:AppDnFormDecomp}
d\mathbf{n} =  2\mathbf{K_A} - \mathbf{n} \wedge \mathbf{a}.
\end{equation}
From \eqref{eq:AppPFormDecomp} and \eqref{eq:AppDnFormDecomp} one can show that:
\begin{equation}\label{eq:AppDFDecomp}
d\mathbf{F} = [\tilde d \mathbf{\tilde F} + 2 \mathbf{K_A} \wedge \mathbf{F^n}]
+ \mathbf{n} \wedge  [- \Lie{n} \mathbf{\tilde F} - \mathbf{a} \wedge \mathbf{F^n} - \tilde d \mathbf{F^n} ].
\end{equation}
Using \eqref{eq:AppDFDecomp} twice and noting that $d^2 \mathbf{F} = 0$ we find the following identities:
\begin{equation}
\begin{split}
& \tilde d ^2 \mathbf{\tilde F} = 2 \mathbf{K_A} \wedge \Lie{n} \mathbf{\tilde F},
\\
&
[\Lie{n},\tilde d] \mathbf{\tilde F} = \mathbf{a} \wedge \Lie{n} \mathbf{\tilde F}.
\end{split}
\end{equation}
Similarly, using \eqref{eq:AppDFDecomp} on $d\mathbf{n}$ and noting that $d^2 \mathbf{n} = 0$ we find the following identities for $K^A_{\mu\nu}$ and $a_\mu$:
\begin{equation}
\begin{split}
&\tilde d \mathbf{K_A} = \mathbf{K_A} \wedge \mathbf{a},\\
&\tilde d \mathbf{a}  = 2 \Lie{n} \mathbf{K_A},
\end{split}
\end{equation}
or in index notation:
\begin{equation}
\begin{split}
& \wn_{[\alpha} (K_A)_{\beta\gamma]} = (K_A)_{[\alpha\beta}\, a_{\gamma]},
\\
& \wn_{[\alpha} a_{\beta]} = \Lie{n} (K_A)_{\alpha\beta}
\stackrel{2d}{=}
\tilde\epsilon_{\alpha\beta} (\Lien K_A + K_S K_A).
\end{split}
\end{equation}
Finally, using \eqref{eq:AppDFDecomp} on the field strength tensor $F_{\mu\nu}$ we find the following identities for the electric and magnetic fields (these are just the homogeneous Maxwell equations):
\begin{align}
& \tilde{d} \mathbf{B} +   2 \mathbf{K_A} \wedge \mathbf{E} = 0,
\\
& \Lien \mathbf{B} + \mathbf{a} \wedge \mathbf{E} + d \mathbf{E} = 0.
\end{align}
In the case of $2+1$ dimensions, the first identity is trivial, and the second reduces to:
\begin{equation}\label{eq:Maxwell}
(\Lien + k_S) B + \tilde{\epsilon}^{\mu\nu} (\wn_\mu + a_\mu) E_\nu = 0,
\end{equation}
in index notation.

Next we turn to discuss space tangent derivatives and the Riemann tensor. Given a space tangent tensor $\widetilde{T}_{\alpha\beta\ldots}$, we define its space tangent covariant derivative as follows:
\begin{equation}
\wn_\mu \widetilde T_{\alpha\beta\ldots} \equiv P^{\mu'}_\mu P^{\alpha'}_\alpha P^{\beta'}_\beta \ldots \nabla_{\mu'} \widetilde T_{\alpha'\beta'\ldots} .
\end{equation}
Note that both the spatial metric $P_{\mu\nu}$ and the space tangent Levi-Civita tensor are covariantly constant under this derivative:
\begin{equation}
\begin{split}
&\wn_\mu P_{\alpha\beta} = 0 ,\\
&\wn_\mu \tilde \epsilon_{\alpha\beta\gamma\ldots} = 0 .
\end{split}
\end{equation}
From this definition, the following formula holds for the exchange of space tangent derivatives:
\begin{equation}
\begin{split}\label{eq:exch_der}
& \left[\wn_\mu, \wn_\nu\right] \widetilde T_{\alpha\beta\gamma\ldots}  =
 \widetilde R_{\alpha\rho\mu\nu}  {{\widetilde T}^\rho}{}_{\beta\gamma\ldots}  +
\ldots +
 2K_{\mu\nu}^A\Lie{n} \widetilde T_{\alpha\beta\gamma\ldots} ,
\end{split}
\end{equation}
where $\widetilde R_{\alpha\rho\mu\nu}$ is defined as the space tangent tensor:
\begin{equation}
\widetilde R_{\alpha\rho\mu\nu} = P_\alpha^{\alpha'} P_\rho^{\rho'} P_\mu^{\mu'} P_\nu^{\nu'} R_{\alpha'\rho'\mu'\nu'} - 2K^A_{\mu\nu}K_{\alpha\rho}-K_{\mu\alpha}K_{\nu\rho}+K_{\nu\alpha}K_{\mu\rho},
\end{equation}
and $R_{\alpha\rho\mu\nu}$ is the standard Riemann curvature associated with the covariant derivative $\nabla_\mu$ and defined via:
\begin{equation}\label{eq:Der_exch_higher_d}
 \left[\nabla_\mu, \nabla_\nu\right] T_{\alpha\beta\gamma\ldots}  =
  R_{\alpha\rho\mu\nu}  {{T}^\rho}{}_{\beta\gamma\ldots} +
\ldots \ .
\end{equation}
Note that in the case where $n_\alpha$ satisfies the Frobenius condition, $\widetilde R_{\alpha\rho\mu\nu}$ reduces to the intrinsic Riemann curvature of the foliation it induces. However generally this tensor does not have all of the regular symmetries of the Riemann tensor. It is therefore useful to define a modified Riemann tensor:\footnote{We use $\perp T_{\alpha\beta\ldots}$ to denote the space projection of the tensor $T_{\alpha\beta\ldots}$, i.e. $\perp T_{\alpha\beta\ldots} \equiv P_\alpha^{\alpha'} P_\beta^{\beta'} \ldots T_{\alpha' \beta'\ldots}$.}
\begin{equation}\label{eq:AppModifiedRiemannDef}
\begin{split}
\widehat R_{\alpha\rho\mu\nu} &\equiv
\widetilde R_{\alpha\rho\mu\nu}
+2K^A_{\mu\nu}K^S_{\alpha\rho}
+K^A_{\mu\alpha}K^S_{\nu\rho}
+K^S_{\mu\alpha}K^A_{\nu\rho}
-K^A_{\nu\alpha}K^S_{\mu\rho}
-K^S_{\nu\alpha}K^A_{\mu\rho}\\
& = \, \perp R_{\alpha\rho\mu\nu}
-2K^A_{\mu\nu}K^A_{\alpha\rho}
-K^S_{\mu\alpha}K^S_{\nu\rho}
-K^A_{\mu\alpha}K^A_{\nu\rho}
+K^S_{\nu\alpha}K^S_{\mu\rho}
+K^A_{\nu\alpha}K^A_{\mu\rho},
\end{split}
\end{equation}
which satisfies the usual Riemann tensor symmetries except for the second Bianchi identity which is replaced by:
\begin{equation}
\wn_{[\sigma|}\widehat R_{\alpha\rho|\mu\nu]} =
2 \left( (\wn_\rho+a_\rho) K^S_{\alpha[\sigma|} \right) K^A_{|\mu\nu]}
+ 2 \wn_{[\sigma|} \left(K^A_{|\mu|\alpha} K^S_{|\nu]\rho}\right) - \alpha \leftrightarrow \rho,
\end{equation}
where the antisymmetrization is everywhere on the $\sigma$, $\mu$ and $\nu$ indexes.
We then define the equivalents of the Ricci tensor and scalar for this modified Riemann tensor $\widehat R_{\alpha\rho\mu\nu}$ as follows:
\begin{equation}
\begin{split}
&\widehat R_{\rho\nu} \equiv R^{\mu}{}_{\rho\mu\nu}  = P^{\alpha\mu} \widehat R_{\alpha\rho\mu\nu} ,\\
&\widehat R \equiv \widehat R^{\nu}_{\nu} = P^{\rho\nu} \widehat R_{\rho\nu}.
\end{split}
\end{equation}

In $2+1$ dimensions the modified Riemann tensor contains only one independent component which we choose to be the scalar $\widehat R$. We can then write $\widehat R_{\alpha\beta\gamma\delta}$ in terms of $\widehat R$:
\begin{equation}
\begin{split}
& \widehat R_{\alpha\beta\gamma\delta} \stackrel{2d}{=} \frac{\widehat R}{2} (P_{\alpha\gamma}P_{\beta\delta} - P_{\alpha\delta}P_{\beta\gamma}),
\end{split}
\end{equation}
and the formula \eqref{eq:exch_der} takes the form:
\begin{equation}
\begin{split}
& \left[\wn_\mu, \wn_\nu\right] \widetilde T_{\alpha\beta\gamma\ldots}
 \stackrel{2d}{=}
 \widehat R
P_{\alpha[\mu} \widetilde T_{\nu]\beta\gamma\ldots}
- 2K_A(
 \tilde \epsilon_{\mu\nu}K^S_{\alpha\rho}
+\tilde \epsilon_{[\mu|\alpha} K^S_{|\nu]\rho}
- \tilde \epsilon_{[\mu|\rho} K^S_{|\nu]\alpha} )\,
 {{\widetilde T}^\rho}{}_{\beta\gamma\ldots}
\\ &
~~~~~~~~~~~~~~~~~~~~~~~~
 + \ldots
+ 2K_A \tilde\epsilon_{\mu\nu} \Lie{n} \widetilde T_{\alpha\beta\gamma\ldots}.
\end{split}
\end{equation}

In the case where the Frobenius condition is satisfied, $\widetilde R_{\alpha\rho\mu\nu}$ and $\widehat R_{\alpha\rho\mu\nu}$ coincide, and equation \eqref{eq:AppModifiedRiemannDef} reduces to one of the Gauss-Codazzi relations for the foliation induced by $n_\alpha$. We can also find generalizations for the other Gauss-Codazzi relations as follows:
\begin{align}
\begin{split}
\perp \left(n^\alpha R_{\alpha\rho\mu\nu} \right)
= - \wn_\mu K_{\nu\rho} + \wn_\nu K_{\mu\rho} +2K_{\mu\nu}^A a_\rho,
\end{split}
\\
\begin{split}
\perp \left(n^\alpha n^\mu R_{\alpha\rho\mu\nu} \right)
= -\Lie{n} K^S_{\nu\rho} +K_\rho{}^\mu K_{\nu\mu} +\wn_{(\nu} a_{\rho)} +a_\nu a_\rho ,
\end{split}
\end{align}
which are derived from the various projections of equation \eqref{eq:Der_exch_higher_d}.

Using the generalized Gauss-Codazzi relations we can derive the following formula for exchanging a space tangent derivative and a Lie derivative in the direction of $n^\alpha$:
\begin{equation}
\begin{split}\label{eq:time_space_exchange}
&\left[\wn_\mu ,\Lie{n}\right] \widetilde T_{\alpha\beta\gamma\ldots} = -
a_\mu \Lie{n}  \widetilde T_{\alpha\beta\gamma\ldots}
\\
&~~~~~~~~~ - \left[(\wn_\nu+a_\nu)K_{\mu\alpha}^S - (\wn_\alpha+a_\alpha)K_{\mu\nu}^S - (\wn_\mu+a_\mu)K_{\alpha\nu}^S\right] \widetilde T^\nu{}_{\beta\gamma\ldots} +\ldots\ .
\end{split}
\end{equation}
Applying a Lie derivative to equation \eqref{eq:exch_der} and using formula $\eqref{eq:time_space_exchange}$ twice, we can derive the following identity for the Lie derivative of $\widehat R_{\alpha\beta\mu\nu}$ (this is a consequence of the $d+1$ dimensional second Bianchi identity):
\begin{equation}
\begin{split}
\Lie{n} \widehat R_{\alpha\beta\mu\nu} =& \widehat R_{[\alpha|\rho\mu\nu}K^{S}{}^\rho{}_{|\beta]}
 + \widehat R_{[\mu|\rho\alpha\beta} K^S {}^\rho {}_{|\nu]}
\\
&+ \left[ \left(\wn_{(\mu|} +a_{(\mu|}\right) \left(\wn_{|\beta)} +a_{|\beta)} \right) + K^A_{(\mu|\rho}K^S{}^\rho{}_{|\beta)} \right] K^S_{\nu\alpha}
\\
&- \left[ \left(\wn_{(\mu|} +a_{(\mu|}\right) \left(\wn_{|\alpha)} +a_{|\alpha)} \right) + K^A_{(\mu|\rho}K^S{}^\rho{}_{|\alpha)} \right] K^S_{\nu\beta}
\\
&- \left[ \left(\wn_{(\nu|} +a_{(\nu|}\right) \left(\wn_{|\beta)} +a_{|\beta)} \right) + K^A_{(\nu|\rho}K^S{}^\rho{}_{|\beta)} \right] K^S_{\mu\alpha}
\\
&+ \left[ \left(\wn_{(\nu|} +a_{(\nu|}\right) \left(\wn_{|\alpha)} +a_{|\alpha)} \right) + K^A_{(\nu|\rho}K^S{}^\rho{}_{|\alpha)} \right] K^S_{\mu\beta}
\ .
\end{split}
\end{equation}
We can use this identity to derive a similar expression for the Lie derivative of the scalar $\widehat{R}$:
\begin{equation}
\Lie{n}{\widehat{R}} = -2K^S{}^{\alpha\mu}\widehat{R}_{\alpha\mu} +2 (\wn_\alpha +a_\alpha)(\wn_\rho +a_\rho)K^S{}^{\alpha\rho}
-2 (\wn_\alpha +a_\alpha)(\wn^\alpha +a^\alpha)K^S.
\end{equation}

Finally, we have the following formulas for integration by parts in terms of space tangent derivatives and Lie derivatives in the direction of $n^\mu$:
\begin{align}
\begin{split}
\int \sqrt{-g} \wn_\mu \tilde J^\mu  = - \int \sqrt{-g} a_\mu \tilde J^\mu
,
\qquad
\int \sqrt{-g} \Lie{n} \phi = - \int \sqrt{-g} K_S \phi   ,
\end{split}
\end{align}
where $\tilde J^\mu$ is a space tangent vector and $\phi$ is a scalar.

\subsection{Milne Boost Transformation Rules}
\label{App:BoostTrans}

In this subsection we detail the transformations of the various basic tangent tensors under infinitesimal Milne boosts as derived from the definitions in \eqref{BoostTrans}, and in subsection \ref{app:general_ids}. From \eqref{BoostTrans} we have:

\begin{align}
& \delw n^\mu = W^\mu,
 &&\delw n_\mu = 0, \notag \\
& \delw g^{\mu\nu} = -W^\mu n^\nu-W^\nu n^\mu,
 && \delw g_{\mu\nu} = W_\mu n_\nu+W_\nu n_\mu, \\
& \delw P^{\mu\nu} = 0 ,
&& \delw P_\mu^\nu = n_\mu W^\nu, \notag
\\
& \delw P_{\mu\nu} =W_\mu n_\nu +W_\nu n_\mu \notag.
\end{align}

From these transformations, and the expressions for the Levi-Civita connection and the acceleration, we obtain:
\begin{align}
\begin{split}
\delw \Gamma^\gamma_{\alpha\beta} = & \frac{1}{2} g^{\gamma\delta}
 \left[\nabla_\alpha(W_\beta n_\delta) + \nabla_\alpha (W_\delta n_\beta) +\nabla_\beta(W_\alpha n_\delta) \right.\\
& ~~~~~ \left.+\nabla_\beta (W_\delta n_\alpha) - \nabla_\delta(W_\alpha n_\beta ) - \nabla_\delta (W_\beta n_\alpha) \right],
\\
P^\gamma_{\gamma'} P_\alpha^{\alpha'} P_\beta^{\beta'} \delw \Gamma^{\gamma'}_{\alpha'\beta'} =
& \ W^\gamma K^S_{\alpha\beta} +
W_\alpha (K_A)_\beta{}^\gamma +
W_\beta (K_A)_\alpha{}^\gamma ,
\end{split}
\\
\begin{split}
\delw a_\mu = & \
2 W^\nu K_{\nu\mu}^A + (W^\nu a_\nu) n_\mu.
\end{split}
\end{align}

For future convenience, we define the space projected boost transformation of a space tangent tensor as follows:
\begin{equation}
\delwt \widetilde{T}_{\alpha\beta\gamma\ldots}
= P_\alpha^{\alpha'} P_\beta^{\beta'} \ldots \delw
 \widetilde{T}_{\alpha'\beta'\gamma'\ldots},
\end{equation}
where $ \widetilde{T}_{\alpha\beta\gamma\ldots}$ is a space tangent tensor.
Since $\delw P^{\mu\nu}=0$, this projected boost transformation operator satisfies the Leibniz product rule:
\begin{equation}
\delwt \left[\widetilde T_{\alpha\beta\ldots\mu\nu\ldots} \widetilde S_{\gamma\delta\ldots}{}^{\mu\nu\ldots}\right] = \delwt \widetilde T_{\alpha\beta\ldots\mu\nu\ldots} \widetilde S_{\gamma\delta\ldots}{}^{\mu\nu\ldots} + \widetilde T_{\alpha\beta\ldots\mu\nu\ldots} \delwt \widetilde S_{\gamma\delta\ldots}{}^{\mu\nu\ldots},
\end{equation}
where $\widetilde T_{\alpha\beta\ldots\mu\nu\ldots}$ and $\widetilde S_{\gamma\delta\ldots\mu\nu\ldots} $ are space tangent tensors, and for a scalar expression $\phi$ we have:
\begin{equation}
\delw \phi = \delwt \phi.
\end{equation}
Therefore for a scalar $\phi$ built from contractions of space tangent tensors, we may safely substitute $\delw$ with $\delwt$ in our analysis.

The following space projected boost transformation rules can be derived from the definitions and identities in subsection \ref{app:general_ids}:
\begin{align}
& \delwt P_{\mu\nu} = 0 ,
\\
& \delwt a_\mu = - 2 K_{\mu\nu}^A W^\nu  \stackrel{2d}{=}
-2 K_A \tilde \epsilon_{\mu\nu} W^\nu,
\\
\begin{split}
& \delwt \wn_\mu \widetilde T_{\alpha\beta\gamma\ldots}
=
 \wn_\mu \delw \widetilde T_{\alpha\beta\gamma\ldots}
+
\left[K_{\mu\alpha}^A W_\rho - K_{\alpha\rho}^A W_\mu - K_{\mu\rho}^A W_\alpha \right] \widetilde T^\rho{}_{\beta\gamma\ldots} + \ldots
\\
& \qquad \qquad ~~~
\stackrel{2d}{=}
 \wn_\mu \delw \widetilde T_{\alpha\beta\gamma\ldots}
+ K_A
\left[\tilde\epsilon_{\mu\alpha} W_\rho - \tilde\epsilon_{\alpha\rho} W_\mu - \tilde\epsilon_{\mu\rho} W_\alpha \right] \widetilde T^\rho{}_{\beta\gamma\ldots} + \ldots ,
\end{split}
\\
&
\delwt K^A_{\alpha\beta} = 0
\qquad
\stackrel{2d}{\longrightarrow}
\qquad
\delwt K^A = 0,
\\
&
\delwt \Lie{n}  T_{\alpha\beta\gamma\ldots}
=
 \Lie{n} \delwt \widetilde T_{\alpha\beta\gamma\ldots}
+W^\rho \wn_\rho  \widetilde T_{\alpha\beta\gamma\ldots}
+ \left[(\wn_\alpha+a_\alpha) W^\rho \right]\widetilde T_{\rho\beta\gamma\ldots} + \ldots,
\\
&
\delwt K^S_{\alpha\beta} = \frac{1}{2} (\wn_\alpha+a_\alpha) W_\beta
+
\frac{1}{2} (\wn_\beta+a_\beta) W_\alpha,
\\
\begin{split}
&\delwt \widehat R_{\alpha\beta\mu\nu} = -\wn_\mu (K^A_{\alpha\beta} W_\nu) +\wn_\nu (K^A_{\alpha\beta} W_\mu ) - \wn_\alpha (K_{\mu\nu}^A W_\beta) + \wn_\beta (K^A_{\mu\nu} W_\alpha)
\\
& ~~~~~~~~~~~~~~~
- K^A_{\alpha\mu} (\wn_{[\beta} +a _{[\beta}) W_{\nu]} + K^A_{\alpha\nu} (\wn_{[\beta} +a_{[\beta}) W_{\mu]}
\\
& ~~~~~~~~~~~~~~~
+ K^A_{\beta\mu} (\wn_{[\alpha} +a _{[\alpha}) W_{\nu]} - K^A_{\beta\nu} (\wn_{[\alpha} +a_{[\alpha}) W_{\mu]},
\end{split}
\\
\begin{split}
&\delwt \widehat R = -4 \wn_\alpha (K_A^{\alpha\beta} W_\beta)
-2 K_A^{\alpha\beta} (\wn_\alpha +a_\alpha) W_\beta
\\
&
~~~~~
\stackrel{2d}{=}
-4\tilde\epsilon^{\alpha\beta} \wn_\alpha K_A W_\beta
-6\tilde\epsilon^{\alpha\beta} K_A \wn_\alpha  W_\beta
-2\tilde\epsilon^{\alpha\beta} K_A a_\alpha W_\beta ,
\end{split}
\\
& \delwt \tilde \epsilon_{\alpha\beta\gamma \ldots} = 0.
\end{align}
Finally, for the transformation rules of the gauge field, the electric field and the magnetic field we have the following:
\begin{align}
&\delw A_\mu = - W_\mu, \\
&\delw F_{\mu\nu} = - 2 \del_{[\mu} W_{\nu]}, \\
\begin{split}
&\delw E_\mu = W^\nu F_{\mu\nu} -2 n^\nu \del_{[\mu} W_{\nu]} = W^\nu F_{\mu\nu} +\Lien W_\mu, \\
&\delwt E_\mu = W^\nu B_{\mu\nu} +\Lien W_\mu,
\end{split}
\\
\begin{split}
& \delw B_{\mu\nu} = 2 n_{[\mu} W^{\mu'} B_{\mu' \nu]}
-2 \wn _{[\mu} W_{\nu]},
\\
& \delwt B_{\mu\nu} = -2 \wn _{[\mu} W_{\nu]} \quad
\stackrel{2d}{\longrightarrow}
\quad
\delwt B = -\tilde \epsilon^{\mu\nu} \wn_\mu W_\nu.
\end{split}
\end{align}

\subsection{Weyl Transformation Rules}
\label{app:WeylTrans}
In this subsection we detail the transformations of the various basic tangent tensors under (infinitesimal) anisotropic Weyl transformations. These transformations can be derived from the definitions in \eqref{eq:RelSymmWeylTranform}, and in subsection \ref{app:general_ids}. Starting from \eqref{eq:RelSymmWeylTranform}, we have:
\begin{equation}
\begin{split}
&\dels P_{\mu\nu} = 2\sigma P_{\mu\nu}, \qquad
\dels n^\mu = -z\sigma n^\mu, \qquad
\dels n_\mu = z\sigma n_\mu,
\\
& \dels g_{\mu\nu} = 2\sigma g_{\mu\nu} +2(1-z)\sigma n_\mu n_\nu, \qquad
\dels P^\alpha{}_{\alpha'} = 0.
\end{split}
\end{equation}
Next, from these transformations and the expression for the Levi-Civita connection and the Lie derivative we obtain the following:
\begin{equation}
\begin{split}
&
\dels \Lie{n} \widetilde T_{\alpha\beta\gamma\ldots} =
-z\sigma \Lie{n} \widetilde T_{\alpha\beta\gamma\ldots} +
 \Lie{n} \delta \widetilde T_{\alpha\beta\gamma\ldots},
\\
&
\dels \tilde\Gamma^\nu {}_{\mu\alpha}
= P_\alpha^\nu \wn_\mu \sigma  +P_\mu^\nu  \wn_\alpha \sigma  - P_{\mu\alpha} \wn^\nu \sigma ,
\\
&
\dels (\wn_\mu \widetilde T_{\alpha\beta\gamma\ldots}) = \wn_\mu  (\dels \widetilde T_{\alpha\beta\gamma\ldots})
- I[\widetilde T] \wn_\mu \sigma \widetilde T_{\alpha\beta\ldots}
\\
&~~~~~~~~~~~~~~~~~~~~~~~~
- ( \wn_\alpha \sigma ) \widetilde T_{\mu\beta\gamma \ldots} +\wn_\rho \sigma P_{\mu\alpha} \widetilde T^\rho{}_{\beta\ldots} - \ldots,
\end{split}
\end{equation}
where $I[\widetilde T]$ is the number of indexes of $ \widetilde T_{\alpha\beta\gamma\ldots} $ and we have defined the space projected Levi-Civita connection:
\begin{equation}
\tilde\Gamma^\nu {}_{\mu\alpha} = P_{\nu'}^\nu P^{\mu'}_\mu P^{\alpha'}_\alpha \tilde\Gamma^{\nu'} {}_{\mu'\alpha'}.
\end{equation}
From these formulas, as well as the definitions and identities of subsection \ref{app:general_ids}, the following Weyl transformation rules can be derived:
\begin{equation}
\begin{split}\label{Weyl_trans_laws}
&
\dels K_{\mu\nu}^S = (2-z) \sigma K_{\mu\nu}^S +P_{\mu\nu} \Lie{n} \sigma,
\\
&
\dels a_\mu = z\wn_\mu\sigma,
\\
&
\dels K_{\mu\nu}^A = z \sigma K_{\mu\nu}^A,
\\
&
\dels K^A = (z-2) \sigma K^A,
\\
&
\dels \tilde \epsilon_{\alpha\beta\ldots} = d \sigma \tilde \epsilon_{\alpha\beta\ldots},
\\
&
\dels \widehat R_{\alpha\rho\mu\nu} = 2\sigma \widehat R_{\alpha\rho\mu\nu}
+ P_{\alpha\nu} \wn_{(\rho} \wn_{\mu)} \sigma
- P_{\alpha\mu} \wn_{(\rho} \wn_{\nu)} \sigma
\\
&
~~~~~~~~~~~~~~~~~~~~~~~~~~~+
P_{\rho\mu} \wn_{(\alpha} \wn_{\nu)} \sigma
-
P_{\rho\nu} \wn_{(\alpha} \wn_{\mu)} \sigma,
\\
&
 \dels \widehat{R}_{\alpha\mu} = (2-d) \wn_{(\alpha} \wn_{\mu)} \sigma -P_{\alpha\mu} \widetilde \Box \sigma,
\\
&
\dels \widehat R = -2 \sigma \widehat R -2(d-1) \widetilde \Box \sigma.
\end{split}
\end{equation}
Finally, for the transformations of the gauge field, the electric field and the magnetic field we have:
\begin{align}
\begin{split}\label{eq:WeylTransGaugeSector}
&\dels A_\mu = (2-z) \sigma A_\mu, \\
&\dels F_{\mu\nu} = (2-z) \sigma F_{\mu\nu} -2 (2-z) A_{[\mu}\del_{\nu]} \sigma, \\
& \dels E_\mu = (2-2z) \sigma E_\mu -2 (2-z) n^\nu A_{[\mu}\del_{\nu]} \sigma
\\ & \qquad ~
= (2-2z) \sigma E_\mu - (2-z) [\tilde A_\mu \Lien \sigma - A_0 \wn_\mu \sigma], \\
& \dels B_{\mu\nu} = (2-z) \sigma B_{\mu\nu} - 2(2-z) \tilde A_{[\mu} \wn_{\nu]} \sigma.
\end{split}
\end{align}

Note that for $z \neq 2$,  when applying a Weyl transformation to a gauge invariant expression one can obtain an non-gauge-invariant expression, since the $U(1)$ and the scale symmetries no longer commute in this case.
For the case of $2+1$ dimensions with $z=2$, these transformation rules reduce to:
\begin{equation}
\dels B = -d\sigma B = -2\sigma B, \qquad \dels E_\mu = (2-2z)\sigma E_\mu = -2\sigma E_\mu .
\end{equation}

\section{The Case with Frobenius and Galilean Boost Invariance}\label{app:FrobBoost}
In the following appendix we detail the calculations behind the results of the cohomological analysis for the case with the Frobenius condition satisfied and with Galilean boost invariance. The calculations are organized according to various sectors as explained in subsection
\ref{subsec:WithFrobeniusWithBoost}.
In this case we set $K_A=0$ since the Frobenius condition is satisfied and we include the gauge field contributions. The equations for classification by sectors  \eqref{rest:constraint1}, \eqref{rest:constraint1b} read:
\begin{equation}
\begin{split}
& n_T =  n_{K_S} +n_\mathcal{L}  +n_B +2n_E,
\\
& n_S =  n_a + n_\nabla + 2 n_R  - n_E,
\\
& 2n_T +n_S = 4.
\end{split}
\end{equation}
We use the notations of subsection  \ref{ssb:prescription} for the various expressions in the cohomological analysis.

\subsection{The (2,0,0) Sector}\label{FrobBoost2_0_0}
This sector contains the following ghost number zero, TPD and gauge invariant, independent  expressions:
\begin{equation}
\begin{alignedat}{3}
 & \phi_1 =  K_S^{\mu\nu} K^S_{\mu\nu},\qquad &
 & \phi_2 =  K_S^2, \qquad &
 &\phi_3 =  \Lien K_S,
 \\
 & \phi_4 = B^2 ,\qquad &
 & \phi_5 = E_\mu a^\mu, \qquad &
 & \phi_6 = \wn_\mu E^\mu.
\end{alignedat}
\end{equation}
If we now define the independent boost-ghost number one expressions:
\begin{equation}
\begin{alignedat}{4}
& \chi_1 = W^\mu a^\nu K^S_{\mu\nu}, \quad &
& \chi_2 = W^\mu a_\mu K_S, \qquad &
& \chi_3 = K_S \wn_\mu W^\mu, \qquad &
& \chi_4 = K_S^{\alpha\beta} \wn_\alpha W_\beta, \\
& \chi_5 = W_\mu \wn^\mu K_S, \quad &
& \chi_6 = W^\alpha \wn^\beta K^S_{\alpha\beta}, \qquad &
& \chi_7 = a^\mu \Lien W_\mu, \qquad &
& \chi_8 = W^\mu \Lien a_\mu, \\
& \chi_9 = \Lien \wn_\mu W^\mu, \quad &
&\chi_{10} = B \temn a_\mu W_\nu ,\quad &
&\chi_{11} = B\temn \wn_\mu W_\nu ,\quad &
&\chi_{12} = \temn W_\nu \wn_\mu B,
\end{alignedat}
\end{equation}
then the Milne boost transformations read  $\delwt \phi_i = B_{ij} \chi_j$, where $B_{ij}$ is given by:
\settowidth{\mycolwd}{$\,-z\,$} 
\begin{equation}
B_{ij} =
\left(
\begin{array}{*{12}{@{}I{\mycolwd}@{}}}
0 & 2 & 2 & 0 & 0 & 0 & 0 & 0 & 0 & 0 & 0 & 0 \\
2 & 0 & 0 & 2 & 0 & 0 & 0 & 0 & 0 & 0 & 0 & 0  \\
-2 & 0 & 0 & 0 & 1 & 0 & 1 & 1 & 1 & 0 & 0 & 0 \\
0 & 0 & 0 & 0 & 0 & 0 & 0 & 0 & 0 & 0 & -2 & 0 \\
0 & 0 & 0 & 0 & 0 & 0 & 1 & 0 & 0 & 1 & 0 & 0 \\
2 & -1 & 0 & 2 & -1 & 2 & -1 & 0 & 1 & 0 & 1 & 1 \\
\end{array}
\right).
\end{equation}
It is easy to check that no boost invariant combinations exist in this sector.

\subsection{The (2,0,1) Sector}\label{FrobBoost2_0_1}
This sector contains the following ghost number zero, TPD and gauge invariant, independent  expressions:
\begin{equation}
\phi_1 = K_S B, \qquad \phi_2 = \Lie{n} B, \qquad
\phi_3 = \temn E_\mu a_\nu.
\end{equation}
Note that $\temn \wn_\mu E_\nu$ is related to the others by the Maxwell equation \eqref{eq:Maxwell}.
The boost-ghost number one expressions read:
\begin{equation}
\begin{alignedat}{3}
& \chi_1 = B \wn_\alpha W^\alpha, \qquad &
& \chi_2 = B a^\alpha W_\alpha, \qquad &
& \chi_3 = K_S \temn \wn_\mu W_\nu,
\\
& \chi_4 = \temn \Lie{n} \wn_\mu W_\nu ,\qquad &
& \chi_5 = W^\alpha \wn_\alpha B, \qquad&
&\chi_6 = \temn a_\nu \Lie{n} W_\mu.
\end{alignedat}
\end{equation}
The Milne boost transformations read  $\delwt \phi_i = B_{ij} \chi_j$, where $B_{ij}$ is given by:
\begin{equation}
B_{ij} =
\left(
\begin{array}{*{6}{@{}I{\mycolwd}@{}}}
1 & 1 & -1 & 0 & 0 & 0  \\
0 & 0 & 1 & -1 & 1 & 0   \\
0 & 1 & 0 & 0 & 0 & 1  \\
\end{array}
\right),
\end{equation}
and there are no boost invariant expressions in this sector.

\subsection{The (1,2,0) Sector} \label{FrobBoost1_2_0}
This sector contains the following ghost number zero expressions:
\begin{equation}
\begin{alignedat}{4}
& \phi_1 = K_S a^2, \qquad &
& \phi_2 = K^S_{\alpha\beta} a^\alpha a^\beta, \qquad &
& \phi_3 = K_S \widehat{R} , \qquad &
& \phi_4 = K_S \wn_\alpha a^\alpha ,
 \\
& \phi_5 = K_S^{\alpha\beta} \wn_\alpha a_\beta , \qquad &
& \phi_6 = a^\alpha \wn_\alpha K_S , \qquad & &\phi_7 = a_\alpha \wn_\beta K_S^{\alpha\beta} , \qquad&
& \phi_8 = \widetilde\Box K_S ,\\
 & \phi_9 = \wn_\alpha \wn_\beta K_S^{\alpha\beta} , \qquad &
 & \phi_{10} = a^\alpha \Lie{n} a_\alpha,
\qquad &
&\phi_{11} = \Lie{n} \wn_\alpha a^\alpha,
\qquad &
& \phi_{12} = \epsilon_{\alpha\beta} a^\alpha \wn^\beta B.
\end{alignedat}
\end{equation}
The boost-ghost number one expressions read:
\begin{equation}
\begin{alignedat}{4}
& \chi_1 =  \widehat R \wn_\mu W^\mu, \quad &
& \chi_2 = W^\mu \wn_\mu \widehat R, \quad &
& \chi_3 = \widehat R W^\mu a_\mu, \quad &
& \chi_4 = \widetilde \Box \wn_\mu W^\mu,
\\
& \chi_5 = \wn_\mu W^\mu \wn_\nu a^\nu,
\quad &
&\chi_6 =  \wn_\mu W_\nu \wn^\mu a^\nu ,\quad ~&
& \chi_7 = W^\mu \wn_\mu \wn_\nu a^\nu, \quad ~ &
& \chi_8 = a^\mu \wn_\mu \wn_\nu W^\nu,
\\
& \chi_9  = a^\mu \widetilde \Box W_\mu , \quad &
& \chi_{10} = a^2 \wn_\mu W^\mu ,\quad &
& \chi_{11} = a^\mu a^\nu \wn_\mu W_\nu ,\quad &
& \chi_{12} = \wn_\mu a^\mu W^\nu a_\nu ,
\\
& \chi_{13} = \wn_\mu a_\nu W^\mu a^\nu,\quad &
& \chi_{14} = a^2 W_\mu a^\mu.
\end{alignedat}
\end{equation}
The matrix $B_{ij}$ for the Milne boost transformations reads:
\settowidth{\mycolwd}{$\,-z\,$} 
\begin{equation}
B_{ij} = \left(
\begin{array}{*{14}{@{}I{\mycolwd}@{}}}
0 & 0 & 0 & 0 & 0 & 0 & 0 & 0 & 0 & 1 & 0 & 0 &0 &1 \\
0 & 0 & 0 & 0 & 0 & 0 & 0 & 0 & 0 & 0 & 1 & 0 &0 &1 \\
1 & 0 & 1 & 0 & 0 & 0 & 0 & 0 & 0 & 0 & 0 & 0 &0 &0\\
0 & 0 & 0 & 0 & 1 & 0 & 0 & 0 & 0 & 0 & 0 & 1 &0 &0\\
0 & 0 & 0 & 0 & 0 & 1 & 0 & 0 & 0 & 0 & 0 & 0 &1 &0\\
0 & 0 & 0 & 0 & 0 & 0 & 0 & 1 & 0 & 0 & 1 & 0 &1 &0\\

0 & 0 & \frac{1}{4} & 0 & 0 & 0 & 0 & \frac{1}{2} & \frac{1}{2} & \frac{1}{2} & \frac{1}{2} & \frac{1}{2} &\frac{1}{2} &0\\
0 & 0 & \frac{1}{2} & 1 & 0 & 2 & 1 & 0 & 1 & 0 & 0 & 0 &0 &0\\
\frac{1}{2} & \frac{1}{2} & \frac{1}{2} & 1 & 1 & 1 & 1 & 1 & 0 & 0 & 0 & 0 &0 &0\\

0 & 0 & 0 & 0 & 0 & 0 & 0 & 0 & 0 & 0 & 1 & 0 &1 &1\\
0 & 0 & 0 & 0 & 0 & 0 & 1 & 0 & 0 & 0 & 0 & 0 &0 &0\\
0 & 0 & -\frac{1}{2} & 0 & 0 & 0 & 0 & -1 & 1 & 0 & 0 & 0 &0 &0\\
\end{array}
\right),
\end{equation}
and no boost invariant combinations exist.

\subsection{The (1,2,1) Sector}
\label{FrobBoost1_2_1}
This sector contains the following ghost number zero expressions:
\begin{equation}
\begin{alignedat}{4}
& \phi_1 = \temn a_\mu \wn_\nu K_S, \quad~ &
& \phi_2 = \tilde K_{\mu\nu}^S a^\mu a^\nu, \quad ~ &
& \phi_3 =  \tilde K_{\mu\nu}^S \wn^\mu a^\nu, \qquad &
& \phi_4 =  a_\mu \wn_\nu \tilde K_S^{\mu\nu}, \\
& \phi_5 =  \wn_\mu \wn_\nu \tilde K_S^{\mu\nu}, &
& \phi_6 =  \temn a_\mu \Lie{n} a_\nu, \quad ~ &
& \phi_7 =  B a^2, &
& \phi_8 =  B \wn_\alpha a^\alpha,
\\
& \phi_9 = \tb B, &
& \phi_{10} = B \widehat R, &
& \phi_{11} = a^\alpha \wn_\alpha B.
\end{alignedat}
\end{equation}
The boost-ghost number one expressions read:
\begin{equation}
\begin{alignedat}{3}
& \chi_1 = \temn W_\mu a_\nu a^2, \qquad &
& \chi_2 = \temn W_\mu a_\nu \wn_\alpha a^\alpha, \qquad &
& \chi_3 = \temn W_\mu a^\alpha\wn_\nu a_\alpha,
\\
& \chi_4 = \temn \wn_\mu W_\nu a^2, \qquad &
& \chi_5 = \temn \wn_\mu W_\alpha a^\alpha a_\nu,
\qquad &
& \chi_6 = \widehat R \temn a_\mu W_\nu,
\\
& \chi_7 = \temn \wn_\alpha \wn_\mu W_\nu a^\alpha, \qquad &
& \chi_8 = \temn \wn^\alpha \wn_\mu W_\alpha a_\nu,\qquad &
&\chi_9 = \temn \wn_\mu W_\nu  \wn_\alpha a^\alpha,
\\
& \chi_{10} = \temn \wn_\mu W_\alpha \wn_\nu a^\alpha, \qquad &
&\chi_{11} = \temn W_\mu \wn_\nu \wn_\alpha a^\alpha
, \qquad &
& \chi_{12} = \temn \wn_\mu \widehat R W_\nu,
\\
& \chi_{13} = \temn \widehat R \wn_\mu W_\nu, \qquad &
& \chi_{14} = \temn \widetilde \Box \wn_\mu W_\nu.
\end{alignedat}
\end{equation}
The matrix $B_{ij}$ for the Milne boost transformations reads:
\settowidth{\mycolwd}{$\,-z\,$} 
\begin{equation}
B_{ij} = \left(
\begin{array}{*{14}{@{}I{\mycolwd}@{}}}
 0 & -1 & 1 & 0 & -1 & -\frac{1}{2} & 0 & -1 & 0 & 0 & 0 & 0 & 0 & 0 \\
 -\frac{1}{2} & 0 & 0 & \frac{1}{2} & -1 & 0 & 0 & 0 & 0 & 0 & 0 & 0 & 0 & 0 \\
 0 & \frac{1}{2} & -1 & 0 & 0 & 0 & 0 & 0 & \frac{1}{2} & -1 & 0 & 0 & 0 & 0 \\
 0 & -\frac{1}{2} & 0 & \frac{1}{2} & 0 & \frac{1}{4} & \frac{1}{2} & -\frac{1}{2} & 0 & 0 & 0 & 0 & 0 & 0 \\
 0 & 0 & 0 & 0 & 0 & \frac{1}{4} & \frac{1}{2} & \frac{1}{2} & 1 & 0 & -\frac{1}{2} & \frac{1}{2} & 0 & 0 \\
 0 & -1 & 1 & 0 & -1 & 0 & 0 & 0 & 0 & 0 & 0 & 0 & 0 & 0 \\
 0 & 0 & 0 & -1 & 0 & 0 & 0 & 0 & 0 & 0 & 0 & 0 & 0 & 0 \\
 0 & 0 & 0 & 0 & 0 & 0 & 0 & 0 & -1 & 0 & 0 & 0 & 0 & 0 \\
 0 & 0 & 0 & 0 & 0 & 0 & 0 & 0 & 0 & 0 & 0 & 0 & 0 & -1 \\
 0 & 0 & 0 & 0 & 0 & 0 & 0 & 0 & 0 & 0 & 0 & 0 & -1 & 0 \\
 0 & 0 & 0 & 0 & 0 & 0 & -1 & 0 & 0 & 0 & 0 & 0 & 0 & 0 \\
\end{array}
\right),
\end{equation}
and once again no boost invariant combinations exist.

\section{The Case without Frobenius and with Galilean Boost Invariance}\label{app:B_noF}

In the following appendix we detail the calculations behind the results of the cohomological analysis for the case in which  the Frobenius condition is not satisfied and with Galilean boost invariance.
The calculations are organized according to various sectors as explained in subsection
\ref{subsec:WithoutFrobeniusWithBoost}.
In this case we include contributions from the gauge field since we have Galilean boost invariance and include $K_A$ since the Frobenius condition is not satisfied.
The equations for classification by sectors  \eqref{rest:constraint1}, \eqref{rest:constraint1b} become:
\begin{equation}
\begin{split}
& n_T =  n_{K_S} - n_{K_A} +n_\mathcal{L}  +n_B +2n_E,
\\
& n_S =  2 n_{K_A} + n_a + n_\nabla + 2 n_R  - n_E,
\\
& 2n_T +n_S = 4.
\end{split}
\end{equation}
This case contains an infinite number of sectors, but as explained in subsection  \ref{subsec:WithoutFrobeniusWithBoost} we focus on the sectors with $n_D=n_T+n_S<4$ and the parity even sector with $n_D = 4$.
We again follow everywhere the notations of subsection  \ref{ssb:prescription}. All results agree  with the expectations from the null reduction as described in subsection \ref{subsubsec:nullRed} and in  \cite{Jensen:2014hqa}.

\subsection{The (2,0,0) Sector}\label{app:NFWB2_0_0}

This sector contains the following ghost number zero, TPD and gauge invariant, independent  expressions:
\begin{equation}
\begin{alignedat}{3}
& \phi_1 = K_S^2, \qquad &
& \phi_2 = Tr(K_S^2), \qquad &
& \phi_3 = \Lie{n} K_S,
\\
& \phi_4 = B^2, \qquad &
& \phi_5 = \wn_\mu E^\mu, \qquad &
& \phi_6 = a_\mu E^\mu.
\end{alignedat}
\end{equation}
The boost-ghost number one expressions are given by:
\begin{equation}
\begin{alignedat}{3}
& \chi_1 = K_S a^\mu W_\mu, \qquad &
& \chi_2 = K_S^{\alpha\beta} a_\alpha W_\beta, \qquad &
& \chi_3 = a^\mu \Lie{n} W_\mu,
\\
& \chi_4 = W^\mu \Lie{n} a_\mu, \qquad &
& \chi_5 = \wn_\alpha K_S^{\alpha\beta} W_\beta, \qquad &
& \chi_6 = \wn_\mu K_S W^\mu,
\\
& \chi_7 = K_S^{\alpha\beta} \wn_\alpha W_\beta, \qquad &
& \chi_8 = K_S \wn_\alpha W^\alpha, \qquad &
& \chi_9 = \Lie{n} \wn_\alpha a^\alpha,
\\
& \chi_{10} = B \temn a_\mu W_\nu, \qquad &
& \chi_{11} = B \temn \wn_\mu W_\nu, \qquad &
& \chi_{12}  = \temn \wn_\mu B W_\nu,
\\
& \chi_{13} = K_A \temn E_\mu W_\nu.\qquad &
\end{alignedat}
\end{equation}
The Milne boost transformations read  $\delwt \phi_i = B_{ij} \chi_j$, where $B_{ij}$ is given by:
\settowidth{\mycolwd}{$\,-z\,$} 
\begin{equation}
B_{ij} =
\left(
\begin{array}{*{13}{@{}I{\mycolwd}@{}}}
 2 & 0 & 0 & 0 & 0 & 0 & 0 & 2 & 0 & 0 & 0 & 0 & 0 \\
 0 & 2 & 0 & 0 & 0 & 0 & 2 & 0 & 0 & 0 & 0 & 0 & 0 \\
 0 & -2 & 1 & 1 & 0 & 0 & 0 & 0 & 1 & 0 & 0 & 0 & 0 \\
 0 & 0 & 0 & 0 & 0 & 0 & 0 & 0 & 0 & 0 & -2 & 0 & 0 \\
 -1 & 2 & -1 & 0 & 2 & -1 & 2 & 0 & 1 & 0 & 1 & 1 & 2 \\
 0 & 0 & 1 & 0 & 0 & 0 & 0 & 0 & 0 & 1 & 0 & 0 & -2 \\
\end{array}
\right).
\end{equation}
It is easy to check that no boost invariant combinations exist in this sector.

\subsection{The (2,0,1) Sector}\label{app:NFWB2_0_1}
This sector contains following three ghost number zero expressions:
\begin{equation}
\phi_1 = BK_S, \qquad \phi_2 = \temn a_\mu E_\nu, \qquad \phi_3 = \temn \wn_\mu E_\nu.
\end{equation}
The boost-ghost number one expressions are given by:
\begin{equation}
\begin{alignedat}{3}
& \chi_1 = K_S  \temn a_\mu W_\nu, \qquad &
& \chi_2 = \tilde K_S^{\alpha \beta} a_\alpha W_\beta, \qquad &
& \chi_3 = \temn a_\mu \Lie{n} W_\nu
\\
& \chi_4 = \temn \Lie{n} a_\mu W_\nu, \qquad &
& \chi_5 = \wn_\alpha \tilde K_S^{\alpha\beta} W_\beta, \qquad &
& \chi_6 = \temn \wn_\mu K_S W_\nu,
\\
& \chi_7 = \tilde K_S^{\alpha\beta} \wn_\alpha W_\beta, \qquad &
& \chi_8 = K_S \temn \wn_\mu W_\nu, \qquad &
& \chi_9 = \temn \Lie{n} \wn_\mu W_\nu,
\\
& \chi_{10} = B a_\mu W^\mu, \qquad &
& \chi_{11} = W^\mu \wn_\mu B, \qquad &
& \chi_{12} = B \wn_\mu W^\mu,
\\
& \chi_{13} = K_A E^\mu W_\mu,
\end{alignedat}
\end{equation}
and the matrix $B_{ij}$ for the Milne boost transformations is given by:
\settowidth{\mycolwd}{$\,-z\,$} 
\begin{equation}
B_{ij} =
\left(
\begin{array}{*{13}{@{}I{\mycolwd}@{}}}
 0 & 0 & 0 & 0 & 0 & 0 & 0 & -1 & 0 & 1 & 0 & 1 & 0 \\
 0 & 0 & 1 & 0 & 0 & 0 & 0 & 0 & 0 & -1 & 0 & 0 & -2 \\
 0 & 0 & -1 & 0 & 0 & 0 & 0 & 0 & 1 & 0 & -1 & -1 & 2 \\
\end{array}
\right).
\end{equation}
No boost invariant combinations exist in this sector.

\subsection{The (1,2,0) Sector}\label{app:NFWB1_2_0}
This sector contains the following ghost number zero expressions:
\begin{equation}
\begin{alignedat}{4}
& \phi_1 = K_S a^2, \quad &
& \phi_2 = K^S_{\alpha\beta} a^\alpha a^\beta, \quad &
& \phi_3 = K_S \widehat{R} , \quad &
& \phi_4 = K_S \wn_\alpha a^\alpha ,
 \\
& \phi_5 = K_S^{\alpha\beta} \wn_\alpha a_\beta , \quad &
& \phi_6 = a^\alpha \wn_\alpha K_S , \quad & & \phi_7 = a_\alpha \wn_\beta K_S^{\alpha\beta} , \quad &
& \phi_8 = \widetilde\Box K_S ,
 \\
& \phi_9 = \wn_\alpha \wn_\beta K_S^{\alpha\beta} , \quad &
& \phi_{10} = a^\alpha \Lie{n} a_\alpha,
\quad &
& \phi_{11} = \Lie{n} \wn_\alpha a^\alpha, \quad &
& \phi_{12} = \teab  a_\alpha \wn_\beta B,
\\
& \phi_{13} = K_A \temn E_\mu a_\nu ,\quad ~&
& \phi_{14} = \temn \wn_\mu K_A E_\nu, \quad&
& \phi_{15} = B \Lie{n} K_A ,\quad&
& \phi_{16} = K_A \Lie{n} B,
\\
& \phi_{17}  = K_A B K_S.
\end{alignedat}
\end{equation}
Note that we have used the Maxwell equation \eqref{eq:Maxwell} to relate some dependent terms.
The boost-ghost number one expressions read:
\begin{equation}
\begin{alignedat}{3}
& \chi_1 =  \widehat R \wn_\mu W^\mu, \qquad &
& \chi_2 = W^\mu \wn_\mu \widehat R, \qquad &
& \chi_3 = \widehat R W^\mu a_\mu,
\\
& \chi_4 = \widetilde \Box \wn_\mu W^\mu, \qquad &
&\chi_5 = \wn_\mu W^\mu \wn_\nu a^\nu,
\qquad &
&\chi_6 =  \wn_\mu W_\nu \wn^\mu a^\nu ,
\\
& \chi_7 = W^\mu \wn_\mu \wn_\nu a^\nu, \qquad &
& \chi_8 = a^\mu \wn_\mu \wn_\nu W^\nu, \qquad &
& \chi_9  = a^\mu \widetilde \Box W_\mu ,
\\
& \chi_{10} = a^2 \wn_\mu W^\mu ,\qquad &
& \chi_{11} = a^\mu a^\nu \wn_\mu W_\nu ,\qquad &
& \chi_{12} = \wn_\mu a^\mu W^\nu a_\nu ,
\\
&
\chi_{13} = \wn_\mu a_\nu W^\mu a^\nu,\qquad &
& \chi_{14} = a^2 W_\mu a^\mu,\qquad &
& \chi_{15} = B K_A W^\mu a_\mu,
\\
& \chi_{16} = B K_A \wn_\mu W^\mu,\qquad&
&\chi_{17} = B W^\mu \wn_\mu K_A,\qquad&
&\quad \chi_{18} = K_A W^\mu \wn_\mu B,
\\
&\chi_{19} = \temn K_A a_\mu W_\nu K_S,
\qquad &
& \chi_{20} = \temn K_S K_A \wn_\mu W_\nu,\qquad &
&\chi_{21} = \temn W_\mu K_S \wn_\nu K_A,
\\
& \chi_{22} =  \temn W_\mu K_A \wn_\nu K_S, \qquad &
& \chi_{23} = K_A \tilde K_S^{\mu\nu} a_\mu W_\nu, \qquad &
& \chi_{24} = K_A  \tilde K_S^{\mu\nu} \wn_\mu W_\nu,
\\
& \chi_{25} = K_A \wn_\mu  \tilde K_S^{\mu\nu}  W_\nu, \qquad&
& \chi_{26} = \tilde K_S^{\mu\nu} \wn_\mu   K_A  W_\nu, \qquad &
& \chi_{27} = K_A^2 E_\mu W^\mu,
\\
& \chi_{28} = K_A \temn W_\mu \Lie{n} a_\nu,
\qquad &
& \chi_{29} = K_A \temn a_\nu \Lie{n} W_\mu,
\qquad &
& \chi_{30} = \temn a_\mu W_\nu \Lie{n} K_A,
\\
& \chi_{31} = \temn K_A \Lie{n} \wn_\mu W_\nu, \qquad &
& \chi_{32} = \temn W_\mu \Lie{n} \wn_\nu K_A, \qquad&
&\chi_{33} = \temn \Lie{n} W_\mu  \wn_\nu K_A,
\\
&\chi_{34} = \temn \wn_\mu W_\nu \, \Lie{n} K_A.
\end{alignedat}
\end{equation}
Since this sector contains a larger number of expression it will be more convenient to write the Milne boost transformations explicitly (instead of the matrix $B_{ij}$). These read:
\begin{align*}
&\delw \phi_{1} = \chi_{10} + \chi_{14} - 4 \chi_{19},
\\ & \delw \phi_{2} = \chi_{11} + \chi_{14} - 2 \chi_{19} + 4 \chi_{23},
\\ & \delw \phi_{3} = \chi_{1} + \chi_{3} - 2 \chi_{19} - 6 \chi_{20} + 4 \chi_{21},
\\ & \delw \phi_{4} = \chi_{5} + \chi_{12} + 2 \chi_{19} - 2 \chi_{20} + 2 \chi_{21},
\\ & \delw \phi_{5} =
 \chi_{6} + \chi_{13} + 2 \chi_{19} - 2 \chi_{20} + \chi_{21} + 2 \chi_{23} + 2 \chi_{24} +
  2 \chi_{26} + \chi_{30} - \chi_{34},
\\ & \delw \phi_{6} = \chi_{8} + \chi_{11} + \chi_{13} + 2 \chi_{19} + 2 \chi_{22} + 2 \chi_{30},
\\
\begin{split}
&\delw \phi_{7} =
 \frac{1}{4}  \chi_{3} + \frac{1}{2} \chi_{8} + \frac{1}{2} \chi_{9} + \frac{1}{2} \chi_{10} + \frac{1}{2} \chi_{11} +
  \frac{1}{2} \chi_{12} + \frac{1}{2} \chi_{13} + \frac{1}{2} \chi_{19}
\\ &   ~~~~~~~~~~~
   + \chi_{22} - 3 \chi_{23} + 2 \chi_{25} + \chi_{29},
\end{split}
\\ \numberthis
\begin{split}
&  \delw \phi_{8} =
 \frac{1}{2} \chi_{3} + \chi_{4} + 2 \chi_{6} + \chi_{7} + \chi_{9} - \chi_{19} - 2 \chi_{21} - 4 \chi_{22}
 \\
 & ~~~~~~~~~~~
 +
  2 \chi_{23} - 2 \chi_{28} - 2 \chi_{30} - 2 \chi_{32},
\end{split}
\\
\begin{split}
&
\delw \phi_{9} =
 \frac{1}{2} \chi_{1} + \frac{1}{2} \chi_{2} + \frac{1}{2} \chi_{3} + \chi_{4} + \chi_{5} + \chi_{6} + \chi_{7} + \chi_{8} -
  \chi_{21} - 2 \chi_{22}
\\ & ~~~~~~~~~~~
 + 2 \chi_{23} - 2 \chi_{24} - 4 \chi_{25} - 2 \chi_{26} - \chi_{28} -
  \chi_{29} - 3 \chi_{31} + 2 \chi_{33} - 2 \chi_{34},
\end{split}
\\
&
 \delw \phi_{10} =
 \chi_{11} + \chi_{13} + \chi_{14} + 4 \chi_{23} + 2 \chi_{28} + 2 \chi_{29} - 2 \chi_{30},
\\
\begin{split}
& \delw \phi_{11} =
 \chi_{7} - 2 \chi_{19} + 2 \chi_{20} - 2 \chi_{21} - 2 \chi_{28} - 2 \chi_{29} + 2 \chi_{30}
\\ & ~~~~~~~~~~~
 -
  2 \chi_{31} + 2 \chi_{32} + 2 \chi_{33} - 2 \chi_{34},
\end{split}
\\
 &
  \delw \phi_{12} = -\frac{1}{2} \chi_{3} - \chi_{8} + \chi_{9} - 2 \chi_{18} - \chi_{19} - 2 \chi_{23} -
  2 \chi_{29},
\\ & \delw \phi_{13} = \chi_{15} + 2 \chi_{27} + \chi_{29},
\\ & \delw \phi_{14} = -\chi_{17} - \chi_{33},
\\ & \delw \phi_{15} = \chi_{17} - \chi_{34},
\\ & \delw \phi_{16} = \chi_{18} + \chi_{20} - \chi_{31},
\\ & \delw \phi_{17} = \chi_{15} + \chi_{16} - \chi_{20},
\end{align*}
and we find no boost invariant expressions in this sector.

\subsection{The (1,2,1) Sector}\label{app:NFWB1_2_1}
This sector contains the following independent ghost number zero expressions:
\begin{equation}
\begin{alignedat}{4}
& \phi_1 = \temn a_\mu   \wn_\nu K_S,
\quad &
& \phi_2 = \tilde K^S_{\alpha\beta} a^\alpha a^\beta, \quad &
& \phi_3 = \tilde K_S^{\alpha\beta} \wn_\alpha a_\beta, \quad &
& \phi_4 =  a^\alpha \wn^\beta \tilde K^S_{\alpha\beta},
\\
& \phi_5 = \wn_\alpha \wn_\beta \tilde K_S^{\alpha\beta}, \quad &
& \phi_6 = \temn a_\mu  \Lie{n} a_\nu,
\quad &
& \phi_7 = B a^2, \quad &
& \phi_8 = B \wn_\mu a^\mu,
\\
& \phi_9 = \widetilde \Box B, \quad &
& \phi_{10} = B \widehat R, \quad &
& \phi_{11} =a^\mu \wn_\mu B, \quad &
& \phi_{12} = K_S^2 K_A,
\\
& \phi_{13} = K_S^{\alpha \beta} K^S_{\alpha\beta} K_A, \quad ~ &
& \phi_{14} =  K_A \Lie{n} K_S,
\quad ~ &
& \phi_{15} =  K_S \Lie{n}K_A,
\quad ~ &
& \phi_{16} =  \Lie{n}^2 K_A ,
\quad
\\
& \phi_{17} =  K_A B^2, \quad &
& \phi_{18} = a_\mu E^\mu K_A, \quad &
& \phi_{19} = K_A \wn_\mu E^\mu, \quad &
& \phi_{20} =  E^\mu \wn_\mu K_A.
\end{alignedat}
\end{equation}
The boost-ghost number one expressions read:\begin{align*}
&
\begin{alignedat}{3}
&\chi_1 = \temn a^2 W_\mu a_\nu
, \qquad &
&\chi_2 =\temn \wn_\alpha a^\alpha  W_\mu a_\nu , \qquad &
& \chi_3 =  \temn W_\mu a^\alpha \wn_\nu a_\alpha,
\\
& \chi_4  = \temn a^2  \wn_\mu W_\nu
, \qquad &
& \chi_5 = \temn a^\alpha a_\nu \wn_\mu W_\alpha , \qquad &
& \chi_6 =
\widehat R \temn a_\mu W_\nu,
\\
& \chi_7 =  \temn a^\alpha \wn_{\alpha}\wn_{\mu} W_\nu, \qquad &
& \chi_8 = \temn a_\nu \wn^\alpha \wn_\mu W_\alpha  , \qquad &
& \chi_9 =  \temn \wn_\alpha a^\alpha \wn_\mu W_\nu,
\\
& \chi_{10} = \temn\wn_\mu W^\alpha \wn_{\nu} a_{\alpha} ,\qquad &
& \chi_{11} = \temn W_\mu \wn_\nu \wn_\alpha a^\alpha, \qquad &
& \chi_{12} =  \temn \wn_\mu R W_\nu,
\end{alignedat}
\\
&
\begin{alignedat}{3}
& \chi_{13} =
\widehat R  \temn \wn_\mu W_\nu,
\qquad &
& \chi_{14} = \temn  \widetilde \Box \wn_\mu W_\nu, \qquad &
& \chi_{15} =  \temn K_A B a_\mu W_\nu,
\\
& \chi_{16} =  K_A K_S W_\mu a^\mu,
\qquad &
& \chi_{17} = K_A W_\mu K_S^{\mu\nu} a_\nu, \qquad &
& \chi_{18} =  \Lie{n} K_A W_\mu a^\mu,
\\
& \chi_{19} = K_A W^\mu \Lie{n} a_\mu ,  \qquad &
& \chi_{20} = K_A a^\mu \Lie{n} W_\mu,
\qquad &
& \chi_{21} = \temn K_A B \wn_\mu W_\nu,
\\
&
\chi_{22} = \temn  B W_\mu \wn_\nu K_A, \qquad &
&\chi_{23} =\temn K_A  W_\mu \wn_\nu B,
\qquad &
& \chi_{24} =  K_A K_S \wn_\mu W^\mu,
\end{alignedat}
\numberthis
\\
&
\begin{alignedat}{3}
& \chi_{25} =  K_S W^\mu \wn_\mu K_A ,
\qquad &
& \chi_{26} = K_A W^\mu \wn_\mu K_S
, \qquad &
& \chi_{27} =  K_A W_\nu \wn_\mu K_S^{\mu\nu},
\\
& \chi_{28} =   K_A \wn_\mu W_\nu K_S^{\mu\nu} , \qquad &
& \chi_{29} =  \wn_\mu K_A K_S^{\mu\nu} W_\nu
, \qquad &
& \chi_{30} =   \Lie{n} K_A \wn_\mu W^\mu,
\\
& \chi_{31} =  W^\mu \Lie{n} \wn_\mu K_A,
\qquad &
& \chi_{32} =  \wn^\mu K_A \Lie{n} W_\mu , \qquad &
& \chi_{33} = K_A \Lie{n} \wn_\mu W^\mu,
\\
& \chi_{34} = \temn K_A^2  W_\mu E_\nu.
\end{alignedat}
\end{align*}
The boost transformations are given by:
\begin{align*}
&
\delw \phi_{1} = -\chi_{2} + \chi_{3} - \chi_{5} - \frac{1}{2} \chi_{6} - \chi_{8} +
  2 \chi_{17} - 2 \chi_{20} - 2 \chi_{26},
\\ & \delw \phi_{2} = - \frac{1}{2}\chi_{1}+
 \frac{1}{2}  \chi_{4} - \chi_{5} + 2 \chi_{16} - 4 \chi_{17},
\\ & \delw \phi_{3} =  \frac{1}{2} \chi_{2} - \chi_{3} +  \frac{1}{2}  \chi_{9} - \chi_{10} - 2 \chi_{17} -
  \chi_{18} + \chi_{25} - 2 \chi_{28} - 2 \chi_{29} -
  \chi_{30},
\\
&
\delw \phi_{4} = -  \frac{1}{2} \chi_{2} +  \frac{1}{2} \chi_{4} +  \frac{1}{4} \chi_{6} +  \frac{1}{2} \chi_{7} -  \frac{1}{2} \chi_{8} - \chi_{16} + 3 \chi_{17} + \chi_{18} +
  \chi_{20} + \chi_{26} - 2 \chi_{27},
\\
\begin{split}
& \delw \phi_{5} =  \frac{1}{4}  \chi_{6} + \frac{1}{2} \chi_{7} +  \frac{1}{2}  \chi_{8} + \chi_{9} -  \frac{1}{2} \chi_{11} +
   \frac{1}{2} \chi_{12} +  \frac{1}{2} \chi_{13} +  \frac{1}{2} \chi_{14} - 2 \chi_{16}
   \\&~~~~~~~~~~~~~ +
  3 \chi_{17} - \chi_{18} - \chi_{20} - \chi_{25} -
  4 \chi_{26} + 8 \chi_{27} + 6 \chi_{28} + 2 \chi_{29}
  \\&~~~~~~~~~~~~~+
  2 \chi_{30} + \chi_{31} + 2 \chi_{32} +
  3 \chi_{33},
\end{split}
\\
&
 \delw \phi_{6} = -\chi_{2} + \chi_{3} - \chi_{5} + 4 \chi_{16} -
  4 \chi_{17} + 4 \chi_{18} - 2 \chi_{19} +
  2 \chi_{20},
\\ & \delw \phi_{7} = -\chi_{4} - 4 \chi_{15},
\\ & \delw \phi_{8} =  -\chi_{9} +
  2 \chi_{15} - 2 \chi_{21} + 2 \chi_{22},
\\ & \delw \phi_{9} = -\chi_{14} -   2 \chi_{23},\numberthis
\\ & \delw \phi_{10} =  -\chi_{13} - 2 \chi_{15} - 6 \chi_{21} +
  4 \chi_{22},
\\ & \delw \phi_{11} = -\chi_{7} + 2 \chi_{23},
\\ & \delw \phi_{12} =   2 \chi_{16} + 2 \chi_{24},
\\ & \delw \phi_{13} = 2 \chi_{17} + 2 \chi_{28},
\\ & \delw \phi_{14} =  -2 \chi_{17} + \chi_{19} +
  \chi_{20} + \chi_{26} + \chi_{33},
\\ & \delw \phi_{15} = \chi_{18} + \chi_{25} + \chi_{30},
\\ & \delw \phi_{16} = -\chi_{18} - 2 \chi_{29} +
  2 \chi_{31} + \chi_{32},
\\ & \delw \phi_{17} =  -2 \chi_{21},
\\ & \delw \phi_{18} =  \chi_{15} + \chi_{20} + 2 \chi_{34},
\\ & \delw \phi_{19} = -\chi_{16} + 2 \chi_{17} -
  \chi_{20} + \chi_{21} - \chi_{23} - \chi_{26} + 2 \chi_{27} +
  2 \chi_{28} + \chi_{33} - 2 \chi_{34},
\\ &\delw \phi_{20} = -\chi_{22} + \chi_{32},
\end{align*}
and once again no boost invariant combinations exist in this sector.

\subsection{The (0,4,0) Sector}\label{app:NFWB_4_0_0}
This sector contains the following independent ghost number zero expressions:
\begin{equation}\label{eq:ThePhis}
\begin{alignedat}{3}
&\phi_1 = \widehat R^2, \quad  &
& \phi_2 = \widehat R a^2, \quad  &
& \phi_3 = a^\alpha \wn_\alpha \widehat R,
\\
& \phi_4 = \widehat R \wn_\alpha a^\alpha, \quad &
& \phi_5 = \widetilde \Box \widehat R, \quad&
& \phi_6 = a^4,
\\
&\phi_7 = a^\alpha \wn_\alpha (a^2), \quad&
&\phi_8 = a^2 \wn_\alpha a^\alpha , \quad &
& \phi_9 = (\wn_\alpha a^\alpha)^2, \quad
\\
& \phi_{10} = \wn_{(\alpha} a_{\beta)}\wn^{(\alpha} a^{\beta)}, \quad  &
&\phi_{11} = a^\alpha \wn_\alpha \wn_\beta a^\beta, \quad &
&\phi_{12} = \widetilde\Box \wn_\alpha a^\alpha,
\\
& \phi_{13} = K_A B \widehat R, \quad &
& \phi_{14} = K_A B a^2, \quad &
& \phi_{15} = K_A B \wn_\mu a^\mu,
\\
& \phi_{16} = K_A a^\mu \wn_\mu B , \quad &
& \phi_{17} = B a^\mu \wn_\mu K_A,\quad &
& \phi_{18} = \wn_\mu K_A \wn^\mu B,
\\
& \phi_{19} = K_A \widetilde \Box B, \quad &
& \phi_{20} = B \widetilde \Box K_A,\quad &
& \phi_{21} = \tilde K^S_{\alpha\beta} K_A a^\alpha a^\beta,
\\
&\phi_{22} = K_A a_\mu \temn  \wn_\nu K_S,\quad &
&\phi_{23} = K_S \temn a_\mu \wn_\nu K_A,
\quad &
& \phi_{24} = \tilde K_S^{\alpha\beta} K_A \wn_\alpha a_\beta,
\\
& \phi_{25} = K_A a^\alpha \wn^\beta \tilde K^S_{\alpha\beta}, \quad &
& \phi_{26} = \tilde K_S^{\alpha\beta} a_\alpha \wn_\beta  K_A,
\quad &
& \phi_{27} = \temn \wn_\mu K_A \wn_\nu K_S,
\\
& \phi_{28} = K_A \wn_\alpha \wn_\beta \tilde K_S^{\alpha\beta},
\qquad &
& \phi_{29} = \wn_\alpha K_A \wn_\beta \tilde K_S^{\alpha\beta},\quad &
& \phi_{30} = \wn_\alpha \wn_\beta K_A \tilde K_S^{\alpha\beta},
\\
& \phi_{31} = \temn a_\mu K_A \Lie{n} a_\nu,\quad &
& \phi_{32} = \temn \Lie{n} \wn_\mu K_A a_\nu,
\quad &
& \phi_{33} = \temn \wn_\mu K_A \Lie{n} a_\nu,
\\
& \phi_{34} = K_A^2 \wn_\mu E^\mu, \quad &
& \phi_{35} = K_A E^\mu \wn_\mu K_A, \quad &
& \phi_{36} = a_\mu E^\mu K_A^2,
\\
& \phi_{37} = K_A^2 B^2,
\quad &
&\phi_{38} = K_A \Lie{n}^2 K_A, \quad&
&\phi_{39} = (\Lie{n} K_A)^2,
\\
&\phi_{40} =  K_A^2 \Lie{n} K_S, \quad  &
& \phi_{41} = K_A K_S \Lie{n} K_A, \quad &
&\phi_{42} = K_S^{\alpha \beta} K^S_{\alpha\beta} K_A^2, \\
&\phi_{43} = K_S^2 K_A^2.
\end{alignedat}
\end{equation}
The independent boost-ghost number one expressions in this sector are given by:
\begin{align*}
&
\begin{alignedat}{3}
& \chi_1 = \widehat R K_A \temn W_\mu a_\nu,
\quad &
& \chi_2 = \widehat R K_A \temn \wn_\mu W_\nu, \quad &
& \chi_3 = \widehat R \temn \wn_\mu K_A W_\nu,
\\
&\chi_4  = K_A \temn \wn_\mu R W_\nu, \quad&
& \chi_5 = K_A \temn a^2 W_\mu a_\nu, \quad&
&\chi_6 = K_A \wn_\alpha a^\alpha \temn W_\mu a_\nu,
\\
& \chi_7 = K_A \temn a^\alpha \wn_\alpha W_\mu a_\nu, \quad &
&\chi_8 = \temn a^\alpha \wn_\alpha K_A W_\mu a_\nu, \quad &
& \chi_9 = K_A \temn W_\mu a^\alpha \wn_{(\alpha} a_{\nu)},
\\
&\chi_{10} = \temn a^2 K_A \wn_\nu W_\mu
,\quad&
&\chi_{11} = \temn a^2 W_\mu \wn_\nu K_A, \quad&
& \chi_{12} = K_A \temn a^\alpha \wn_{(\alpha}\wn_{\mu)} W_\nu,
\end{alignedat}
\\
&
\begin{alignedat}{2}
& \chi_{13} = \temn K_A \wn_\alpha \wn^\alpha W_\mu  \, a_\nu, \quad &
& \chi_{14} = \temn W_\mu a^\alpha \wn_{(\alpha} \wn_{\nu)} K_A,
\\
& \chi_{15} = \temn W_\mu a_\nu \wn_\alpha \wn^\alpha K_A, \quad &
& \chi_{16} = \temn K_A W_\mu \wn_\nu \wn_\alpha a^\alpha,
\\
& \chi_{17} = \temn K_A \wn_\alpha a^\alpha \wn_\mu W_\nu, \quad &
& \chi_{18} = \temn K_A \wn_\mu W^\alpha \wn_{(\nu} a_{\alpha)},
\\
& \chi_{19} = \temn a^\alpha \wn_\alpha K_A \wn_\mu W_\nu, \quad &
& \chi_{20} = \temn a^\alpha \wn_\mu W_\alpha \wn_\nu K_A,
\\
&\chi_{21} = \temn \wn^\alpha W_\alpha \wn_\mu K_A \, a_\nu, \quad &
& \chi_{22} = \temn W_\mu \wn_\nu K_A \wn_\alpha a^\alpha ,
\\
& \chi_{23} = \temn W_\nu \wn^\alpha K_A \wn_{(\alpha} a_{\mu)}, \quad &
& \chi_{24} = \temn K_A \widetilde \Box \wn_\mu W_\nu,
\\
& \chi_{25} =  \temn \widetilde \Box \wn_\mu K_A W_\nu, \quad &
& \chi_{26} = \temn \widetilde \Box K_A \wn_\mu W_\nu,
\\
& \chi_{27} = \temn \wn_{(\mu} \wn_{\alpha)} K_A \wn^\alpha W_\nu,\quad &
& \chi_{28} = \temn \wn_\mu K_A \wn_\alpha \wn^\alpha W_\nu,
\\
& \chi_{29} = \temn \wn_\mu K_A \wn_{(\alpha} \wn_{\nu)} W^\alpha , \quad &
& \chi_{30} = K_A^2 B \temn W_\mu a_\nu,
\\
& \chi_{31} = K_A B \temn W_\mu \wn_\nu K_A,\quad &
& \chi_{32} = K_A^2 \temn B \wn_\nu W_\mu,
\end{alignedat}
\numberthis
\\
&
\begin{alignedat}{3}
& \chi_{33} = K_A^2 \temn W_\mu \wn_\nu B,\quad&
& \chi_{34} = K_A \Lie{n} K_A W_\mu a^\mu, \quad&
& \chi_{35} = K_A^2 a^\mu \Lie{n} W_\mu,
\\
&\chi_{36} = K_A^2 W^\mu \Lie{n} a_\mu,\quad&
& \chi_{37} = K_A^2 \Lie{n} \wn_\mu W^\mu, \quad &
& \chi_{38} = K_A W^\mu \Lie{n} \wn_\mu K_A,
\\
&\chi_{39} = W^\mu \Lie{n} K_A \wn_\mu K_A,\quad &
& \chi_{40} = K_A \Lie{n} K_A \wn_\mu W^\mu,\quad&
&\chi_{41} = K_A \wn^\mu K_A \Lie{n} W_\mu,
\\
& \chi_{42} = K_A^2 W_\mu K_S^{\mu\nu} a_\nu, \quad&
& \chi_{43} = K_A^2 K_S W_\mu a^\mu
, \quad&
&\chi_{44} = K_A^2 \wn_\mu W_\nu K_S^{\mu\nu},
\\
&\chi_{45} = K_A^2 K_S \wn_\mu W^\mu ,\quad&
&\chi_{46} = K_A \wn_\mu K_A K_S^{\mu\nu} W_\nu , \quad &
& \chi_{47}  = K_A K_S \wn_\mu K_A W^\mu,
\\
& \chi_{48}  = K_A^2 W_\nu \wn_\mu K_S^{\mu\nu}, \quad &
& \chi_{49} = K_A^2 W^\mu \wn_\mu K_S, \quad &
& \chi_{50} = K_A^3 \temn E_\mu W_\nu.
\end{alignedat}
\end{align*}

Note that, in building this list of independent expressions, one has to take into account dimensionally dependent identities which are derived from the requirement that, for any space tangent tensor in $2+1$ dimensions, the antisymmetrization of 3 or more indexes  vanishes. An example of such an identity is $W^\alpha K_A \temn a_\mu \wn_{(\nu}a_{\alpha)} = \chi_9 - \chi_6$.

The boost transformations of the ghost number zero expressions are given by:
\begin{align*}
\delw \phi_{1}
	= &  4 \chi_{1} - 12 \chi_{2} - 8 \chi_{3},
\\
\delw \phi_{2}
	= &  4 \chi_{1} + 2 \chi_{5} + 6 \chi_{10} + 4 \chi_{11},
\\
\begin{split}
\delw \phi_{3}
	= &  -\frac{6}{4} \chi_{1} - 2 \chi_{4} + 2 \chi_{7} + 2 \chi_{8} + 2 \chi_{9} - 6 \chi_{12} + 4 \chi_{14} - 10 \chi_{19}
\\   &  + 4 \chi_{20} + 6 \chi_{34} + 6 \chi_{35} - 6 \chi_{42} + 2 \chi_{43},
\end{split}
\\
\delw \phi_{4}
	= &  -2 \chi_{1} - 2 \chi_{2} - 2 \chi_{3} + 2 \chi_{6} - 6 \chi_{17} + 4 \chi_{22},
\\
\begin{split}
\delw \phi_{5}
	= &   \chi_{1} + 3 \chi_{3} + 2 \chi_{4} + 2 \chi_{13} + 2 \chi_{15} + 2 \chi_{16} - 4 \chi_{17} + 4 \chi_{18} - 4 \chi_{19} \\
 &  + 4 \chi_{20} + 4 \chi_{21} - 4 \chi_{23} - 6 \chi_{24} -  4 \chi_{25} - 6 \chi_{26} - 8 \chi_{27} - 16 \chi_{28} + 12 \chi_{29}
\\
&  - 4 \chi_{34} +   4 \chi_{36} + 4 \chi_{38} + 4 \chi_{39} + 12 \chi_{40} + 12 \chi_{41} - 4 \chi_{42} +  4 \chi_{45}
\\&
- 12 \chi_{46} + 8 \chi_{47} + 4 \chi_{49},
\end{split}
\\
\delw \phi_{6} = &  8 \chi_{5},
\\
\delw \phi_{7} = &  4 \chi_{7} + 4 \chi_{8} + 8 \chi_{9},
\\
\delw \phi_{8} = &  -2 \chi_{5} + 4 \chi_{6} + 2 \chi_{10} + 2 \chi_{11},
\\
\delw \phi_{9} = &  -4 \chi_{6} - 4 \chi_{17} + 4 \chi_{22},
\\
\delw \phi_{10} = &  -4 \chi_{6} + 4 \chi_{9} - 4 \chi_{17} + 4 \chi_{18} - 4 \chi_{23},
\\
\begin{split}
\delw \phi_{11} = &  -\frac{1}{2} \chi_{1} - 2 \chi_{7} - 2 \chi_{8} - 2 \chi_{9} - 2 \chi_{12}  \\
& + 2 \chi_{14} + 2 \chi_{16} - 4 \chi_{19} + 2 \chi_{20} + 2 \chi_{35} - 2 \chi_{42} - 2 \chi_{43},
\end{split}
\\
\begin{split}
\delw \phi_{12}
	= &  -\chi_{1} + \chi_{3} - 2 \chi_{13} - 2 \chi_{15} - 4 \chi_{16} + 4 \chi_{17} -  4 \chi_{18} + 4 \chi_{19}
	\\&
	- 4 \chi_{20} - 4 \chi_{21} + 4 \chi_{23} - 2 \chi_{24} -  2 \chi_{25} - 2 \chi_{26} - 4 \chi_{27} \\
&   - 6 \chi_{28} + 4 \chi_{29} + 4 \chi_{34} - 4 \chi_{36} - 4 \chi_{38} - 4 \chi_{39}  \\
&  + 4 \chi_{41} + 4 \chi_{42} - 4 \chi_{45} - 4 \chi_{46} - 8 \chi_{47} - 4 \chi_{49},
\end{split}
\\
\delw \phi_{13} = &  -\chi_{2} + 2 \chi_{30} + 4 \chi_{31} + 6 \chi_{32},
\\
\delw \phi_{14} = &  \chi_{10} + 4 \chi_{30},
\\
\delw \phi_{15} = &  -\chi_{17} - 2 \chi_{30} + 2 \chi_{31} + 2 \chi_{32},
\\
\delw \phi_{16} = &  -\frac{1}{4} \chi_{1} - \chi_{12} + 2 \chi_{33} + \chi_{35} - \chi_{42},
\\
\delw \phi_{17} = &  -\chi_{19} + 2 \chi_{31},
\\
\delw \phi_{18} = &  \frac{1}{4} \chi_{3} - \chi_{28} + \chi_{29} + \chi_{41} - \chi_{46},
\\
\delw \phi_{19} = &  -\chi_{24} - 2 \chi_{33}, \\
\delw \phi_{20} = &  -\chi_{26} - 2 \chi_{31}
,
\\
\delw \phi_{21} = &  -\frac{1}{2} \chi_{5} - \chi_{7} + \frac{1}{2} \chi_{10} - 4 \chi_{42} + 2 \chi_{43},
\\
\delw \phi_{22} = &
 \frac{1}{4} \chi_{1} - \chi_{6} - \chi_{7} + \chi_{9} + \chi_{10} - \chi_{12} - \chi_{13} - \chi_{34} -   \chi_{35} + \chi_{42} - \chi_{43} - 2 \chi_{49},
\\
\delw \phi_{23} = &  -\chi_{8} + \chi_{11} - \chi_{21} - 2 \chi_{47},
\\
\delw \phi_{24} = &
\frac{1}{2} \chi_{6} - \chi_{9} + \frac{1}{2} \chi_{17} - \chi_{18} - 2 \chi_{42} + \chi_{43} - 2 \chi_{44} +  \chi_{45} - 2 \chi_{46} + \chi_{47},
\\
\delw \phi_{25} = &  -\frac{1}{4} \chi_{1} - \frac{1}{2} \chi_{6} - \frac{1}{2} \chi_{10} - \frac{1}{2} \chi_{13} + \chi_{34} +   \chi_{35} + 3 \chi_{42} - \chi_{43} - 2 \chi_{48} + \chi_{49},
\\
\delw \phi_{26} = &  - \frac{1}{2} \chi_{11} + \frac{1}{2} \chi_{19} - \chi_{20} - \frac{1}{2} \chi_{21} - 2 \chi_{46} +  \chi_{47},
\\
\delw \phi_{27} = &  - \frac{1}{4} \chi_{3} - \chi_{20} - \chi_{22} - \chi_{23} + \chi_{29} - \chi_{39} -   \chi_{41} + \chi_{46} - \chi_{47},
\\
\numberthis
\begin{split}
\delw \phi_{28} = &   \frac{1}{2} \chi_{2} + \frac{1}{2} \chi_{4} + \chi_{12} + \frac{1}{2} \chi_{13} - \frac{1}{2} \chi_{16} + \chi_{17}  \\
 & + \frac{1}{2} \chi_{24} - \chi_{34} - 2 \chi_{35} + 3 \chi_{37} + \chi_{38} + 2 \chi_{40} + 2 \chi_{41}  \\
 & + 4 \chi_{42} - 2 \chi_{43} + 6 \chi_{44} + 2 \chi_{46} - \chi_{47} + 8 \chi_{48} - 4 \chi_{49},
\end{split}
 \\
\delw \phi_{29} = &   \frac{1}{4} \chi_{3} + \frac{1}{2} \chi_{19} + \frac{1}{2} \chi_{21} - \frac{1}{2} \chi_{22} + \frac{1}{2} \chi_{28} + \chi_{39} +  \chi_{41} + 3 \chi_{46} - \chi_{47},
 \\
\delw \phi_{30} = &  -\chi_{14} + \frac{1}{2} \chi_{15} - \frac{1}{2} \chi_{26} + \chi_{27} - 2 \chi_{46} + \chi_{47},
\\
\delw \phi_{31} = &  -\chi_{6} - \chi_{7} + \chi_{9} + \chi_{10} + 3 \chi_{34} + 2 \chi_{35} -   2 \chi_{36} - 4 \chi_{42} + 3 \chi_{43},\\
\delw \phi_{32} = &  -\chi_{14} + \chi_{15} + \chi_{20} + \chi_{21} - \chi_{34} + 2 \chi_{38},\\
\delw \phi_{33} = &
 \chi_{8} - \chi_{11} - \chi_{20} - \chi_{22} - \chi_{23} + 3 \chi_{39} + 2 \chi_{41} - 4 \chi_{46} +  3 \chi_{47},
 \\
\delw \phi_{34} = &  -\chi_{32} - \chi_{33} - \chi_{35} + \chi_{37} + 2 \chi_{42} - \chi_{43} + 2 \chi_{44} + 2 \chi_{48} - \chi_{49} + 2 \chi_{50},
\\
\delw \phi_{35} = &  -\chi_{31} + \chi_{41},
\\
\delw \phi_{36} = &  -\chi_{30} + \chi_{35} - 2 \chi_{50},\\
\delw \phi_{37} = &  2 \chi_{32},\\
\delw \phi_{38} = &  -\chi_{34} + 2 \chi_{38} + \chi_{41} - 2 \chi_{46},\\
\delw \phi_{39} = &  2 \chi_{39},\\
\delw \phi_{40} = &  \chi_{35} + \chi_{36} + \chi_{37} - 2 \chi_{42} + \chi_{49},\\
\delw \phi_{41} = &  \chi_{34} + \chi_{40} + \chi_{47},\\
\delw \phi_{42} = &  2 \chi_{42} + 2 \chi_{44}
,\\
\delw \phi_{43} = &  2 \chi_{43} + 2 \chi_{45},
\end{align*}

Overall, we find a 4-dimensional space of boost invariant combinations in this sector, as expected from the null reduction arguments outlined in subsection \ref{subsubsec:nullRed}. To allow for an easier comparison with the null reduction, we choose a basis for this space that corresponds to the various (parity even) scalars of the
analogous ($3+1$)-dimensional manifold with $n_D=4$ derivatives. The expressions in this basis are given by:
\begin{equation}
\begin{split}\label{eq:BoostInvariants}
\beta_1^{{}_{R^2}} = &  \phi_{1} - 3 \phi_{2} - 4 \phi_{4} + \frac{9}{4} \phi_{6} + 6 \phi_{8} + 4 \phi_{9} -
 4 \phi_{13} + 6 \phi_{14} + 8 \phi_{15} + 4 \phi_{37},
\\
\beta_2^{{}_{\Box R}} = &  -\frac{3}{2}\phi_{2} + \phi_{3} + \phi_{5} - \frac{3}{2} \phi_{7} - 3 \phi_{10} -  5 \phi_{11} - 2 \phi_{12} - 2 \phi_{16} - 2 \phi_{17} - 4 \phi_{18}
\\
&
-  2 \phi_{19} - 2 \phi_{20} - 6 \phi_{21} + 6 \phi_{22} + 6 \phi_{23} +
 6 \phi_{31} - 6 \phi_{32} - 6 \phi_{39} - 12 \phi_{41} - 6 \phi_{43},
\\
\beta_3^{{}_{W^2}} = &
\phi_{1} + 2 \phi_{4} + \phi_{9} - 16 \phi_{13} - 16 \phi_{15} + 12 \phi_{17}  \\ &
 +12 \phi_{18} + 12 \phi_{23} - 12 \phi_{27} + 24 \phi_{29} + 12 \phi_{33} -
 72 \phi_{35} + 64 \phi_{37} - 36 \phi_{39},
 \\
\beta_4^{{}_{E_4}} = &  \phi_{2} + \phi_{7} - 2 \phi_{8} - 2 \phi_{9} + 2 \phi_{10} + 4 \phi_{14} +
 4 \phi_{15} + 4 \phi_{16} + 4 \phi_{17} + 12 \phi_{21} - 4 \phi_{22}
 \\ &
 - 4 \phi_{23} + 8 \phi_{24} + 8 \phi_{25} + 8 \phi_{26} - 4 \phi_{31} -
 8 \phi_{33} + 8 \phi_{34} + 16 \phi_{35} + 8 \phi_{36}
 \\ &
+ 12 \phi_{39}  - 8 \phi_{40} - 4 \phi_{43},
\end{split}
\end{equation}
where  $ \beta_1^{{}_{R^2}} $ , $ \beta_2^{{}_{\Box R}} $ , $ \beta_3^{{}_{W^2}} $ and $ \beta_4^{{}_{E_4}} $ are proportional to the ($3+1$)-dimensional manifold curvature scalars $ R_{3+1}^2 $ , $ \Box_{3+1} R_{3+1} $ , $ W^2 $ and $ E_4 $ respectively.\footnote{Here, $R_{3+1}$ is the Ricci scalar, $ W^2 $ is the Weyl tensor squared and $ E_4 $ is the Euler  density of the ($3+1$)-dimensional manifold.}

The identification of these combinations with the corresponding ($3+1$)-dimensional scalars was made using the following arguments:
\begin{enumerate}
\item Start by identifying the ($2+1$)-dimensional expression corresponding to the ($3+1$)-dimensional Ricci scalar $ R_{3+1} $. This is a boost invariant expression with $n_D = d_\sigma = 2 $, and therefore belongs to the $(0,2,0)$ sector. It can be easily checked that there is only one independent boost invariant expression in this sector, given by:
\begin{equation}\label{eq:RedRicci}
\beta^{{}_R} = \widehat R - 2\wn_\alpha a^\alpha - 2B K^A -\frac{3}{2} a^2.
\end{equation}
\item $R_{3+1}^2$ is then identified with $ \beta_1^{{}_{R^2}} = (\beta^{{}_R})^2 $.
\item Next, identify $ \Box_{3+1} R_{3+1} $ with the following expression:\footnote{It is easy to check that the operator $ \tb + a^\mu \wn_\mu $ corresponds to the ($3+1$)-dimensional $\Box_{3+1}$, either by noting that $ (\tb + a^\mu \wn_\mu) \phi $ is a total derivative, or using the null reduction directly.}
\begin{equation}
\beta_2^{{}_{\Box R}} = \tb \beta^{{}_R} + a^\mu \wn_\mu \beta^{{}_R},
\end{equation}
where $\beta^{{}_R}$ is given in equation \eqref{eq:RedRicci}.
\item The (3+1)-dimensional Weyl tensor squared $W^2$ is identified with $\beta_3^{{}_{W^2}}$ by noting that it is the only boost-invariant combination which is Weyl invariant (this can be verified using the Weyl transformations detailed later in \eqref{eq:NFWB_0_4_0_WeylTrans}).
\item The ($3+1$)-dimensional Euler density is identified with $\beta_4^{{}_{E_4}}$ by noting that it is the only boost-invariant combination which is both a total derivative and contains no derivatives of order greater than 2.
\end{enumerate}

Next, we turn to the Weyl cohomology itself.
The independent Weyl-ghost number one expressions in this sector read:
\begin{align*}\label{eq:TheAmazingPsis}
&
\begin{alignedat}{3}
&\psi_1 = \widehat R a^\alpha \wn_\alpha \sigma,\qquad&
&\psi_2 = \wn_\alpha \sigma \wn^\alpha \widehat R, \qquad&
&\psi_3 = a^\alpha \wn_\alpha \sigma \wn_\beta a^\beta,
\\
&\psi_4 = a^2 a^\alpha \wn_\alpha \sigma,
\qquad&
&\psi_5 = \wn_\alpha (a^2) \wn^\alpha \sigma, \qquad&
&\psi_6 = \wn_\alpha \sigma \wn^\alpha \wn_\beta a^\beta,
\\
&\psi_7 = \widehat R \tb \sigma, \qquad&
&\psi_8 = \tb \sigma \wn_\alpha a^\alpha, \qquad&
&\psi_9 = \wn_{(\alpha} \wn_{\beta)} \sigma \wn^{(\alpha} a^{\beta)},
\\
& \psi_{10} = \tb \sigma a^2, \qquad&
&\psi_{11} = \wn_{(\alpha} \wn_{\beta)} \sigma a^\alpha a^\beta, \qquad&
&\psi_{12} = a^\alpha \tb \wn_\alpha \sigma,
\end{alignedat}
\\
& \numberthis
\begin{alignedat}{2}
&\psi_{13} =  \tb^2 \sigma, \qquad &
&\psi_{14} = B \wn_\alpha \sigma \wn^\alpha K_A,
\\
&\psi_{15} = K_A \wn_\alpha \sigma \wn^\alpha B, \qquad &
&\psi_{16} = K_A B a^\alpha \wn_\alpha
\sigma,
\\
 & \psi_{17} = K_A B \tb \sigma,&
& \psi_{18} = \tilde K_S^{\alpha\beta} \wn_\alpha \sigma K_A a_\beta,
\\
&\psi_{19} = K_A K_S \teab \wn_\alpha \sigma \,  a_\beta,&
& \psi_{20} = K_A \wn_\alpha \sigma \wn_\beta \tilde K_S^{\alpha\beta},
\\
&\psi_{21} = \tilde K_S^{\alpha \beta} \wn_\alpha \sigma \wn_\beta K_A,&
& \psi_{22} = \teab K_A \wn_\alpha \sigma \wn_\beta K_S,
\\
&\psi_{23} = \te^{\alpha \beta} K_S \wn_\alpha \sigma \wn_\beta K_A, \qquad &
&\psi_{24} = \wn_{(\alpha} \wn_{\beta)} \sigma K_A \tilde K_S^{\alpha \beta},
\\
&\psi_{25} = \teab \Lie{n} \sigma \wn_\alpha K_A a_\beta,&
&\psi_{26} = \teab K_A \wn_\alpha \sigma \Lie{n} a_\beta,
\\
&\psi_{27} = \teab \wn_\alpha \sigma \Lie{n} K_A \, a_\beta,&
&\psi_{28} = \teab \wn_\alpha \sigma \Lie{n}\wn_\beta K_A,
\\
& \psi_{29} = \teab K_A \Lie{n}\wn_\alpha \sigma \, a_\beta,&
& \psi_{30} = \teab \Lie{n}\wn_\alpha \sigma \wn_\beta K_A,
\\
&
\psi_{31} = K_A^2 E^\alpha \wn_\alpha \sigma, \qquad&
&\psi_{32}= K_A \Lie{n} K_A \Lie{n} \sigma,
\\
&
\psi_{33} = K_A^2 \Lie{n}^2 \sigma, \qquad&
&\psi_{34} = K_A^2 K_S \Lie{n} \sigma.
\end{alignedat}
\end{align*}

We define $ I_i = \int \sqrt{-g}\,\sigma \phi_i $ and $ L_i  = \int \sqrt{-g}\,\sigma \psi_i $. We also define the boost invariant integrated ghost number one expressions:
\begin{equation}
\begin{split}
& I_1^{{}_{R^2}} = \int \sqrt{-g}\,\sigma \beta_1^{{}_{R^2}}, \qquad
I_2^{{{}_{\Box R}}} = \int \sqrt{-g}\,\sigma \beta_2^{{{}_{\Box R}}},
\\
& I_3^{{}_{W^2}} = \int \sqrt{-g}\,\sigma \beta_3^{{}_{W^2}}, \qquad
I_4^{{}_{E_4}} = \int \sqrt{-g}\,\sigma \beta_4^{{}_{E_4}}.
\end{split}
\end{equation}
Note that $L_{1-34}$ are not all independent, as they are related using integration by parts via the following formulas:
\begin{align*}
L_{7}  = & -L_1 -L_2,
\\
L_8 = & -L_3-L_6,
\\
L_{9}  = &  -\frac{1}{2} L_{1} - \frac{1}{2} L_{5} - L_{6} - 2 L_{18} + L_{22} + L_{23} +   2 L_{26} + L_{28} ,
\\
 L_{10}  = &  -L_{4} - L_{5} ,
 \\
L_{11}  = &  -L_{3} - L_{4} - L_{5}/2 + 2 L_{19} + 2 L_{27} ,
 \\
 \numberthis
\begin{split}
L_{13}  = &  L_{1} + L_{3} + L_{4} + L_{5} + L_{6} - 2 L_{12} + 4 L_{18}
 \\ &
  -   2 L_{19} - 4 L_{26} - 4 L_{27} - 8 L_{32} - 8 L_{34} ,
\end{split}
  \\
L_{17}  = &  -L_{14} - L_{15} - L_{16} ,
 \\
L_{24}  = &  -L_{18} - L_{20} - L_{21} ,
\\
L_{29}  = & -\frac{1}{2}L_{25} - \frac{1}{2}L_{26} - \frac{1}{2}L_{27} - L_{32} - L_{34}, \\
L_{30}  = &  - \frac{1}{2}L_{28} - L_{32}, \\
L_{33}  = & -2 L_{32} - L_{34}.
\end{align*}

The Weyl transformations of the ghost number zero expressions are given by:
\begin{align*}\label{eq:NFWB_0_4_0_WeylTrans}
&
\begin{alignedat}{1}
\dels \phi_{1} = &  -4 \psi_{7},
\\
\dels \phi_{2} = &  -2 \psi_{10} + 4 \psi_{1},
\\
\dels \phi_{3} = &  -\psi_{1} + 2 \psi_{2} - 2 \psi_{12} + 4 \psi_{18} - 2 \psi_{19} +   4 \psi_{25} + 8 \psi_{29},
  \\
\dels \phi_{4} = &  -2 \psi_{8} + 2 \psi_{7},
\\
\dels \phi_{5} = &  -4 \psi_{2} - 2 \psi_{7} - 2 \psi_{13},
\\
\dels \phi_{6} = & 8 \psi_{4},
\\
\dels \phi_{7} = &  4 \psi_{5} - 2 \psi_{4} + 4 \psi_{11} - 8 \psi_{19} - 8 \psi_{27},
\\
\dels \phi_{8} =  & 4 \psi_{3} + 2 \psi_{10},
\\
\dels \phi_{9} = &  4 \psi_{8}
,\\
\dels \phi_{10} = &  4 \psi_{9} - 2 \psi_{5} + 4 \psi_{27} + 4 \psi_{19} + 2 \psi_{3},\\
\dels \phi_{11} = &
 2 \psi_{6} - 2 \psi_{3} + 2 \psi_{12} - 4 \psi_{25} - 8 \psi_{29} - \psi_{1} -
  4 \psi_{18} + 2 \psi_{19},
\end{alignedat}
\\
\numberthis
&
\begin{alignedat}{2}
& \dels \phi_{12} =   -4 \psi_{6} - 2 \psi_{8} + 2 \psi_{13}, \qquad &
& \dels \phi_{13} =  -2 \psi_{17},
\\
& \dels \phi_{14} =  4 \psi_{16}, \qquad&
& \dels \phi_{15} =  2 \psi_{17},
\\
& \dels \phi_{16} =   2 \psi_{15} - 2 \psi_{16}, \qquad &
&\dels \phi_{17} =  2 \psi_{14},
\\
& \dels \phi_{18} =   -2 \psi_{14}, &
& \dels \phi_{19} =   -4 \psi_{15} - 2 \psi_{17},
\\
& \dels \phi_{20} =  0 , &
& \dels \phi_{21} =  4 \psi_{18},
\\
& \dels \phi_{22} =  2 \psi_{22} + 2 \psi_{19} - 2 \psi_{29}, &
& \dels \phi_{23} =   -2 \psi_{25} + 2 \psi_{23},
\\
& \dels \phi_{24} =   -2 \psi_{18} + 2 \psi_{24}
,&
& \dels \phi_{25} =   2 \psi_{20},
\\
&\dels \phi_{26} =   2 \psi_{21}, &
&\dels \phi_{27} =   2 \psi_{23} - 2 \psi_{30} - 2 \psi_{25},
\\
&\dels \phi_{28} =   -2 \psi_{20}, &
& \dels \phi_{29} =    0,
\end{alignedat}
\\
&
\begin{alignedat}{2}
& \dels \phi_{30} =   -2 \psi_{21}, &
& \dels \phi_{31} =  2 \psi_{26} - 2 \psi_{29},
\\
&\dels \phi_{32} =   -2 \psi_{28},&
& \dels \phi_{33} =  -2 \psi_{30} ,
\\
&\dels \phi_{34} = -2 \psi_{31} ,&
&\dels \phi_{35} = 0 ,
\\
&\dels \phi_{36} =  2 \psi_{31},&
&\dels \phi_{37} =  0 ,
\\
&\dels \phi_{38} =   -2 \psi_{32},&
&\dels \phi_{39} =   0 ,
\\
&\dels \phi_{40} =   -2 \psi_{34} + 2 \psi_{33},\qquad &
&\dels \phi_{41} = 2 \psi_{32} ,
\\
&\dels \phi_{42} =   2 \psi_{34} ,&
&\dels \phi_{43} =   4 \psi_{34}.
\end{alignedat}
\end{align*}

The Weyl transformations of the integrated ghost number one boost invariant expressions are given by:
\begin{equation}
\begin{split}
&\dels I_1^{{}_{R^2}} = 12 L_2 - 18 L_5 - 24 L_6 - 24 L_{14} - 24 L_{15},\\
&\dels I_2^{{}_{\Box R}} = 0, \\
&\dels I_3^{{}_{W^2}} = 0, \\
&\dels I_4^{{}_{E_4}} = 0.
\end{split}
\end{equation}
There are therefore $n_{cc} = 3$ independent cocycles in this sector given by $E_1 = I_2^{{}_{\Box R}}$, $E_2 = I_3^{{}_{W^2}}$ and $E_3 =I_4^{{}_{E_4}} $. This is as expected from the relativistic cohomology in $3+1$ dimensions.

For the calculation of the coboundaries we used the following formulas for integration by parts of the integrated Weyl-ghost number one expressions:
\begin{align*}
\int \sqrt{-g}  \psi_{1} = &  -I_{2} - I_{3} - I_{4} ,
\\
\int \sqrt{-g}  \psi_{2} = &   -I_{3} - I_{5},
\\
\int \sqrt{-g}  \psi_{3} = &  -I_{8} - I_{9} - I_{11} ,
\\
\int \sqrt{-g}  \psi_{4} =  & -I_{6} - I_{7} - I_{8},
\\
\begin{split}
\int \sqrt{-g}  \psi_{5} = &  -(I_{2} + I_{7} + 2 I_{10} + 2 I_{11} + 4 I_{21}
      	- 4 I_{22} - 4 I_{23} - 4 I_{31}
      	\\
      	&~~~~~~~~~~~~~+ 4 I_{32} + 4 I_{39} + 8 I_{41} + 4 I_{43}),
\end{split}
\\
\int \sqrt{-g}  \psi_{6} = &  -I_{11} - I_{12},
\\
\int \sqrt{-g}  \psi_{7} = &  I_{2} + 2 I_{3} + I_{4} + I_{5},
\\
\int \sqrt{-g}  \psi_{8} = &  I_{8} + I_{9} + 2 I_{11} + I_{12},
\\
\begin{split}
\int \sqrt{-g}  \psi_{9} = &   I_{2} + \frac{1}{2} I_{3} + \frac{1}{2} I_{4} + \frac{1}{2} I_{7} + I_{10} + 2 I_{11} + I_{12} 	+ 4 I_{21} - 3 I_{22}
- 3 I_{23} + 2 I_{24}
\\ &
+ 2 I_{25} + 2 I_{26}  - 2 I_{31} + 2 I_{32} - 2 I_{33} - 6 I_{38} - 6 I_{40} - 8 I_{41} - 2 I_{43},
\end{split}
\\
\begin{split}
\int \sqrt{-g}  \psi_{10} = &   I_{2} + I_{6} + 2 I_{7} + I_{8} + 2 I_{10} + 2 I_{11} + 4 I_{21} \\
	& - 4 I_{22} - 4 I_{23} - 4 I_{31} + 4 I_{32} + 4 I_{39} + 8 I_{41} + 4 I_{43},
\end{split}
	\\
\begin{split}
\int \sqrt{-g}  \psi_{11} = &   I_{2}/2 + I_{6} + \frac{3}{2} I_{7} + 2 I_{8} + I_{9} + I_{10} + 2 I_{11} \\
&	+  2 I_{21} - 2 I_{31} - 2 I_{39} - 4 I_{41} - 2 I_{43},
\end{split}
			\numberthis
\\
\begin{split}
\int \sqrt{-g}  \psi_{12} = &  - \frac{3}{2} I_{2} - \frac{1}{2} I_{3} -   \frac{1}{2}I_{4} - I_{6} - 2 I_{7} - 2 I_{8} - I_{9} 	-  2 I_{10} - 4 I_{11} - I_{12}
\\
&	 - 6 I_{21} + 3 I_{22} + 3 I_{23} - 2 I_{24} -  2 I_{25} - 2 I_{26} + 4 I_{31} - 2 I_{32}  \\
&	+ 2 I_{33} + 8 I_{38} +  4 I_{39} + 8 I_{40} + 18 I_{41} + 6 I_{43},
\end{split}
\\
\begin{split}
\int \sqrt{-g}  \psi_{13} = &  I_{2} + I_{6} + 2 I_{7} + 2 I_{8} + I_{9} + 2 I_{10} + 4 I_{11} + I_{12} \\
&	+  4 I_{21} - 4 I_{22} - 4 I_{23} - 4 I_{31} + 4 I_{32} + 4 I_{39} +  8 I_{41} + 4 I_{43},
\end{split}
\\
\int \sqrt{-g}  \psi_{14} = &  -I_{18} - I_{20} - I_{17},
\\
\int \sqrt{-g}  \psi_{15} =  & -I_{19} - I_{18} - I_{16},
\\
\int \sqrt{-g}  \psi_{16} = &  -I_{16} - I_{17} - I_{15} - I_{14},
\\
\int \sqrt{-g}  \psi_{17} = &   I_{15} + 2 I_{16} + 2 I_{17} + 2 I_{18} + I_{19} + I_{20} + I_{14},
\\
\int \sqrt{-g}  \psi_{18} = &  -I_{21} - I_{25} - I_{26} - I_{24},
\\
\int \sqrt{-g}  \psi_{19} =  & I_{23} + I_{22} - 2 I_{41} - 2 I_{43},
\\
\int \sqrt{-g}  \psi_{20} = &  -I_{25} - I_{29} - I_{28}
,\\
\int \sqrt{-g}  \psi_{21} = &  -I_{26} - I_{30} - I_{29},
\\
\int \sqrt{-g}  \psi_{22} = &  -I_{22} - I_{27} - 2 I_{40}
,
\\
\int \sqrt{-g}  \psi_{23} = &  -I_{23} - 2 I_{41} + I_{27},
\\
\int \sqrt{-g}  \psi_{24} = &  I_{21} + I_{24} + 2 I_{25} + 2 I_{26} + I_{28} + 2 I_{29} + I_{30},
\\
\int \sqrt{-g}  \psi_{25} = &  -I_{32} - I_{33},\\
\int \sqrt{-g}  \psi_{26} = &  -I_{33} - 2 I_{38} - 4 I_{41} - 2 I_{40} - 2 I_{43},
\\
\int \sqrt{-g}  \psi_{27} = &  -I_{32} - 2 I_{39} - 2 I_{41}
,
\\
\int \sqrt{-g}  \psi_{28} = &  -2 I_{38} - 2 I_{39} - 2 I_{41},
\\
\int \sqrt{-g}  \psi_{29} = &  I_{32} + I_{33} + 2 I_{38} + 2 I_{39} + 2 I_{40} + 6 I_{41} + 2 I_{43},
\\
\int \sqrt{-g}  \psi_{30} = &  2 I_{38} + 2 I_{39} + 2 I_{41}
,
\\
\int \sqrt{-g}  \psi_{31} =  & -I_{36} - 2 I_{35} - I_{34},
\\
\int \sqrt{-g}  \psi_{32} = &  -I_{41} - I_{39} - I_{38},\\
\int \sqrt{-g}  \psi_{33} = &  2 I_{38} + 2 I_{39} + I_{40} + 4 I_{41} + I_{43},\\
\int \sqrt{-g}  \psi_{34} = &  - I_{40} - 2 I_{41} -I_{43}.
\end{align*}
We find only one ($n_{cb}=1$) independent coboundary which is $F_1 = \dels G_1^{{}_{R^2}} = -12 I_2^{{}_{\Box R}}$ where
$ G_1^{{}_{R^2}}= \int \sqrt{-g}\, \beta_1^{{}_{R^2}}$. We are left with two ($n_{an}=2$) anomalies given by $A^{(0,4,0)}_{{}_{W^2}} = I_3^{{}_{W^2}}$ (which is B-type) and $A^{(0,4,0)}_{{}_{E_4}} = I_4^{{}_{E_4}}$ (which is A-type)
 as expected from the null reduction. The explicit expressions for the anomalies can be found in equation \eqref{Anoamlies_BNF}.

\section{The Case without Frobenius and with No Galilean Boost Invariance}\label{app:NFNB}

In the following appendix we detail the calculations behind the results of the cohomological analysis for the case in which  the Frobenius condition is not satisfied and without Galilean boost invariance.
The calculations are organized according to various sectors as explained in subsection
\ref{subsec:NFNGB}.
In this case we have no gauge field. We have to include $K_A$ since the Frobenius condition is not satisfied.
The equations for classification by sectors  \eqref{rest:constraint1}, \eqref{rest:constraint1b} read:
\begin{equation}
\begin{split}\label{eq:NFNB_class}
& n_T =  n_{K_S} - n_{K_A} +n_\mathcal{L} ,
\\
& n_S =  2 n_{K_A} + n_a + n_\nabla + 2 n_R ,
\\
& 2n_T +n_S = 4.
\end{split}
\end{equation}
Like the previous case, this case also contains an infinite number of sectors. We focus on the sectors with $n_D=n_T+n_S<4$ and the parity even sector with $n_D = 4$.

\subsection{The (2,0,0), (2,0,1) and (1,2,0) Sectors}\label{app:NFNB_2_0_0_2_0_1_1_2_0}
An immediate consequence of equation \eqref{eq:NFNB_class} is that for the sectors with two time derivatives and no space derivatives $ n_{K_A} = 0 $. These sectors are therefore identical to the corresponding sectors in the Lifshitz with Frobenius case, which was analysed in \cite{Arav:2014goa}.
Therefore the sector (2,0,0) contains one anomaly, given by:
\begin{equation}
A_1^{(2,0,0)} = \int \sqrt{-g}\,\sigma \left[ \tr(K_S^2) - \frac{1}{2} K_S^2 \right],
\end{equation}
and the (2,0,1) sector is empty.
The (1,2,0) sector is also unchanged from the Frobenius case as there are no TPD invariant expressions involving $K_A$ in this sector. There are therefore no anomalies in this sector. This conclusion holds for any value of $z$ as explained in \cite{Arav:2014goa}.

\subsection{The (1,2,1) Sector}\label{app:NFNB_1_2_1}

As explained in \cite{Arav:2014goa} the cohomological analysis for sectors with a single time derivative can be performed for a general value of $z$ (at least when the gauge field associated with the particle number is not involved). This is because these sectors satisfy equation \eqref{rest:constraint1b} for any value of $z$, and the independent expressions in these sectors remain the same for all values of $z$. We call these sectors universal. Since the (1,2,1) sector is one of these universal sectors, we keep $z$ as a general parameter in the following analysis.

This sector contains the following ghost number zero expressions:
\begin{equation}
\begin{alignedat}{3}
& \phi_1 = \tilde{K}_S^{\alpha\beta} a_\alpha a_\beta ,\qquad &
& \phi_2 = \wn_\alpha \wn_\beta \tilde{K}_S^{\alpha\beta},\qquad &
& \phi_3 = \teab a_\alpha \Lien a_\beta ,\\
& \phi_4 = \tilde K_S^{\alpha\beta} \wn_\alpha a_\beta, \qquad &
& \phi_5 = \wn_\alpha \tilde K_S^{\alpha\beta} a_\beta, \qquad &
&\phi_6 = \teab \wn_\alpha \tilde K_S a_\beta ,\\
& \phi_7 = K_A \tr( K_S^2 ) ,\qquad &
& \phi_8 = K_A K_S^2 , \qquad &
& \phi_9 = K_A \Lien K_S ,\\
& \phi_{10} = (\Lien K_A) \, K_S ,\qquad &
& \phi_{11} = \Lien^2 K_A.
\end{alignedat}
\end{equation}

The Weyl ghost number one expressions are:
\begin{equation}
\begin{alignedat}{3}
&\psi_1 = \wn_\mu \sigma\, \tilde K_S^{\mu\alpha} a_\alpha, \qquad &
& \psi_2 = \wn_\mu \sigma\, \wn_\alpha \tilde K_S^{\mu\alpha},\qquad &
&\psi_3 = \wn_\mu \sigma\, \te^{\mu\alpha} K_S a_\alpha ,
\\
&\psi_4 = \wn_\mu \sigma\, \te^{\mu\alpha} \wn_\alpha K_S, \qquad &
& \psi_5 = \wn_\mu \sigma\, \te^{\mu\alpha} \Lien a_\alpha,\qquad &
& \psi_6 = \Lien\sigma\, K_A K_S ,
\\
& \psi_7 = \Lien\sigma\, \Lien K_A, \qquad & & \psi_8 = \wn_\mu \wn_\nu \sigma\, \tilde K_S^{\mu\nu}, \qquad &
&\psi_9 = \wn_\mu \Lien\sigma\, \te^{\mu\alpha} a_\alpha ,\\
&\psi_{10} = \Lien^2 \sigma\, K_A.
\end{alignedat}
\end{equation}

We define $ I_i = \int \sqrt{-g}\,\sigma \phi_i $ and $ L_i  = \int \sqrt{-g}\,\sigma \psi_i $.
Note that $L_{8-10}$ are not independent terms, as they are related to $L_{1-7}$ via integration by parts:
\begin{equation}
\begin{split}
& L_8 = -L_1 -L_2 ,\\
& L_9 = -\frac{1}{2} L_5 - L_6 -L_7, \\
& L_{10} = -L_6-L_7.
\end{split}
\end{equation}

The Weyl transformations of the integrated ghost number one expressions are given by $ \delta^W_\sigma I_i = - M_{ij} L_j $ where:
\settowidth{\mycolwd}{$\,-z-2\,$} 
\begin{equation}
M_{ij} = \left(
\begin{array}{*{7}{@{}I{\mycolwd}@{}}}
2z & 0 & 0 & 0 & 0 & 0 & 0 \\
z-2 & -z & 0 & 0 & 0 & 0 & 0 \\
0 & 0 & 0 & 0 & \frac{3}{2}z & z & z \\
-z-2 & -z & 0 & 0 & 0 & 0 & 0 \\
2-z & z & 0 & 0 & 0 & 0 & 0 \\
0 & 0 & -z & -z & -1 & -2 & -2 \\
0 & 0 & 0 & 0 & 0 & 2 & 0 \\
0 & 0 & 0 & 0 & 0 & 4 & 0 \\
0 & 0 & 0 & 0 & 0 & -z-2 & -2 \\
0 & 0 & 0 & 0 & 0 & z-2 & 2 \\
0 & 0 & 0 & 0 & 0 & 2-z & -2
\end{array}
\right).
\end{equation}

There are $n_{cc}=5$ independent cocycles in this sector, given by:
\begin{equation}
\begin{split}
E_1 &= I_2 + I_5 ,\\
E_2 &= I_1 + I_4 + I_5, \\
E_3 &= -2I_7 + I_8, \\
E_4 &= 2I_7 + I_9 + I_{10}, \\
E_5 &= I_{10} + I_{11}.
\end{split}
\end{equation}

For the calculation of the coboundaries we used the following formulas for integration by parts of the integrated Weyl-ghost number one expressions:
\begin{align*}
&\int \sqrt{-g} \psi_1 = -I_1 - I_4 - I_5,\\
&\int \sqrt{-g} \psi_2 = -I_2 -I_5 ,\\
&\int \sqrt{-g} \psi_3 = -I_6 - 2I_8 -2I_{10},\\
&\int \sqrt{-g} \psi_4 = I_6-2I_9 ,\\
&\int \sqrt{-g} \psi_5 = -2I_8-2I_9-4I_{10}-2I_{11},\\ \numberthis
&\int \sqrt{-g} \psi_6 = -I_{8} -I_{9} - I_{10}, \\
&\int \sqrt{-g} \psi_7 = -I_{10} - I_{11},\\
&\int \sqrt{-g} \psi_8 = I_1 + I_2 +I_4 + 2I_5,\\
&\int \sqrt{-g} \psi_9 = 2I_8 +2 I_9+ 4I_{10}+2I_{11},\\
&\int \sqrt{-g} \psi_{10} = I_8+I_9+2I_{10}+I_{11},\\
\end{align*}
The coboundary space is of dimension $ n_{cb}=4 $, and we choose the basis:
\begin{equation}
\begin{split}
F_1 &= \dels G_1 = -2z E_2, \\
F_2 &= \dels G_2 = z E_1 + (2-z) E_2 ,\\
F_3 &= \dels G_3 = -4z E_3 -4z E_4 -4z E_5, \\
F_6 &= \dels G_6 = (2z+4) E_3 + (2z+4) E_4 + 4 E_5,
\end{split}
\end{equation}
where $G_i = \int \sqrt{-g}\, \phi_i$.
We therefore conclude there is $n_{an}=1$ possible anomaly in this sector, which is B-type (trivial descent), and given by (up to coboundary terms):
\begin{equation}
A_1^{(1,2,1)} = -\frac{1}{2} E_3 = \int \sqrt{-g}\, \sigma K_A \left[\tr(K_S^2) - \frac{1}{2} K_S^2 \right].
\end{equation}
Note that this is different from the Frobenius Lifshitz case (see \cite{Arav:2014goa}), where there are no possible anomalies in the corresponding sector.

\subsection{The (0,4,0) Sector}\label{app:NFNB_0_4_0}
For the analysis of this sector we can reuse parts of the analysis of appendix \ref{app:NFWB_4_0_0}. The analysis is similar except we do not restrict to boost invariant expressions and the gauge field is no longer present.
The ghost number zero expressions are given by $\phi_1 - \phi_{12}$, $\phi_{21} - \phi_{33}$ and $\phi_{38}-\phi_{43}$ of equation \eqref{eq:ThePhis}. The Weyl-ghost number one expressions are given by $\psi_1 - \psi_{13}$ $\psi_{18}-\psi_{30}$ and $\psi_{32} - \psi_{34}$ of equation \eqref{eq:TheAmazingPsis}. We keep the same  numbering here. We again use $I_i = \int \sqrt{-g} \sigma \phi_i$ to denote the integrated Weyl-ghost number one expressions.

We find that there are $n_{cc} = 16$ cocycles in the Weyl cohomology given by:
\begin{align*}
E_1 = & \, 2 I_{42} - I_{43} ,\\
E_2 = & \, -I_{38} - I_{41} ,\\
E_3 = & \, -I_{39} ,\\
E_4 = & \, -I_{23} + I_{27} -   I_{33} ,\\
E_5 = & \, -I_{26} - I_{30} ,\\
E_6 = & \, -I_{29}, \\
E_7 = & \, -I_{25} - I_{28} ,\\
E_8 = & \,  I_{1} - I_{2} - 2 I_{3} + I_{4} - I_{5} - I_{8} - 2 I_{11} - I_{12}, \\
E_9 = & \, -I_{1} -   2 I_{4} - I_{9}, \\ 		
		\numberthis
\begin{split}
E_{10} = & \, -2 I_{1} - I_{2} - 4 I_{4} - 2 I_{6} - 3 I_{7} -   2 I_{10} - 4 I_{21}
\\
& + 4 I_{22} + 12 I_{23} - 8 I_{27} + 4 I_{31} -   8 I_{32} - 4 I_{43} ,\\
E_{11} = & \, -2 I_{23} + 2 I_{27} + I_{32} - 2 I_{41} ,
\end{split}
\\
\begin{split}
E_{12} = & \,  2 I_{1} + I_{2} + 4 I_{4} + 2 I_{6} + 3 I_{7} + 2 I_{10}
\\
&
+ 4 I_{21} -   4 I_{22} - 4 I_{23} - 4 I_{31} + 4 I_{32} - 4 I_{40} ,
\end{split}
\\
E_{13} = & \,  I_{21} + I_{24} + I_{25} - I_{30} ,\\
E_{14} = & \, -I_{1} + 2 I_{2} + 2 I_{3} - I_{6} -   I_{7} - I_{12}, \\
E_{15} = & \, I_{1} - I_{2} - I_{3} + I_{4} + I_{6} + I_{7} - I_{11},
  \\
E_{16} =  & \, -I_{6} - I_{7} - I_{8}.
\end{align*}
There are $n_{cb}=12$ independent coboundaries:
\begin{align*}
F_1 =& \ -4 I_{2} - 8 I_{3} - 4 I_{4} - 4 I_{5},
\\
\begin{split}
F_2 = & \, -6 I_{2} - 4 I_{3} - 4 I_{4} -   2 I_{6} - 4 I_{7} - 2 I_{8} - 4 I_{10} - 4 I_{11}
\\
&
- 8 I_{21} + 8 I_{22} +   8 I_{23} + 8 I_{31} - 8 I_{32} - 8 I_{39} - 16 I_{41} - 8 I_{43},
\end{split}
\\
\begin{split}
F_3 = & \, 4 I_{2} + 2 I_{4} - 2 I_{5} + 2 I_{6} + 4 I_{7} + 4 I_{8} + 2 I_{9} +   4 I_{10} + 8 I_{11} \\
&
+ 2 I_{12} + 8 I_{21} - 8 I_{22} - 8 I_{23} -   8 I_{31} + 8 I_{32} + 8 I_{39} + 16 I_{41} + 8 I_{43} ,
\end{split}
\\
F_4 = & \,
-8 I_{6} -   8 I_{7} - 8 I_{8},
\\
\begin{split}
F_5 = & \,
-2 I_{2} + 6 I_{6} + 4 I_{7} + 10 I_{8} + 4 I_{9} -   4 I_{10} - 8 I_{21} + 8 I_{22} + 8 I_{23}
\\
& + 8 I_{31} - 8 I_{32} -   8 I_{39} - 16 I_{41} - 8 I_{43} ,
\end{split}
\\
\numberthis
\begin{split}
F_6 = & \,
 6 I_{2} + 2 I_{3} + 2 I_{4} + 4 I_{7} - 2 I_{8} - 2 I_{9} + 8 I_{10} +
  \\
& 10 I_{11} + 4 I_{12} + 24 I_{21} - 16 I_{22}- 16 I_{23}
  + 8 I_{24}  +   8 I_{25} + 8 I_{26}
   \\   			
& - 16 I_{31} + 12 I_{32} - 8 I_{33}- 24 I_{38} -   24 I_{40} - 32 I_{41} - 8 I_{43},
\end{split}
 \\
F_7 =  & \, -4 I_{21} - 4 I_{24} -   4 I_{25} - 4 I_{26},
\\
F_8 =  & \,  2 I_{23} - 2 I_{27} - 2 I_{32} - 2 I_{33} - 4 I_{38} - 4 I_{39} -   8 I_{40} - 16 I_{41} - 8 I_{43},
\\
F_9 =  & \,
-2 I_{23} + 2 I_{27} + 2 I_{32} +   2 I_{33} - 4 I_{41}, \\
F_{10} = & \,
 4 I_{21} + 4 I_{24} + 6 I_{25} + 6 I_{26} + 2 I_{28} + 4 I_{29} +   2 I_{30} ,\\
F_{11} =  & \, -2 I_{25} - 2 I_{28} - 2 I_{29}, \\
F_{12} =  & \, -2 I_{23} + 2 I_{27} +   2 I_{32} + 2 I_{33} - 4 I_{38} - 4 I_{39} - 8 I_{41}.
\end{align*}
We are therefore left with $n_{an} = 4$  anomalies, all of which are B-type:
\begin{equation}
\begin{split}
A_1 = & \,
I_{42} - \frac{1}{2} I_{43}, \\
A_2 = & \, I_{38} + I_{41} ,\\
A_3 = & \, I_{26} + I_{30}, \\
A_4 = & \, I_1 +2I_4 +I_9.
\end{split}
\end{equation}
The explicit expressions can be found in equation \eqref{NBNFanom_0_4_0}.


\begin{thebibliography}{999}


\bibitem{Arav:2014goa}
  I.~Arav, S.~Chapman and Y.~Oz,
  ``Lifshitz Scale Anomalies'',
  JHEP {\bf 1502}, 078 (2015)
  [\href{http://arxiv.org/abs/1410.5831}{ hep-th:1410.5831}].

\bibitem{Baggio:2011ha}
  M.~Baggio, J.~de Boer and K.~Holsheimer,
  ``Anomalous Breaking of Anisotropic Scaling Symmetry in the Quantum Lifshitz Model'',
  JHEP {\bf 1207}, 099 (2012)
  [\href{http://arxiv.org/abs/arXiv:1112.6416}{arXiv:1112.6416}].



\bibitem{Griffin:2011xs}
  T.~Griffin, P.~Horava and C.~M.~Melby-Thompson,
  ``Conformal Lifshitz Gravity from Holography'',
  JHEP {\bf 1205}, 010 (2012)
  [\href{http://arxiv.org/abs/arXiv:1112.5660}{arXiv:1112.5660}];
  T.~Griffin, P.~Horava and C.~M.~Melby-Thompson,
  ``Lifshitz Gravity for Lifshitz Holography,''
  Phys.\ Rev.\ Lett.\  {\bf 110} (2013) 8,  081602
  [\href{http://arxiv.org/abs/1211.4872v1}{hep-th/1211.4872}].



\bibitem{Jensen:2014hqa}
  K.~Jensen,
  ``Anomalies for Galilean fields'',
  [\href{http://arxiv.org/abs/1412.7750}{hep-th/1412.7750}].



\bibitem{Auzzi}
  R.~Auzzi, S.~Baiguera and G.~Nardelli,
  ``On Newton-Cartan trace anomalies'',
  [\href{http://arxiv.org/abs/1511.08150v2}{hep-th:1511.08150v2}].



\bibitem{Zamolodchikov:1986gt}
  A.~B.~Zamolodchikov,
  ``Irreversibility of the Flux of the Renormalization Group in a 2D Field Theory'',
  [\href{http://www.jetpletters.ac.ru/ps/1413/article_21504.pdf}{JETP Lett.\  {\bf 43}, 730 (1986)}]

\bibitem{Komargodski:2011vj}
  Z.~Komargodski and A.~Schwimmer,
  ``On Renormalization Group Flows in Four Dimensions'',
  JHEP {\bf 1112}, 099 (2011)
  [\href{http://arxiv.org/pdf/1107.3987v2.pdf}{hep-th/1107.3987}].

\bibitem{Dymarsky:2013pqa}
  A.~Dymarsky, Z.~Komargodski, A.~Schwimmer and S.~Theisen,
  ``On Scale and Conformal Invariance in Four Dimensions'',
  JHEP {\bf 1510}, 171 (2015)
  [\href{http://arxiv.org/abs/1309.2921}{hep-th:1309.2921}].


\bibitem{Jensen:2014aia}
  K.~Jensen,
  ``On the coupling of Galilean-invariant field theories to curved spacetime'',
  [\href{http://arxiv.org/abs/1408.6855}{hep-th/1408.6855}].


\bibitem{Jensen:2014wha}
  K.~Jensen and A.~Karch,
  ``Revisiting non-relativistic limits'',
  [\href{http://arxiv.org/abs/1412.2738}{hep-th/1412.2738}].

\bibitem{Son:2005rv}
  D.~T.~Son and M.~Wingate,
  ``General coordinate invariance and conformal invariance in nonrelativistic physics: Unitary Fermi gas'',
  Annals Phys.\  {\bf 321}, 197 (2006)
  [\href{http://xxx.lanl.gov/abs/cond-mat/0509786}{cond-mat/0509786}].

\bibitem{Son:2008ye}
  D.~T.~Son,
  ``Toward an AdS/cold atoms correspondence: A Geometric realization of the Schr\"odinger symmetry'',
  Phys.\ Rev.\ D {\bf 78}, 046003 (2008)
  [\href{http://arxiv.org/abs/0804.3972}{hep-th:0804.3972}].

\bibitem{Son:2013rqa}
  D.~T.~Son,
  ``Newton-Cartan Geometry and the Quantum Hall Effect'',
  [\href{http://arxiv.org/abs/1306.0638}{cond-mat.mes-hall:1306.0638}].

\bibitem{Geracie:2014nka}
  M.~Geracie, D.~T.~Son, C.~Wu and S.~F.~Wu,
  ``Spacetime Symmetries of the Quantum Hall Effect'',
  Phys.\ Rev.\ D {\bf 91}, 045030 (2015)
  [\href{http://arxiv.org/abs/1407.1252}{cond-mat.mes-hall:1407.1252}].

\bibitem{Bergshoeff:2014uea}
  E.~A.~Bergshoeff, J.~Hartong and J.~Rosseel,
  ``Torsional NewtonÐCartan geometry and the Schr\"odinger algebra'',
  Class.\ Quant.\ Grav.\  {\bf 32}, no. 13, 135017 (2015)
  [\href{http://arxiv.org/abs/1409.5555}{hep-th:1409.5555}].

\bibitem{Hartong:2015wxa}
  J.~Hartong, E.~Kiritsis and N.~A.~Obers,
  ``Field Theory on Newton-Cartan Backgrounds and Symmetries of the Lifshitz Vacuum,''
  JHEP {\bf 1508}, 006 (2015)
[\href{http://arxiv.org/abs/1502.00228}{hep-th:1502.00228}].



\bibitem{Frankel}
 T.~Frankel, "The Geometry of Physics: An Introduction", Cambridge University Press 3rd edition (2012).


	
\bibitem{Deser:1993yx}
S.~Deser and A.~Schwimmer,
``Geometric classification of conformal anomalies in arbitrary dimensions'',
Phys.\ Lett.\ B {\bf 309}, 279 (1993)
[\href{http://arxiv.org/abs/hep-th/9302047}{hep-th/9302047}].

\end{thebibliography}
\end{document}